\documentclass[aps,pre,twocolumn,superscriptaddress,showpacs]{revtex4-1}

\usepackage{natbib}
\usepackage[english]{babel}
\usepackage[dvips]{graphics}
\usepackage{graphicx,epsfig}
\usepackage{amsmath}
\usepackage{amssymb}
\usepackage{color}
\usepackage{multirow}
\usepackage[normalem]{ulem}
\usepackage{booktabs}
\usepackage{lineno}
\usepackage{notes2bib}

\newcommand{\with}{\textnormal{with}}

\newcommand{\tLiNCs}{$^s_\pi$}

\newcommand{\tLiNCBu}{$^u_{\pi-0}$}

\newcommand{\tLiNCBIu}{$^u_{\pi-1}$}

\newcommand{\tLiNCBIIu}{$^u_{\pi-2}$}

\newcommand{\tLiNCBIIIu}{$^u_{\pi-3}$}

\newcommand{\tLiCNBu}{$^u_{0-0}$}
\newcommand{\tLiCNBs}{$^s_{0-0}$}

\newcommand{\reacLiCN}{LiNC/LiCN}

\newcommand{\OP}{\textnormal{PO}}
\begin{document}

\title{Semiclassical basis sets for the computation of 
molecular vibrational states}
\author{F. Revuelta}
\email[E--mail address: ]{fabio.revuelta@upm.es}
\affiliation{Grupo de Sistemas Complejos,
Escuela T\'ecnica Superior de Ingenier\'ia Agron\'omica, 
Alimentaria y de Biosistemas, 
Universidad Polit\'ecnica de Madrid,
28040 Madrid, Spain}
\affiliation{Instituto de Ciencias Matem\'aticas (ICMAT),
28049 Cantoblanco--Madrid, Spain}
\author{E. Vergini}
\email[E--mail address: ]{vergini@tandar.cnea.gov.ar}
\affiliation{
Departamento de F\'isica, Comisi\'on Nacional de Energ\'ia At\'omica,
Av.~del Libertador 8250, 1429 Buenos Aires, Argentina}
\author{R. M. Benito}
\email[E--mail address: ]{rosamaria.benito@upm.es}
\affiliation{Grupo de Sistemas Complejos,
Escuela T\'ecnica Superior de Ingenier\'ia Agron\'omica, 
Alimentaria y de Biosistemas, 
Universidad Polit\'ecnica de Madrid,
28040 Madrid, Spain}
\author{F. Borondo}
\email[E--mail address: ]{f.borondo@uam.es}
\affiliation{Instituto de Ciencias Matem\'aticas (ICMAT),
28049 Cantoblanco--Madrid, Spain}
\affiliation{Departamento de Qu\'{\i}mica,
Universidad Aut\'onoma de Madrid,
28049 Cantoblanco--Madrid, Spain}
\date{\today}
%
\begin{abstract}
In this paper, we extend a method recently reported
[\textit{Phys. Rev. E} \textbf{87}, 042921 (2012)] for the 
calculation of the eigestates of classically highly chaotic systems 
to cases of mixed dynamics, 
i.e.~those presenting regular and irregular motions at the same energy.
The efficiency of the method, which is based on the use of a
semiclassical basis set of localized wave functions,
is demonstrated by applying it to the determination of the vibrational
states of a realistic molecular system, namely the LiCN molecule.
\end{abstract}
\maketitle

\section{Introduction} \label{sec.intro}

The quantum description of physical and chemical processes customarily
pivots around the determination of the eigenenergies and  
eigenfunctions of the system.
Except in the particular case of separable Hamiltonians,
one has to resort to numerical computation for this important task,
and numerous procedures have been designed for this end~\cite{Berezin91, 
Marchildon02, Corey93}.
Moreover, this problem is particularly demanding in the classical limit, 
i.e.~$\hbar \rightarrow 0$, where the density of states is high, 
and also in realistic systems, which usually exhibit a classically 
chaotic dynamical behavior even for modest values of the excitation energy. 
When this happens in time--reversal systems, most eigenfunctions present a very 
complex nodal pattern, that can only be adequately described 
by using large basis sets, usually making computations 
extremely time--consuming.

In this respect, semiclassical methods \cite{Brack97} can be very helpful,
both at the computational level and also providing valuable help 
in the understanding of the correspondence between classical and 
quantum mechanics.
These methods are based on the classical underlying properties of the system,
and constitute a cornerstone in the study of classically chaotic systems.
In the presence of chaos, the traditional Wentzel--Kramers--Brillouin~(WKB) 
or Einstein--Brillouin--Keller~(EBK) approximations cannot be applied 
due to the absence of the invariant tori~\cite{Brack97, Gutzwiller90} 
that provide the support for the corresponding wave functions.
Nevertheless, classical periodic orbits (POs) have a profound 
impact on the (quantum) density of states of the system, 
as shown by Gutzwiller in 1971 with his celebrated (semiclassical)
trace formula~\cite{Gutzwiller90}.
Unfortunately, the application of this expression to the calculation 
of highly excited states is very limited, due to the exponential 
proliferation in the number of POs as energy increases.

The importance of unstable POs for some individual
eigenfunctions of classically chaotic systems is clear after
the seminal work of Heller on \textit{scarring}~\cite{Heller84}.
In that paper, the author coined the term \textit{scar}
to refer to an enhanced localization (over the statistically 
expected value~\cite{Shnirelman74})
of the quantum probability in some eigenfunctions 
along periodic trajectories.
Actually, scars are associated with Bohr--Sommerfeld (BS) quantized short POs.
However, this is a necessary but not sufficient condition for their
appearance which, as a consequence, cannot be predicted.
Scars have been studied theoretically in 
quantum billiards \cite{Kwon11},
anharmonic molecular potentials~\cite{Bacic86, Tennyson86, Farantos87, 
Ezra89, Henderson90,  Arranz97,*Arranz98, Arranz10, 
Borondo06, Benitez13, Parraga13, Revuelta15, Henderson92},
or quantum maps~\cite{Casati95}.
Also, they have been observed in the laboratory in
different microwave~\cite{Nockel97, Stockmann06},
optical fibers~\cite{Michel07}, microcavities \cite{Lee02,*Kwak15},
solid state devices~\cite{Wilkinson96}, graphene \cite{Huang09,*Xu13} 
or ultracold atoms experiments~\cite{lar13}.

Several important results on scarring have been reported in
the literature.
For example, Bogomolny demonstrated how scars are in general 
`distributed' among groups of individual eigenfunctions,
and scarred functions can also be produced in the semiclassical 
limit by averaging of a number of neighbor eigenfunctions 
around the BS quantized energies~\cite{Bogomolny88} 
(see also Ref.~\onlinecite{Polavieja94}).
Later, Berry~\cite{Berry89} demonstrated by using Wigner 
functions that this localization does not only take place 
in configuration space but also in phase space.
Prado and Keating~\cite{Prado01} showed that the scarring localization 
is enhanced in the presence of bifurcations in systems with mixed dynamics,
giving rise to the so called \textit{superscars}.
Going beyond the influence of POs in the quantum mechanics of chaotic systems, 
the effect of the recurrences over homoclinic and heteroclinic quantized 
circuits has also been reported in the literature~\cite{Tomsovic93,
*Tomsovic97,*Wisniacki01,*Wisniacki04,*Wisniacki05,*Wisniacki06}.
Finally, scarring in open systems has been described in 
the literature~\cite{Wisniacki08,*Novaes09}.

Several methods have been proposed to construct localized wave
functions over unstable POs, usually known as ``scar functions''.
For example, Polavieja \textit{et al.} averaged groups of eigenfunctions 
by performing a short--time quantum evolution~\cite{Polavieja94},
and Vergini and coworkers~\cite{Vergini00a,*Vergini00b} combined PO resonances 
by minimizing energy dispersion, including then the semiclassical 
dynamics around the scarring PO up to the Ehrenfest time~\cite{Vergini01}.
More recently, Sibert \textit{et al.}~\cite{Sibert08} and 
Revuelta \textit{et al.}~\cite{Revuelta12, Revuelta15}
applied these ideas to systems with smooth potentials, 
and Vagov \textit{et al.}~\cite{Vagov09} extended the asymptotic 
boundary layer method to calculate stable microresonator 
localized modes over unstable POs.

Scar functions have a very interesting and useful property, 
aside from their spatial localization:
they also present a very low dispersion in energy.
We have recently used this fact to construct an extremely efficient 
basis set for the diagonalization of the Hamiltonian matrix
in a coupled quartic oscillator with a high degree of chaoticity.
As demonstrated in Ref.~\onlinecite{Revuelta13},
the system eigenfunctions can be obtained from a very small 
number of scar functions, i.e. POs, thus getting around 
somehow the exponential growth fate of Gutzwiller theory.
This is based in the replacement of the longer POs 
by the interaction of the shorter ones.
This reduces dramatically the basis size, which in our method 
only increases linearly with the number of accurately 
calculated eigenfunctions.

The aim of this paper is to demonstrate the feasibility 
of extending the method reported in Ref.~\onlinecite{Revuelta13}
to systems of chemical interest with a mixed phase space, 
where trajectories with regular and irregular motions
coexist at the same energy. 
For this purpose, we show how to construct an efficient semiclassical 
basis set formed by localized wave functions, using the method 
originally reported in Refs.~\onlinecite{Revuelta12,Revuelta15}
that is used to compute the vibrational eigenstates
of the LiNC/LiCN isomerizing system.

The paper is organized as follows. 
In Sec.~\ref{sec.system}, we introduce the system under study. 
In Sec.~\ref{sec.method} we describe the method that we have developed 
for the computation of the eigenenergies and eigenfunctions of a
system presenting coexisting regions of regular and irregular motion, 
using a basis set of scar wave functions localized along stable 
(the so called ``tube'' functions)
and unstable POs (``scar'' functions). 
Then, in Sec.~\ref{sec.results} we present the results 
that have been obtained and the corresponding discussion. 
Finally, in Sec.~\ref{sec.concl} we summarize the main 
conclusions of this work and the outlook for further research.

\section{System}\label{sec.system}
In this section we briefly describe the characteristics 
of the dynamical system that we have chosen to study,
i.e.~the LiNC/LiCN isomerizing molecule,
that are relevant for this work. 
We first discuss the effective vibrational Hamiltonian
and the potential energy surface of the system in 
Subsec.~\ref{subsec.hamil}. 
Then, Subsec.~\ref{subsec.chaos} is devoted to the
discussion of the dynamical characteristics of the vibrations
of this molecule. 
In particular, we examine the chaoticity of the system as a
function of the energy using Poincar\'e surfaces of section (SOS).
Finally, we conclude the section by presenting in 
Subsec.~\ref{subsec.bd} the bifurcation-continuation diagram 
of the most relevant POs of the system taking the 
excitation energy as parameter. 
These POs will be used later in the construction of a
semiclassical basis set for the computation of the vibrational
eigenstates of the molecule (see discussion in
Sec.~\ref{sec.results} below).

\subsection{Hamiltonian}\label{subsec.hamil}
The system under study is the LiNC/LiCN isomerizing
molecule which has been extensively studied in the past,
especially in connection with quantum chaos 
\cite{Bacic86, Tennyson86, Ezra89, Henderson90, Arranz97,*Arranz98,
Arranz10, Borondo06, GM08, Murgida10, Benitez13, GM14, Murgida15, Revuelta15}.
%
\begin{figure}
\includegraphics[width=\columnwidth]{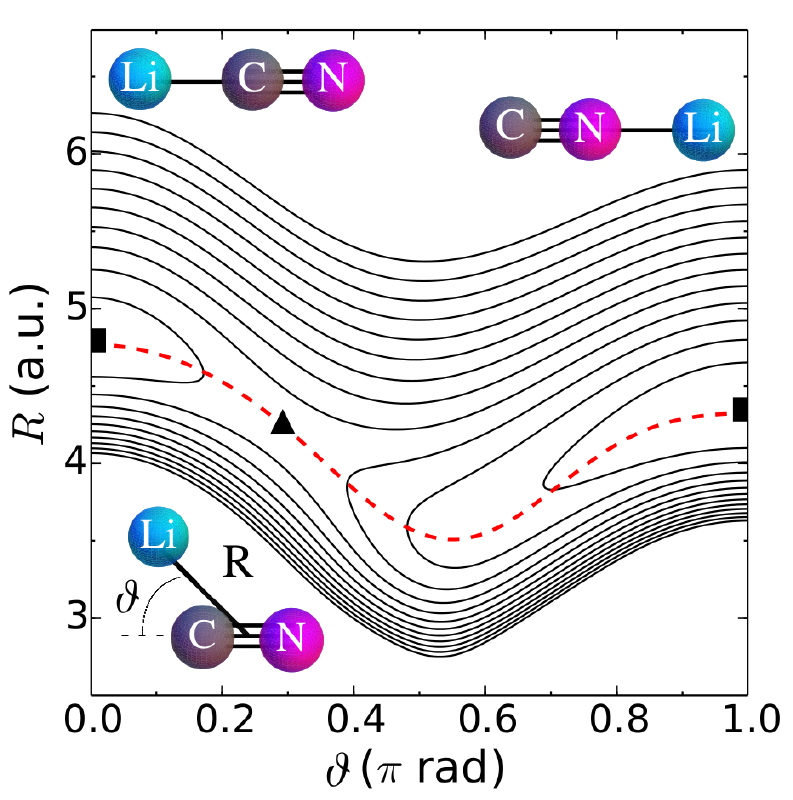}
\caption{Potential energy surface for the LiNC/LiCN
molecular system represented as black contour lines 
separated~1000~cm$^{-1}$ in the Jacobi coordinates defined 
in the inset at the bottom-left corner.
It presents two wells associated to the two existing stable 
linear isomers, LiNC and LiCN.
Their positions are indicated with black squares,
and their geometries sketched in the insets at the top.
The minimum energy path connecting these two wells passing 
through the saddle point, represented as a black triangle,
has been plotted superimposed in dashed red line.
}
\label{fig.1}
\end{figure}
This system exhibits a very floppy motion in the angular coordinate and, 
as a consequence, chaos sets in at very moderate values of the 
excitation energy.

The corresponding vibrational motion can be adequately 
modeled with the following 
rotationless ($J=0$) Hamiltonian
%
\begin{equation}
{\cal H}=\frac{P_R^2}{2\mu_1}+\frac{P_r^2}{2\mu_2}+
\frac{1}{2}\left(\frac{1}{\mu_1 R^2} +
\frac{1}{\mu_2 r^2} \right) P_\vartheta^2 + 
{\cal V}(R,r,\vartheta)
\label{eq.0}
\end{equation}
in Jacobi coordinates, where $R$ and $\vartheta$ describe the 
Li--CN stretching and Li--C--N bending motions, respectively, 
as sketched in the bottom--left corner of Fig.~\ref{fig.1},
while~$r$ accounts for the C--N motion.
The associated reduced masses 
are~$\mu_1~=~m_{\rm Li} m_{\rm CN}/m_{\rm LiCN}$ 
and~$\mu_2=m_{\rm C} m_{\rm N}/m_{\rm CN}$,
with~$m_{\rm LiCN}=m_{\rm Li}+m_{\rm C}+m_{\rm N}$
and~$m_{\rm CN}=m_{\rm C}+m_{\rm N}$.
For all practical purposes, the motion in the~$r$
coordinate plays no role due to the strength
of the C--N triple bound, as reported by some of us
elsewhere~\cite{GM14}.
Thus, one can keep frozen the~$r$ coordinate at
its equilibrium value, $r_e=2.186$~a.u., 
since the associated frequency is very high,
and then decouples very effectively from the rest 
of the modes in the molecule~\cite{GM14}.
Consequently, we can use the following
equivalent two--degrees--of--freedom
Hamiltonian
%
\begin{equation}
{\cal H}=\frac{P_R^2}{2\mu_1}+
\frac{1}{2}\left(\frac{1}{\mu_1 R^2} +
\frac{1}{\mu_2 r_e^2} \right) P_\vartheta^2 + 
V(R,\vartheta) ,
\label{eq.1}
\end{equation}
which still is able to retain all the complexity 
of the molecule under study, thus yielding at the same time
results that are not only qualitative
but also quantitative similar~\cite{GM14}.


The two--dimensional potential energy surface, 
$V (R, \vartheta)$, has been taken from 
the literature~\cite{Essers82}, and it is shown in 
Fig.~\ref{fig.1} as a contours plot. 
Here, we have plotted for simplicity only the fundamental 
domain~$\vartheta \in (0,\pi)$ rad which results from the 
rotational symmetry.
As can be seen, the potential presents two wells at~$\vartheta=0$
and $\pi$ rad, respectively.
They correspond to the two stable linear isomers, LiCN and LiNC,
existing for the molecule; their geometries are sketched
at the top of the figure.
These two isomers are separated by a modest energy barrier of
only~$E_{\rm SP}~\sim$~3454.0~cm$^{-1}$ at the saddle point of 
the potential energy surface, 
where~$(R,\vartheta)_{\rm SP}= (4.22 \mbox{ a.u.}, 0.918 \mbox{ rad})$.
The equilibrium point at the top of this barrier generates at higher 
energies an unstable PO that obviously plays a central role for the 
reactivity of the system~\cite{Revuelta15}.
Finally, the minimum energy path (MEP) connecting the two potential minima 
has been plotted superimposed in the figure as a dashed red line.

\subsection{Chaos in the LiNC/LiCN system}\label{subsec.chaos}

The dynamics of our model for the vibrations of the LiNC/LiCN
molecule can be efficiently monitored by using Poincar\'e SOS, 
taking the MEP, $R_e(\vartheta)$, as the sectioning 
surface~\cite{Ezra89}. 
This choice maximizes the dynamical information obtained 
for the motion in the angular coordinate.
However, this does not define an area preserving map satisfying 
the Louiville theorem~\cite{LL10}.
This inconvenience can be easily overcome by making the following 
canonical transformation
\begin{eqnarray}
    \rho   =&R-R_e(\vartheta), &\quad \psi=\vartheta, \nonumber \\
    P_\rho =&P_R,           &\quad P_\psi=P_\vartheta+ 
                                  P_R[dR_e(\vartheta)/d\vartheta]. 
 \label{eq.PSOS} 
\end{eqnarray}

\begin{figure}
  \includegraphics[width=\columnwidth]{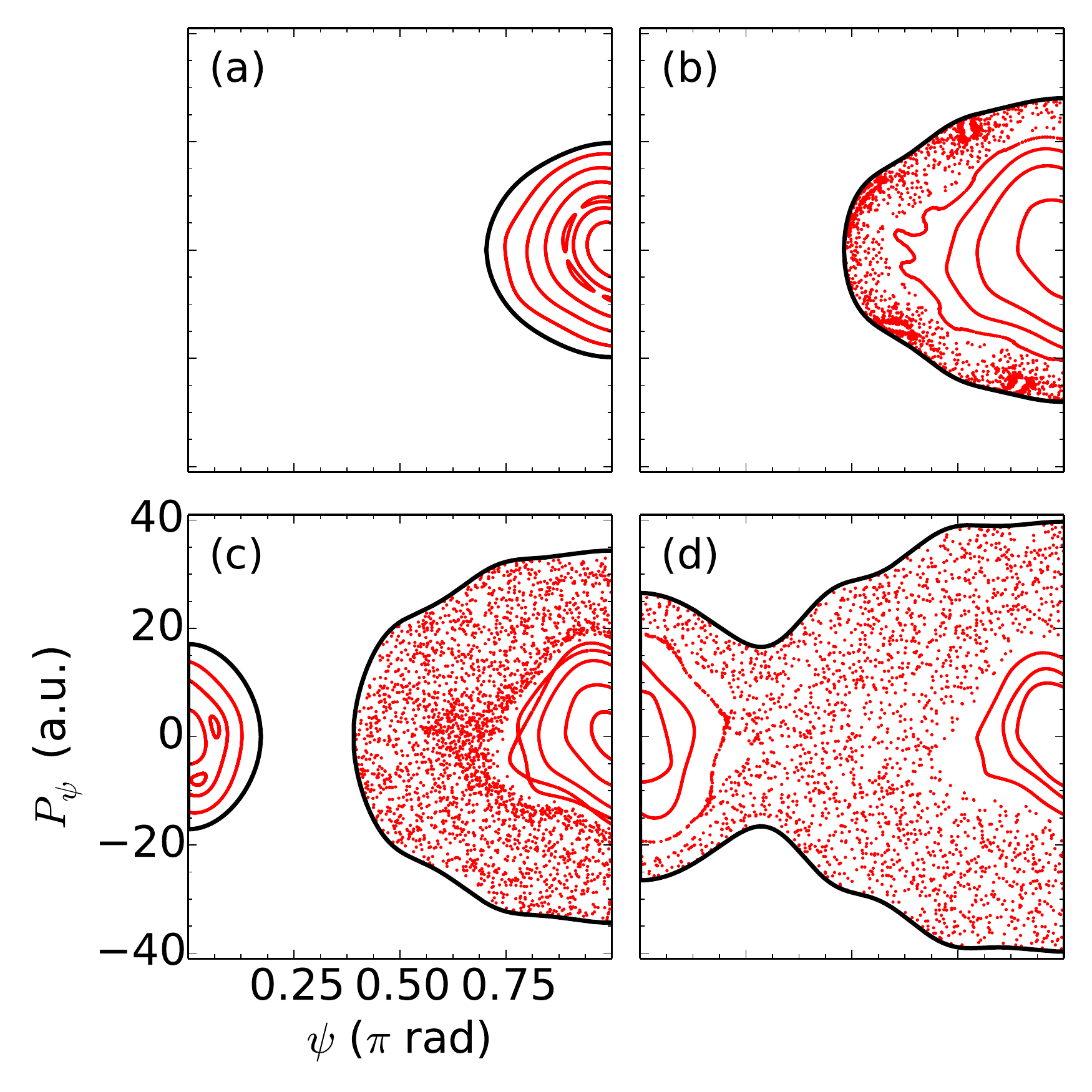}
  \caption{Composite Poincar\'e surface of sections for the 
  LiNC/LiCN vibrational dynamics computed along the minimum energy path 
  shown in Fig.~\ref{fig.1}, i.e.~$\rho=0$ 
  [see Eq.~\eqref{eq.PSOS} at 
  different values of the excitation energy:
      (a)~1000~cm$^{-1}$,
      (b)~2000~cm$^{-1}$,
      (c)~3000~cm$^{-1}$, and
      (d)~4000~cm$^{-1}$.}
      \label{fig.2}
\end{figure}

Some representative results, computed by numerically solving the equations 
of motion derived from Hamiltonian~\eqref{eq.1}, using the Shampine and
Gordon algorithm~\cite{Shampine75}, for different values of the excitation 
energy, $E$, are shown in Fig.~\ref{fig.2}. 
As can be seen, the chaoticity of the system increases with the energy.
At low energies, for example $E$=1000 cm$^{-1}$ as chosen 
in Fig.~\ref{fig.2}~(a), the vibrational motion takes place 
in the LiNC well and it is regular, being then confined in invariant tori.
As higher energies are considered, e.g. panels (b) and (c), 
the invariant tori progressively start to break down, this paving the 
road for widespread chaotic motion, as dictated by the celebrated
Kolmogorov--Arnold--Moser (KAM) theorem~\cite{Arnold78}.
Comparison of results in panels~(b) and~(c) clearly indicates that 
the dynamics in the LiNC well gets increasingly more chaotic as the
excitation energy grows.
In panel (c), which corresponds to an energy above the level of the 
less stable LiCN minimum well, 
motion also takes place in that region of the phase space.
Also, a conspicuous accumulation of points next to the LiNC regular
regions is observed.
This is due to the existence a cantorus, as thoughrouly discussed 
in Refs.~\onlinecite{Zembekov95,*Zembekov96,*Zembekov97}.
At even higher energies, we end up being above the 
PES saddle energy, i.e.~panel~(d), the two isomer wells are then connected,
this allowing classical isomerization dynamics.
%

\subsection{Periodic orbits for LiNC/LiCN and the 
bifurcation--continuation diagram}\label{subsec.bd}
\begin{figure}
\includegraphics[width=0.97\columnwidth]{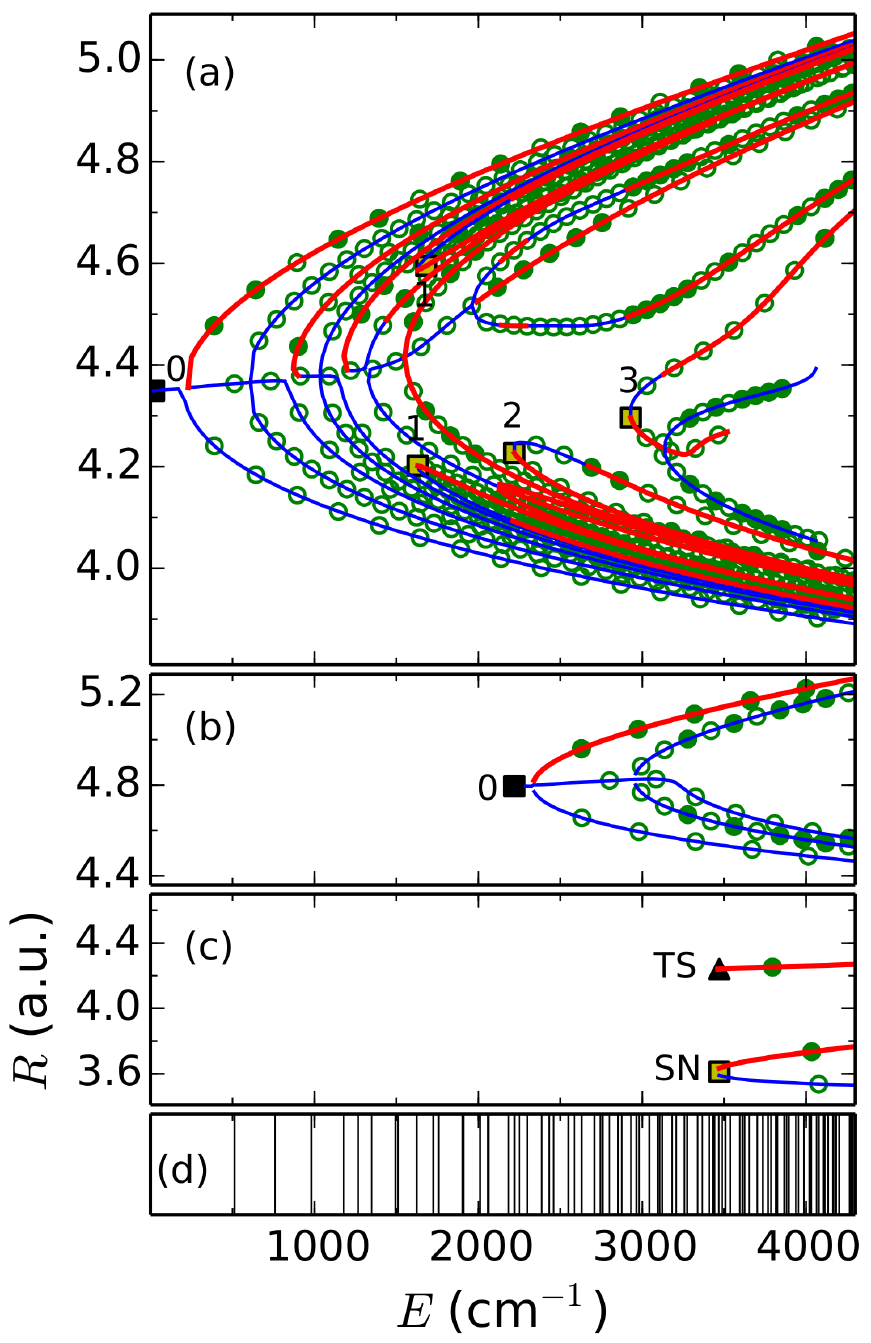}
\caption{Bifurcation--continuation diagram of 
periodic orbits (POs) for the LiNC/LiCN isomerizing system.\\
Panels (a)--(c): Bifurcation--continuation diagrams for:
(a) POs in the LiNC isomer well,
(b) same for LiCN, and
(c) POs ``born'' both in the saddle--node bifurcation discussed 
    in Ref.~\onlinecite{Zembekov95,*Zembekov96,*Zembekov97} 
    (lower double red--blue line),
    and in the potential energy surface saddle 
    (upper single red line). \\
From top to bottom at the highest represented energy of
$E$=4300 cm$^{-1}$, and in the notation used in Fig.~\ref{fig.3} 
and throughout the text:\\
(a)~1A$_{\pi-0}$, 2AB$_{\pi-0}$, 3A$_{\pi-0}$, 4AB$_{\pi-0}$, 5A$_{\pi-0}$, 1AB$_{\pi-1}$,
    1BA$_{\pi-1}$, 6A$_{\pi-0}$, 7AB$_{\pi-0}$, 8AB$_{\pi-0}$, 0$_{\pi-0}$, 8AB$_{\pi-0}$, 
    1A$_{\pi-3}$, 2AB$_{\pi-3}$, 1B$_{\pi-3}$, 2AB$_{\pi-3}$, , 2AB$_{\pi-2}$, 2AB$_{\pi-2}$
    7AB$_{\pi-0}$, 9AB$_{\pi-0}$, 6B$_{\pi-0}$, 9AB$_{\pi-0}$, 6B$_{\pi-0}$, 9AB$_{\pi-0}$, 
    1BA$_{\pi-1}$, 1AB$_{\pi-1}$, 5B$_{\pi-0}$, 4AB$_{\pi-0}$, 3B$_{\pi-0}$, 2AB$_{\pi-0}$,
    and 1B$_{\pi-0}$,\\
(b) 1A$_{0-0}$, 2AB$_{0-0}$, 0$_{0-0}$, 2AB$_{0-0}$, and 1B$_{0-0}$, and \\
(c) TS$^u$, SN$^u$, and SN$^s$. \\
 Thin blue lines indicate stable POs, while unstable POs are 
 referenced by thick red lines. 
 The saddle point has been marked in black triangle,
 the two potential minima in black squares,
 and  the lowest--lying bifurcation point of other
 important families of POs in yellow squares.
 The empty green circles represent the Bohr--Sommerfeld quantized 
 energies determined by Eq.~\eqref{eq.4}.
 The energies of the localized states selected for the construction 
 of the basis set have been highlighted in filled circles
(see discussion in Sec.~\ref{sec.results}. \\
Panel (d): Quantum eigenenergies for the LiNC/LiCN system.
}
\label{fig.3}
\end{figure}

Figure~\ref{fig.3} shows the bifurcation--continuation diagram
with the most relevant POs in the LiNC~[panel (a)] and 
LiCN~[panel (b)] wells, the transition state (TS) at the PES saddle, 
and also those ``born'' in the saddle--node or tangent bifurcation
discussed in Ref.~\onlinecite{Zembekov95,*Zembekov96,*Zembekov97}~[panel (c)].
The POs are characterized in this plots by the initial values
of their strecht coordinate~$R$ as a function of the energy.
Thin blue and thick red lines indicate, respectively, the 
stability and instability of the corresponding orbits.
As can be seen, the number of POs increases with energy 
due to the different biffurcations taking place.
In panels~(a)--(c) only the POs that are symmetryic with
respect to the $\vartheta=\pi$ and 0 rad lines, 
i.e.~isomers LiNC and LiCN, are considered.
We have also highlighted in the figure with empty green 
circles the position of the quantized trajectories, 
i.e. the POs that fullfill the BS rule discussed below 
in Sec.~\ref{subsec.tube}.
Moreover, those that will be used in our construction 
of a basis set for the system have been indicated
with filled green circles (see Subsec.~\ref{subsec.GS}
below).
Notice also how the density of the states of the system
increases with the excitation energy, 
as emphasized in the plot in the bottom panel~(d),
where the quantum energies of the system are represented.

The POs in Fig.~\ref{fig.3}~(a) and~(b) 
have been labeled as ``N X$_{\textnormal{Y-Z}}^\textnormal{W}$'',
~N being an integer identifying the bifurcation at which they first 
appear (in all orbits considered N=1, 2, 3).
Letter~X identifies the branch in the bifurcation diagram, 
being for librations or time--reversal POs
X=A associated with the upper branch 
and X=B with the lower one;
the rotations, i.e. POs that have no time--reversal symmetry, 
and then correspond to both (upper and lower) branches, 
are labeled as X=AB/BA.
The~Y subindex indicates the well where the PO is located:
Y=0 for POs associated with the LiCN isomer, and
Y=$\pi$ for POs of the LiNC isomer. 
Subindex Z=$0, 1, 2, \ldots$ is an integer indicating
the bifurcation where the first PO appears.
The stable/unstable charater of the PO is indicated 
by~W=s/u (for stable or unstable, respectively).
The POs of panel (c) have been labelled as ``TS'' in the case of 
the trajectory located in the neighborhood of the TS or activated 
complex at the PES saddle point, 
and as ``SN$^{s}$'' (``SN$^{u}$'') for the case of the 
stable (unstable) POs ``born'' in the tangent 
bifurcation~\cite{Zembekov95,*Zembekov96,*Zembekov97}.

\begin{figure}
\includegraphics[width=0.96 \columnwidth]{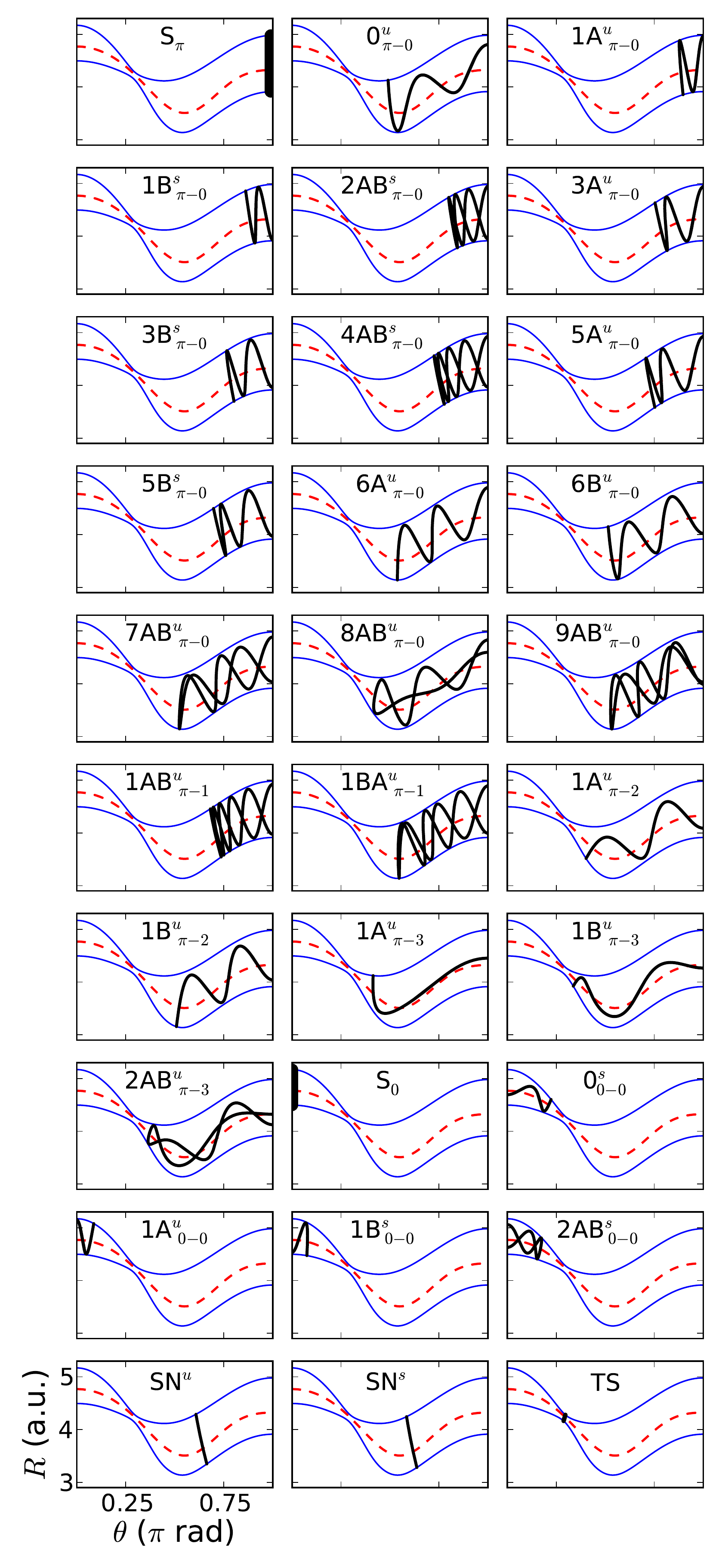}
  \caption{Periodic orbits (black thick lines) of LiNC/LiCN 
  molecular system included in the bifurcation--continuation 
  diagram of Fig.~\ref{fig.3}.
  The minimum energy path and the equipotential 
  lines at~3500~cm$^{-1}$ have been superimposed in 
  dashed red and continuous blue lines, respectively.
}
\label{fig.4}
\end{figure}

All trajectories introduced in Fig.~\ref{fig.3} are presented 
in Fig.~\ref{fig.4} at a particular value of the energy, 
actually $E=$3500~cm$^{-1}$. 
In this figure, we have also included the POs corresponding to the
stretch modes associated to purely vibrational motion of~$R$ in both wells,
which are always stable.
We have labeled them as~S$_\textnormal{Y}$'', where the subindex Y 
indicates again the well where the PO is localized 
(Y=0 for LiCN and $\pi$ for LiNC).
%

\section{Method}\label{sec.method}
In this section we describe the method that we have developed 
for the construction of an efficient semiclassical basis set.
The section is divided in two parts.
First, we describe in Subsec.~\ref{subsec.tubescar}
how to compute localized wave functions along POs. 
Second, in Subsec.~\ref{subsec.GS} we discuss
how the previous localized wave functions are
selected for the construction of our basis set.

\subsection{Computation of localized wave functions}
  \label{subsec.tubescar}
In this subsection, we briefly describe the
method to construct localized wave functions along POs. 
For this purpouse, we distinguish between two different
kinds of states depending on whether the PO is stable or unstable:
for stable POs we will use the so called ``tube'' 
wave functions described in Subsec.~\ref{subsec.tube}, 
while for unstable POs we will construct the ``scar'' 
wave functions that are presented in Subsec.~\ref{subsec.scar}.
More details can be found in 
Refs.~\onlinecite{Revuelta12,Revuelta13,Revuelta15}.

\subsubsection{The ``tube'' wave functions}
  \label{subsec.tube}
Our ``tube'' wave functions are defined as 
%
\begin{equation}
  \psi^{\rm{tube}}_n (R,\vartheta) = \int_0^T \; dt \;
   e^{-iE_n t/\hbar} \phi(R,\vartheta,t),
\label{eq.2}
\end{equation}
where $T$ is the period of the PO, and $E_n$ the corresponding BS
quantized energy (see discussion below).
As can be seen, it consists of a time average of a 
suitably defined wave packet $\phi(R,\vartheta,t)$, whose dynamics
is forced to stay in the neighborhood of the ``scarring'' PO.
This dynamics, given at time~$t$ by the phase space point 
$(R_t,\vartheta_t,P_{R,t},P_{\vartheta,t})$,
is assumed to be that of a frozen Gaussian~\cite{Heller76, Littlejohn86}
centered on the trajectory as
%
%
\begin{equation}
  \begin{array}{rcl}
   \phi(R,\vartheta,t) & = & \exp \{-\alpha_R(R-R_t)^2-\alpha_\vartheta
                             (\vartheta-\vartheta_t)^2 + \\
   \displaystyle       & &  \frac{i}{\hbar} \left[P_{R,t} (R-R_t)+
                             P_{\vartheta,t} (\vartheta-\vartheta_t)\right]
                             + i \gamma_t \} .
  \end{array}
 \label{eq.3}
\end{equation}
%
%
Here, we take $\alpha_R=16.114$ a.u.$^{-2}$ and $\alpha_\vartheta=14.123$~rad$^{-2}$ 
that approximately coincide with the ``width'' in configuration 
space of the LiNC quantum ground state.
The time function $\gamma_t=S_t/\hbar - \mu_t \pi/2$ is the phase 
accumulated during the propagation, which is actually the sum of two terms, 
a first one of dynamical origin given by  
$S_t/\hbar=\int_0^t d\tau \; (P_{R,\tau} \dot{R}_{\tau} + 
P_{\vartheta,\tau} \dot{\vartheta_{\tau}})/\hbar$, 
and a second contribution proportional to~$\mu_t$, 
which equals the number of half turns that the neighbouring 
trajectories describe around the scarring PO.
This second term, which is always more complicated to compute, 
can be evaluated by using a set of \emph{local} coordinates along 
the PO and studying the time evolution of the corresponding 
transversal stability matrix~\cite{Eckhardt91}.
It should be noticed that $\mu_t$ is not a canonical invariant,
and as a result its magnitude depends on the definition chosen 
for the angle swept by the manifolds.
Very often, only the value of this magnitude after a full period of the PO,
$\mu_T$, usually known as the \emph{winding number} is needed.
In this case, the function is canonically invariant and is equal, 
for unstable POs, to the Maslov~\cite{Maslov91}
index appearing in Gutzwiller's trace formula~\cite{Creagh90, Robbins91}.
More importantly, the required phase becomes much easier to calculate,
since it is simply equal to $\pi/2$ times the number of turning points
plus self-conjugated points in the PO.

  \label{subsec.bs}
In order to maximize the localization along the PO, 
the tube functions are defined at the energies, $E_n$, 
fullfilling  the BS quantization rule 
%
\begin{equation}
  \gamma = \frac{S(E_n)}{\hbar}-\mu \frac{\pi}{2}
         = 2\pi n, \qquad n=0,1,2,\ldots,
 \label{eq.4}
\end{equation}
where $n$ is an integer number giving the number of nodes in the wave 
function along the PO, and $\gamma, S$, and $\mu$ being defined over 
one period of the PO, i.e. $\gamma=\gamma_T, S=S_T$, and $\mu=\mu_T$.

%
Notice that many orbits of Fig.~\ref{fig.4} are symmetric respect
to $\vartheta =0$ or $\vartheta= \pi$ rad, while the considered wave
functions are symmetric with respect to these values. This means
that the tube functions associated with symmetric POs
have an even number of excitations, i.~e.~$n$ is even. Thus,
in order to simplify notation, the~$n$ number used to identify
these tube functions equals half the number of excitations.
%
\subsubsection{The ``scar'' wave functions}\label{subsec.scar}
The tube functions introduced in Eq.~\eqref{eq.2} 
can be constructed both over stable or unstable POs. 
However, in the latter case it is convenient to introduce 
an improvement by defining what we call ``scar'' functions 
which incorporate short time dynamical information 
on the homoclinic structure of the PO invariant manifolds~\cite{Vergini00a,*Vergini00b}.

These scar functions are computed by first propagating the 
corresponding tube wave functions and then performing a 
finite--time Fourier transformation at the BS quantized energies,
in the following way
%
\begin{eqnarray}
\psi^{\rm scar}_n (R,\vartheta) = 
     \qquad \qquad \qquad \qquad \qquad \qquad \qquad \qquad \nonumber& \\
      \int_{-T_E}^{+T_E} dt\;
      \cos \left(\frac{\pi t}{2T_E} \right) \;
       e^{-i (\hat{\cal H}-E_n)t/\hbar} \; \psi^{\rm tube}_n(R,\vartheta), \quad &
 \label{eq.5}
\end{eqnarray}
%
where 
\begin{equation}
   T_E=\frac{1}{2\lambda} \ln \left( \frac{A}{\hbar} \right)
   \label{eq.Te}
\end{equation}
is the so--called Ehrenfest time, which can be only defined for unstable POs
and depends on two parameters:
the stability exponent of the PO~\cite{LL10},~$\lambda$,
and the area of a characteristic SOS,~$A$.
This time can be (semiclassically) understood as the lapse of time 
that a Gaussian wave packet needs to spread over this characteristic
Poincar\'e SOS area of the system.
Also, a cosine window is used in the definition (\ref{eq.5}) in order
to minimize the dispersion in energy of the scar functions~\cite{Vergini08}.
Among other methods, wavelets provide an efficient method to
perform the time evolution appearing in Eq.~\eqref{eq.5},
with a precision of at least six decimal places~\cite{Sparks06}.

%
\begin{figure}
\includegraphics[width=\columnwidth]{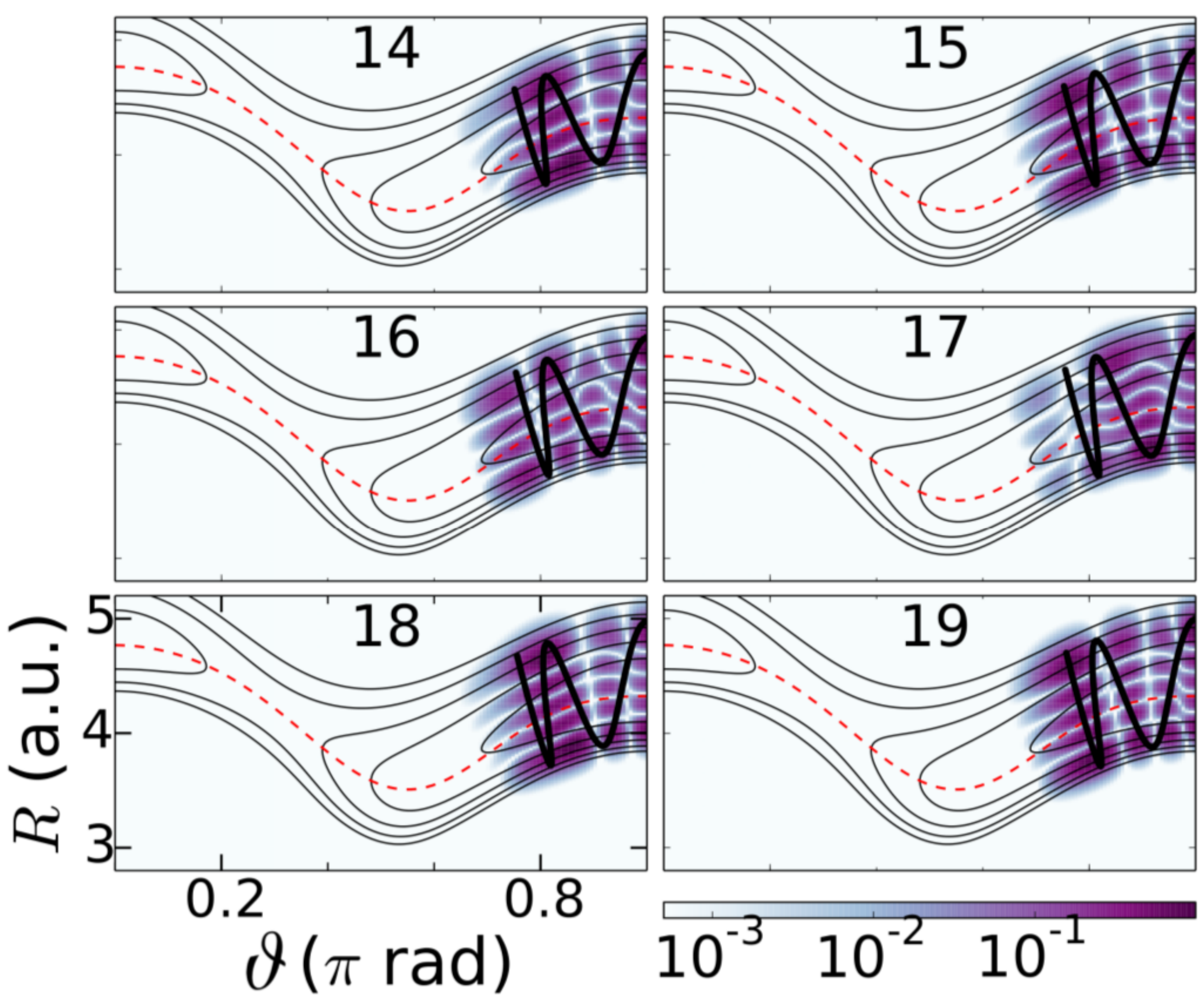}
\caption{Some examples of scar functions~\eqref{eq.3}
 for LiNC/LiCN.
The unstable periodic orbit 3A$^u$ (think black line),
the minimum energy path (dashed red line),
and the contour plots of the 
potential energy surface have been plotted superimposed.
The number in the center of each panel gives the 
integer appearing in Bohr--Sommerfeld quantization 
rule~\eqref{eq.4}.
}
\label{fig.5}
\end{figure}
Figure~\ref{fig.5} shows some examples of very highly excited scar
functions along the quantized unstable POs 3A$^u$ of Fig.~\ref{fig.4},
corresponding the quantum numbers $n=14-19$
and BS energies between~3091.48~cm$^{-1}$ and~3982.40~cm$^{-1}$.
In all examples shown in the figure the characteristic area, $A$, 
appearing in Eq.~\eqref{eq.Te} has been 
estimated as the integral $\int dR \; P_R$ computed along the 
line $\vartheta=\pi$~rad at the quantized excitation energy,
since the PO lives in the vicinity of that region.
All these functions, as well as any other throughout the paper, 
have been computed by setting~$\hbar~=~1$~a.u.
%
%
As can be seen, the probability density is not well
localized over the PO because of the complex topology of the 
trajectory, this somehow reducing the scarring phenomenon 
due to the quantum dynamics implied by finite~$\hbar$.
Also, notice that these functions are both excited in the
in the~$R$ and~$\vartheta$ directions, and seem to have a rather
simple pattern~\cite{Revuelta15}.
Consequently, one can easily adscribe quantum numbers
accounting for the number of excitations (or nodes) in each direction.
For example, the scar functions shown at the top row of Fig.~\ref{fig.5},
labeled as~14 and~15,
correspond to states with~3 excitations in each direction,
and then~$(n_R, n_\vartheta) = (3, 3)$.
The two scar functions presented in the middle row have different
quantum numbers: while the one on the left (labeled as 16) 
has~$(n_R, n_\vartheta) = (4, 3)$, the one on the right (17) 
corresponds to~$(n_R, n_\vartheta) = (2, 4)$.
Finally, the scar functions (18 and 19) shown in the bottom row are
associated with~$(n_R, n_\vartheta) = (2, 4)$ and~$(n_R, n_\vartheta) 
= (3, 4)$, respectively.
Nevertheless, it is simpler to label this functions as we have done by 
counting the number of nodes that they have along the (desymmetrized) PO.
This number~$n = 14 - 19$, which has also been shown in each panel, 
equals the integer fullfilling the BS rule~\eqref{eq.4}, for~$\mu=16$.

\subsection{Selective Gram-Schmidt method for the construction 
  of the basis set}
\label{subsec.GS}
As energy increases, the exponential proliferation 
of classical POs in floppy molecules leads to a
dramatic increment in the number of quantized POs
over which our localized wave functions can be defined.
Thus, a selective procedure for the best suited localized states
must be developed in order to construct an efficient basis set
for the computation of vibrational states of this kind of 
systems that keeps the eigenvalue problem at moderate sizes.
Afterwrads, the Hamiltonian matrix associated with Eq.~\eqref{eq.1}
can be computed, and then diagonalized using standard procedures.

This subsection is divided in two parts. 
First, Subsec.~\ref{subsec.basisset} describes the algorithm 
developed for the construction of our basis set, 
which is called \emph{selective Gram-Schmidt method (SGSM)}.
Second, we discuss in Subsec.~\ref{subsec.local} the procedure that
we have developed for presenting our results in a way that provides
a clear physical insight into them.

\subsubsection{Definition of the basis set}
\label{subsec.basisset}
To define our basis set, we have generalized the usual
Gram-Schmidt method (GSM)~\cite{Lang02}, and developed a new 
\emph{selective} Gram-Schmidt method (SGSM).
This SGSM is the second pillar of our method, 
and it is able to choose a basis set of linearly independent 
functions in a vectorial space from a larger (overcomplete)
 set of functions, 
that can be used to efficiently compute the chaotic 
eigenfunctions of our system~\cite{Revuelta13}.

The SGSM starts from an initial set of $N$ localized 
(tube and scar) functions,~$\vert \psi_j^{(0)} \rangle$, 
from which the procedure selects the minimum number of them,
$N_b \le N$, necessary to adequately describe the Hilbert 
space defined by the eigenfunctions whose
energies are contained in a given energy window, 
that is, the SGSM defines a basis set in that window.
The elements of this basis set~$\vert \psi_{j_i}^{(0)} \rangle$,
where subindex $i$ orders the elements according to their
semiclassical relevance (see discussion below),
are automatically selected with the aid of the conventional GSM.
Thus, associated with the basis~$\vert \psi_{j_i}^{(0)} \rangle$, 
we construct an auxiliary basis~$\vert \varphi_{i} \rangle$, 
formed by the orthogonalization of~$\vert \psi_{j_i}^{(0)} \rangle$. 
For example, if we set
$$\vert \varphi_1 \rangle = \vert \psi_{j_1}^{(0)} \rangle$$
then, a second auxiliary function $\vert \varphi_2 \rangle$ 
is given by $$\vert \varphi_2 \rangle = \frac{\vert \psi_{j_2}^{(1)}
\rangle}{\vert \psi_{j_2}^{(1)} \vert},$$ where $j_2 \ne j_1$ and
$$\vert \psi_{j_2}^{(1)} \rangle =  \vert \psi_{j_2}^{(0)} \rangle -
\langle \varphi_1 \vert \psi_{j_2}^{(0)} \rangle \vert \psi_{j_2}^{(0)} \rangle,$$
and so on.

In our SGSM method, the selection procedure of the basis functions 
with a given symmetry for the calculation of the eigenenergies, $E$, 
up to a given energy
%
\begin{equation}
        E < E_\text{ref},
\label{eq.6}
\end{equation}
is done automatically by using a definite set of rules,
which are based on a \emph{selection parameter},~$\eta$.
For a given localized function~$\eta$ is defined as
%
\begin{equation}
  \eta_j = \rho_j [\sigma_j^2+(\delta E_j)^2]^{1/2} .
 \label{eq.7}
\end{equation}
This parameter depends on three terms. 
First, it depends on the density of states,~$\rho_j$, 
at the quantization BS energy~$E_j$,
which is only relevant when the energy window is large.
Second, it also depends on the tube/scar function's dispersion,
given by
\begin{equation}
     \label{eq.sigmaj}
     \sigma_j = \sqrt{ \langle \psi_j^{(0)} \vert {\cal \hat H}^2 \vert \psi_j^{(0)} \rangle 
                     - \langle \psi_j^{(0)} \vert {\cal \hat H} \vert \psi_j^{(0)} \rangle },
\end{equation}
where~${\cal \hat H}$ is the quantum version of the 
classical Hamiltonian~\eqref{eq.1}.
Third,~$\eta$ depends on a new parameter, $\delta E_j$, defined as
%
\begin{equation}
 \delta E_j=\left\{ \begin{array}{ll}
                     0,          & {\rm if} \; E_j \le  E_\text{ref} \\
                     E_j  - E_\text{ref}, & {\rm if} \; E_j > E_\text{ref}
                    \end{array} 
            \right. .
\end{equation}
The function~$\delta E_j$ is included in Eq.~\eqref{eq.7} in order
to improve the numerical accuracy by reducing boundary effects. 
When large energy windows are considered,
$\delta E_j$ has a small influence on the results, 
and then it can even be neglected. 
It is thus clear, that the parameter~$\eta$ introduced in Eq.~(\ref{eq.7}) 
can be also defined using other criteria, that account for example for
the stability or the period of the POs~\cite{Revuelta13}. 
In this work, however, all these coefficients have been dropped out
for simplicity. 
On the other hand, this has been done because we want to use a single 
selection parameter for all orbits, no matter if they are stable or unstable. 
Recall that the stability exponent is complex for stable POs, 
and then~$\eta$ would no longer be real.
On the other hand, the inclusion of the period in Eq.~(\ref{eq.7}), 
as done in the Ref.~\onlinecite{Revuelta13},
renders less accurate results.
This last result is a consequence of the barriers existing 
in the LiNC/LiCN system, 
which confine the POs in certain regions of phase space.
At low energies, this confinement is caused by the invariant tori.
At higher energies, the dynamical barrier close to 
$\vartheta=0.611$ rad [see accumulation of points next to the 
LiNC regular region in Fig.~\ref{fig.2} (b)] acts as an effective 
quantum separatrix in phase space~\cite{Revuelta15}.
Furthermore, we also have the PES barrier separating the two isomers.
On the contrary, in generic highly chaotic systems the unstable POs
densely cover the system phase space.

The SGSM is then defined, in an algorithmic way, as follows:
\begin{itemize}
\item \textbf{0.} With the method described in Subsec.~\ref{subsec.tubescar}, 
we compute all the localized states, $\vert \psi_j^{(0)}\rangle$, 
whose BS quantized energies, $E_j$, fullfill Eq.~\eqref{eq.4}
for the POs shown in Fig.~\ref{fig.3}(c),
and Eq.~\eqref{eq.4} for the POs in Fig.~\ref{fig.3}(a) and~(b)
(cf. also~Fig.~\ref{fig.4}), and are contained at the same 
time in the enlarged energy window defined by
%
\begin{equation}
     E_j < E_\text{ref} + 2 \sigma_j,
\label{eq.9}
\end{equation}
where $\sigma_j$ is given by Eq.~\eqref{eq.sigmaj}.
For the stable POs, normalized tube functions are computed,
whereas for the unstable ones the scar functions are constructed.
This is the most time demanding step of our procedure.
It should be remarked here that for the system under study,
similar results would be obtained using solely the tube wave functions.
Moveover, they are also adequate for systems with a higher
degree of chaoticity~\cite{Revuelta13}.
However, we have decided to use the scar wave functions
over the unstable POs as they have a lower dispersion in energy,
rendering thus slightly better results.
Let us finally remark that it can be \emph{a priori}
expected that the overlap of the tube and 
scar functions outside the enlarged window~\eqref{eq.9} with the desired system
eigenfunctions is negligible, due to the fact that they were
constructed minimizing their energy dispersion.

\item \textbf{1.} From the initial set of localized functions,
$\vert \psi_{j}^{(0)} \rangle$, we select a smaller number of them,
$N_b \le N$, forming a basis set that is \emph{optimal} for our purposes,
as the number of accurately computed eigefunctions 
scales linearly with~$N_b$.
Notice that the number of tube and scar functions calculated 
for this purpose, $N$, should always be greater or equal to
%
\begin{equation}
  N_b = N_{\rm sc}(E_{ref} +2 \sigma_{\rm sc}) + c_b \sigma_{\rm sc} \rho,
\label{eq.10}
\end{equation}
where, $N_{\rm sc}(E)$,~$\sigma_{\rm sc}$ are, respectivelly, 
semiclassical approximations to the number of states with an energy 
smaller than~$E$ and to the scar function dispersion~\cite{Vergini08},
and the term $c_b \sigma_{\rm sc} \rho$, that enlarges the window size,
is introduced to reduce border effects.
If this is not the case, more (longer) POs, and consequently more 
localized functions, must be included in the basis at this step,
as described in step~0.

The first element of our basis set is the tube or scar function with the
smallest~$\eta_j$ value
%
\begin{equation}
 \qquad  \quad \vert \varphi_1 \rangle =
 \vert \psi_{j_1}^{(0)} \rangle, \qquad \with \;
 \frac{1}{\eta_{j_1}}=\max\left\{\frac{1}{\eta_j}\right\}.
\end{equation}
According to Eq.~\eqref{eq.7}, this choice gives priority to the wave 
functions which are more localized in energy.

\item \textbf{2.a}
The remaining localized functions are then orthogonalized to 
$\vert \psi_{j_1}^{(0)} \rangle$ as 
%
\begin{equation}
  \vert \psi_j^{(1)} \rangle = \vert \psi_j^{(0)} \rangle - \langle \varphi_1
  \vert \psi_j^{(0)} \rangle \vert \varphi_1 \rangle, \quad j \neq j_1 .
 \label{eq.12}
\end{equation}

\item \textbf{2.b} The second element of the basis set is
$\vert \psi_{j_2}^{(0)} \rangle$, where the index $j_2$
($j_2 \ne j_1$) satisfies
\begin{eqnarray}
\frac{\vert \psi_{j_2}^{(1)} \vert^2}{\eta_{j_2}} =
\max \left\{ \frac{\vert \psi_j^{(1)} \vert^2}{\eta_j} \right\}_{j \neq j_1},
\label{eq.13}
\end{eqnarray}
where the norm in the numerator has been introduced in order to
make the basis set elements as different as possible between them.
Indeed, notice that after the orthogonalization of Eq.~\eqref{eq.12}
the more similar $\vert \psi_j^{(0)} \rangle_{j \neq j_1}$ is to 
$\vert \varphi_1 \rangle$,
the smaller the norm of function $\vert \psi_j^{(1)} \vert_{j \neq j_1}$ is.
Then the auxiliary function $|\varphi_2\rangle$ is computed as
%
\begin{eqnarray}
\vert \varphi_2 \rangle = \frac{\vert \psi_{j_2}^{(1)} \rangle}
{\vert \psi_{j_2}^{(1)} \vert}.
\label{eq.14}
\end{eqnarray}

The previous steps, 2.a and 2.b, are repeated for all the remaining 
elements in the initial basis set of localized (tube and scar) functions, 
in such a way that the $n^\text{th}$ step in the procedure is defined as:

\item \textbf{n.a}
New functions are obtained by orthogonalization to the auxiliary function in the
previous step,~$\vert \varphi_{n-1} \rangle$,
%
\begin{eqnarray}
\vert \psi_j^{(n-1)} \rangle = \vert \psi_j^{(n-2)} \rangle -
\langle \varphi_{n-1}
\vert \psi_j^{(n-2)} \rangle \vert \varphi_{n-1} \rangle, \nonumber \\
j \neq j_1, j_2, ..., j_{n-1}. \label{eq.15}
\end{eqnarray}
\item \textbf{n.b} The $n$--th basis element is $\vert \psi_{j_n}^{(0)} \rangle$,
where the~$j_n$ index satisfies
%
\begin{eqnarray}
\frac{\vert \psi_{j_n}^{(n-1)} \vert^2}{\eta_{j_n}} =
\max \left\{ \frac{\vert \psi_j^{(n-1)} \vert^2}{\eta_j} \right\}_{j 
     \neq j_1, j_2,..., j_{n-1}},
\label{eq.16}
\end{eqnarray}
and the next auxiliary function is constructed according to
%
\begin{eqnarray}
  \vert \varphi_n \rangle = \frac{\vert \psi_{j_n}^{(n-1)} \rangle}
  {\vert \psi_{j_n}^{(n-1)} \vert}.
\label{eq.17}
\end{eqnarray}
\item The procedure finishes when the number of selected elements in 
the basis set equals~$N_b$ given by Eq.~(\ref{eq.10}).
\end{itemize}

Afterwards, the corresponding Hamiltonian matrix is
computed in the basis set of localized functions,
or alternatively in the equivalent basis set of auxiliary functions. 
Diagonalization using standard routines~\cite{NR96} finally renders~$N_b$ 
eigenstates in the energy window defined in Eq.~\eqref{eq.6}.

\subsubsection{\emph{Local} representation}
\label{subsec.local}
To get a useful representation of the results obtained
in our localized basis set construction procedure, 
 a \emph{local} representation should be used,
in which each single eigenfunction is reconstructed as
\begin{equation} \label{eq.N}
    \vert N \rangle = \sum_{j=1}^{N_b} C_{Nj} \vert \varphi_j^{\rm loc} \rangle ,
\end{equation}
being~$C_{Nj} = \langle \varphi_j^{\rm loc} \vert N \rangle$.
The procedure to compute the functions~$\varphi_j^{\rm loc}$ is also
based on the GSM, but in this case we give priority to those localized 
(tube/scar) wave functions with larger localization intensities, 
i.~e. with a larger overlap with the eigenfunction~$\vert N \rangle$.
For this purpouse, we proceed as follows:
\begin{itemize}
\item \textbf{1.} The first element of the \emph{local} representation 
is taken as the localized state,~$\vert \psi_j^{(0)} \rangle$, 
with the largest localization intensity, which is defined as
\begin{equation} \label{eq.xn}
    x_j^{(n)} = \vert  \langle \psi_j^{(n)} \vert N \rangle \vert ^2.
\end{equation}
Then, 
\begin{equation}
    \vert \varphi_1^{\rm loc} \rangle = \vert \psi_{j_1}^{(0)} \rangle,
\end{equation}
being~$x_1 \equiv x_{j_1}^{(0)} = \max \{ x_j^{(0)} \}$
the largest localization intensity.
This intensity provides valuable information on the localization of 
the~$\vert N \rangle$ eigenfunction over the quantized orbit associated 
with~$\vert \psi_{j_1}^{(0)} \rangle$.

\item \textbf{2.a} For the identification of the second largest localization
intensity,~$x_2$, one must first orthogonalize the remaining 
localized states~$\vert \psi_j^{(0)} \rangle$ 
to~$\vert \varphi_1^{\rm loc} \rangle$ in the following way
\begin{equation}
     \vert \psi_j^{(1)} \rangle = \vert \psi_j^{(0)} \rangle 
     - \langle \varphi_1^{\rm loc} \vert \psi_j^{(0)} \rangle \vert 
       \varphi_j^{\rm loc} \rangle, \quad j \ne j_1 . 
\end{equation}

\item  \textbf{2.b} The second element of the local representation
is defined as
\begin{equation} \label{eq.ortho}
    \vert \varphi_2^{\rm loc} \rangle = 
      \frac{\vert \psi_{j_2}^{(1)} \rangle}{\vert \psi_{j_2}^{(1)} \vert },
\end{equation}
with~$x_2 \equiv x_{j_2}^{(1)} = \max \{ x_j^{(1)}, j \ne j_1  \}$.

Due to the orthogonalization in~\eqref{eq.ortho},
the intensity~$x_2$ cannot be directly related to the localization
of the~$\vert N \rangle$ eigenfunction over the PO, 
along which~$\vert \psi_{j_2}^{(0)} \rangle$ is constructed.
Nonetheless, the sum~$x_1 + x_2$ 
is the square of the modulus of the
the projection
of~$\vert N \rangle$ onto the subspace defined
by~$\vert \psi_{j_1}^{(0)} \rangle$ and~$\vert \psi_{j_2}^{(0)} \rangle$.

The previous steps~2.a and~2.b are repeated until all~$N_b$ auxiliary
functions are computed, in such a way that the~$n$--th step is defined as:

\item \textbf{n.a} The remaining functions,~$\vert \psi_j^{(n-2)} \rangle$,
are orthogonalized to the last element of the \emph{local} representation            
computed,~$\vert \varphi_{n-1}^{\rm loc} \rangle$, as
\begin{eqnarray}
     \vert \psi_j^{(n-1)} \rangle & = & \vert \psi_j^{(n-2)} \rangle 
     - \langle \varphi_{n-1}^{\rm loc} \vert \psi_j^{(n-2)} \rangle 
        \vert \varphi_{n-1}^{\rm loc} \rangle, \nonumber \\
          j & \ne & j_1, j_2, \ldots, j_{n-1} . 
\end{eqnarray}

\item  \textbf{n.b} The $n$--th element of the \emph{local} representation
is given by
\begin{equation}
    \vert \varphi_n^{\rm loc} \rangle = 
      \frac{\vert \psi_{j_n}^{(n-1)} \rangle}{\vert \psi_{j_n}^{(n-1)} \vert },
\end{equation}
with~$x_n \equiv x_{j_n}^{(n-1)} = \max \{ x_j^{(n-1)}, 
          j \ne j_1, j_2, \ldots, j_{n-1} \}$.
\end{itemize}
Recall here that the sum~$x_1 + x_2 + \ldots + x_n$ is related to the 
projection of~$\vert N \rangle$ onto the subspace defined
by the localized functions~$\vert \psi_{j_1}^{(0)} \rangle,
\vert \psi_{j_2}^{(0)} \rangle, \ldots, , \vert \psi_{j_{n-1}}^{(0)} \rangle$.

\section{Results and discussion}\label{sec.results}
In this section we present some results for the vibrational 
eigenstates of the floppy LiNC/LiCN molecule obtained with our 
basis set of (semiclassical) functions localized on POs 
plus the corresponding discussion.
The section is divided in four parts.
First, in Subsect.~\ref{subsec.spec},
we give full details of our computational procedure, 
and demonstrate that each individual eigenfunction can be essentially 
reconstructed using a very small number of basis elements.
Second, we present in Subsect.~\ref{subsec.x1x2}
the localization intensities of the system eigenfunctions.
Third, in Subsect.~\ref{subsec.PR},
we demonstrate the efficiency of our basis basis set by comparison 
with other standard approximations through the computation of the 
participation ratios.
Finally, we conclude by presenting estimations of the error 
in the eigenenergies and the corresponding eigenfunctions 
in Subsect.~\ref{subsec.errors}.

\subsection{Spectrum of the LiNC/LiCN eigenfunctions in a 
  basis set of functions localized along periodic orbits}
 \label{subsec.spec}
Using the method reported in the Sec.~\ref{sec.method}
we have constructed a basis set formed by solely~90 elements
that is able to accurately describe the 66 low--lying 
eigenfunctions of the LiNC/LiCN isomerizing system.
The structure of all these eigenfunctions in our 
localized basis set is discussed in detail in the
Supplemental Material. 

The construction of our localized basis set
is performed in the following three steps.
First, we set in Eq~\eqref{eq.10} the values 
of~$E_\text{ref}=4100$~cm$^{-1}$ and~$c_b=6$. 
Second, we calculate the quantization energies of each PO, 
which are shown with empty green circles in Fig.~\ref{fig.3}. 
Finally, we construct the tube functions for all these POs 
in the case they are stable, and scar functions for the unstable ones.
This procedure renders a total number of~508 localized wave functions.
From this whole set, our method has defined our final basis set
by selecting the~90 best suited, being~19 of them tube functions 
and the remaining~71 scar functions. 
The BS energies of the selected states have been highlighted
with filled green circles in Fig.~\ref{fig.3}.
As it can be seen, the number of selected states increases 
with energy at a similar rate as the density of eigenfunctions.
This can be clearly understood by comparing the number of states
contained in a given energy window for panels~(a) and~(d). 
For example, if we take an energy window of 200~cm$^{-1}$ 
in Fig.~\ref{fig.3}(d), we can see that it only includes 
one state if the energy is smaller than 1000~cm$^{-1}$ 
(the level spacing for the three low--lying eigenenergies
is~$\sim 230$~a.u.),~2 if the energy is~$\sim 1200$~cm$^{-1}$,~3
in the range~$\sim 1500$~cm$^{-1}$, or~4 for energies~$\sim 2000$~cm$^{-1}$
(see Supplemental Material for further information).
Thus, as energy increases, the density of states increases accordingly.
In Fig.~\ref{fig.3}(a) we can see that the BS energies of the localized 
functions that form our localized basis set follow a similar pattern:
the number of  selected states (filled green points) is very low at small
energies and are quite separated, while they get closer and closer 
for higher energies.
Likewise, a more detailed analysis of Fig.~\ref{fig.3}(a)
shows that the number of selected BS energies included in a window 
of~$200$~a.u. equals the number of eigenenergies just discussed.

\begin{figure}
\includegraphics[width=\columnwidth]{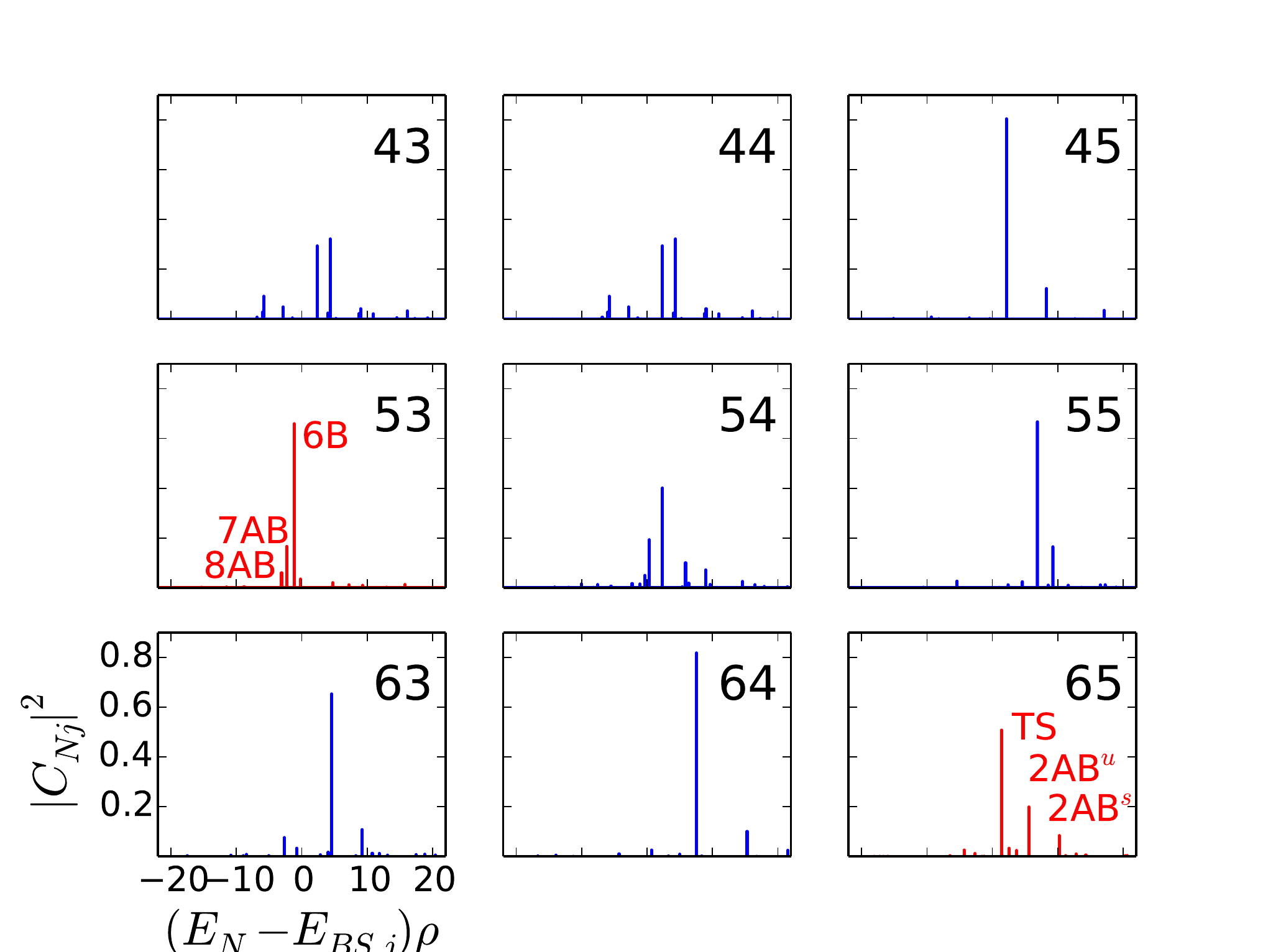}m
\caption{Spectra of some representative LiNC/LiCN eigenfunctions
  in our set of localized basis function.
  The horizontal axis used consists of the energy difference between
  the computed eigenenergy,~$E_N$, and the Bohr--Sommerfeld
  quantized energy,~$E_{BS,j}$, measured in units of the mean level spacing,~$1/\rho$.
  In the case of the eigenfunctions~$\vert 53 \rangle$ and $\vert 65 \rangle$ (red)
  discussed in Subsec.~\ref{subsec.PR},   we have indicated 
  which are the sticks associated with the tube/scar functions 
  contributing the most, ($\vert$6B$^u_{\pi-0}$,21$\rangle$, 
  $\vert$7AB$^u_{\pi-0}$,39$\rangle$, and $\vert$8AB$^u_{\pi-0}$,37$\rangle$
  for~$\vert 53 \rangle$; $\vert$TS$^u\rangle$, 
  $\vert$2AB$^u_{\pi-3}$,38$\rangle$, and 
  $\vert$6AB$^u_{0-0}$,6$\rangle$ for~$\vert 65 \rangle$) to the reconstruction
 (further details, see also Figs.~\ref{fig.10} and~\ref{fig.11},
 and SM).
 }
\label{fig.6}
\end{figure}

As already stated in Subsec.~\ref{subsec.tubescar}, 
the tube/scar functions have a very low dispersion in
energy~\cite{Polavieja94, Vergini00a,*Vergini00b, Vergini01, 
Revuelta12, Revuelta15}.
One can then ask whether there is a similar relationship for 
the eigenfunctions computed in a basis set formed by these 
localized wave functions.
The answer to this question is afirmative, 
as shown by the results in Fig.~\ref{fig.6}, 
where the spectra of some representative eigenfunctions are presented.
In the picture, we have also indicated the most contributing localized 
states, via their quantized POs, to the reconstruction of the 
eigenfunctions~$\vert 53 \rangle$ and~$\vert 65 \rangle$ (red spectra).
This will be discussed in more detail below (cf.~Subsec.~\ref{subsec.PR}).
As it can be seen, the spectrum of each eigenfunction is mainly concentrated
around the corresponding eigenenergy, which is taken as the origin of 
the horizontal axis.
Notice that the spectrum has been represented as a function of
the difference between eigenenergy and BS quantized energy measured 
in units of the mean level spacing,~$1/\rho$, 
since it provides a meaningful scaling.
As already discussed in the previous paragraph, 
the density of states increases with the energy, and, as a consequence, 
the energy difference,~i.~e. the level spacing,
between the eigenfunctions decreases.
Thus, a comparison between two \emph{bare} eigenenergies is
not really very meaningfull:
one must also take into account the density of states in order 
to compare enery diffferences. 
For example, an energy difference of~10~cm$^{-1}$ might be very small 
for the low--lying states, which have a mean level spacing of~$\sim 230$~a.u.,
but being rather large for very excited states, where the number of
eigenergies included in an window of~10~cm$^{-1}$ is dramatically large.
However, when the energy difference is measured in mean level spacing units
by multipliying it by the density of states, it is very simple to say whether
this \emph{relative} energy difference is large or small:
if it is larger than one, it must always be considered large, 
while it can be considered small if it is smaller than, at least, 
one half  of the mean level spacing ($\le 0.5$).
%

\begin{figure}
\includegraphics[width=\columnwidth]{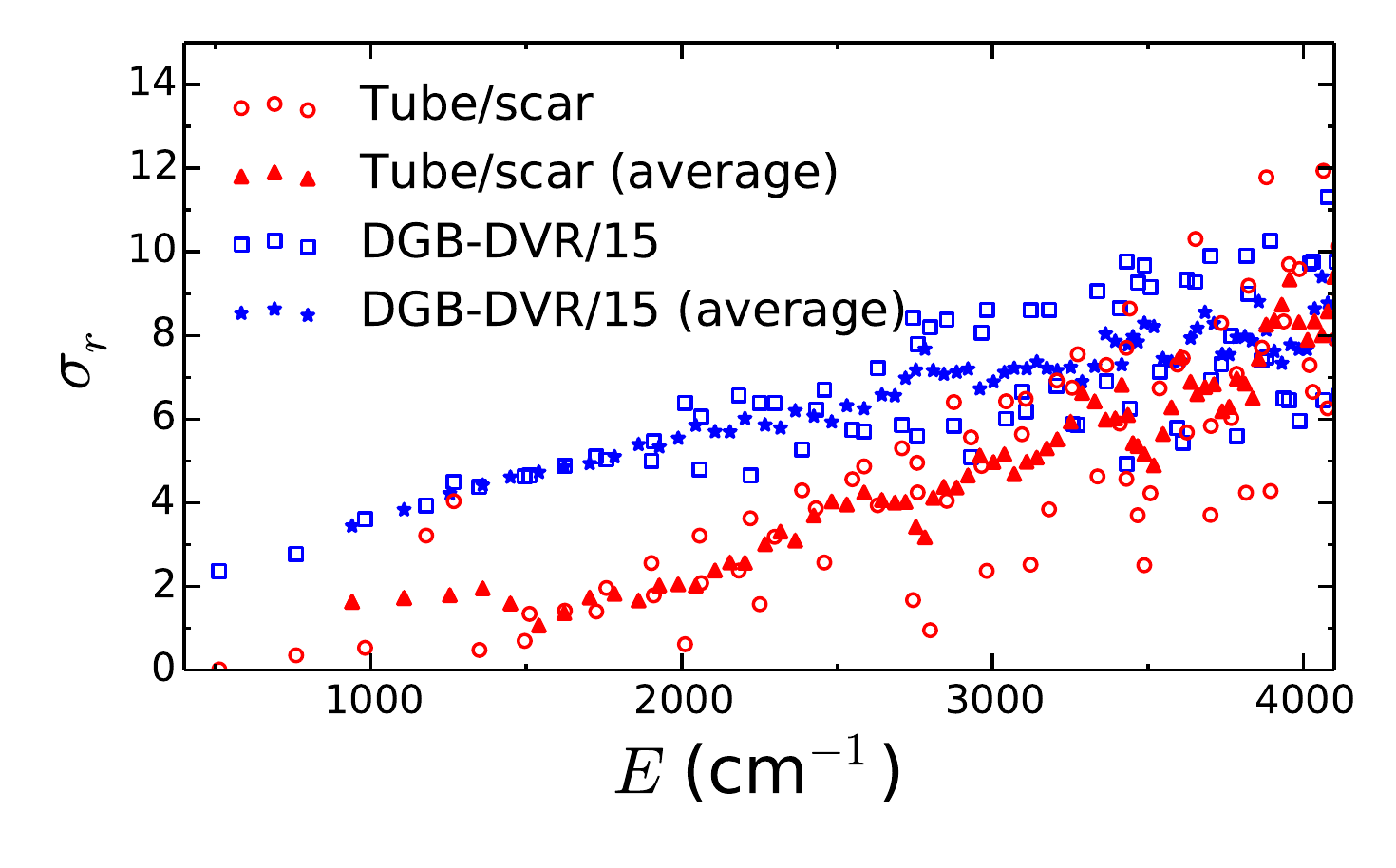}
\caption{Relative spectral dispersion~\eqref{eq.sigmar} 
  for the LiNC/LiCN eigenfunctions obtained with
  our basis set of localized function as a function of the energy 
 (red empty circles) and with a DGB--DVR basis set
  as defined in Ref.~\onlinecite{Bacic86} (blue empty squares).
  In both cases, the average values, computed as a mobile mean of step 5,
  has been also been plotted superimposed (red filled triangles and
  blue filled stars, respectively).
  The DGB--DVR results have been divided by~15 in order that they 
  are defined in the same range as our semiclassical results. 
}
\label{fig.7}
\end{figure}

Figure~\ref{fig.7} shows (with empty red circles) the relative spectral
dispersion of all computed LiNC/LiCN eigenfunctions,~$\sigma_r$, 
defined as
\begin{equation} \label{eq.sigmar}
   \sigma_r=\sigma_N \rho,
\end{equation}
where $\sigma_N$ is the dispersion of eigenfunction~$\vert N \rangle$ 
in our semiclassical basis set.
Then $\sigma_r$ measures the dispersion of eigenfunction~$\vert N \rangle$ 
in mean level spacing units.
In order to better identify the behaviour shown by this magnitude, 
we have also plotted superimposed its average value (filled red triangles), 
computed as a mobile mean of step~5.
As can be seen, the average value of the dispersion increases with the energy;
this being an indication of the necessity of more basis elements for the 
reconstruction of the more excited eigenfunctions.
Still, it should be remarked that the obtained values for dispersion 
of our basis set remain small compared to other standard methods.
In order to demonstrate this assesment, we have superimposed
in Fig.~\ref{fig.7} the relative dispersion for a basis set formed 
by~345 basis elements defined by a combination of the 
Discrete Variable Representation (DVR) for the~$\vartheta$ coordinate 
and a function representation of distributed Gaussian basis (DGB) 
in the radial coordinate~$R$~\cite{Bacic86}.
Recall that this kind of DGB--DVR basis sets have been extensively 
applied to the study of triatomic molecules such as HCP~\cite{Arranz10b},
HNC/HCN~\cite{Bacic91}, H$_2$O~\cite{Bacic88}, 
H$_3^+$~\cite{Bramley94}, KNC/KCN~\cite{Henderson92},  SO$_2$~\cite{Ma99},
HO$_2$~\cite{Arranz10},
or the system under study, LiNC/LiCN~\cite{Bacic86, Bacic91}. 
We have used~345 DGB---DVR basis elements which render
computed eigenenergies with a precision of~0.1~cm$^{-1}$.
Notice that the DGB--DVR results have been divided over~15 in order
that they are defined in the same range as our semiclassical calculations.
Recall that the larger the DGB--DVR basis set,
the larger the dispersion and the corresponding participation ratios
(see discussion in Subsec.~\ref{subsec.PR}).
As can be seen, both the \emph{bare} relative dispersion 
(empty blue squares) and its average value (filled blue triangles) 
are between~15 and~30 times larger than the ones rendered by our 
localized basis set.
Furthermore, as will be see below in Subsec.~\ref{subsec.errors},
this low dispersion of the eigenstates in our localized basis set,
which is always smaller than~12 level spacing units,
also reflects in a small value of the participation ratio, 
this fact further demonstrating the efficiency of our method.
%

\subsection{Localization intensities of the eigenfunctions}
  \label{subsec.x1x2}
\begin{figure}
\includegraphics[ width=\columnwidth]{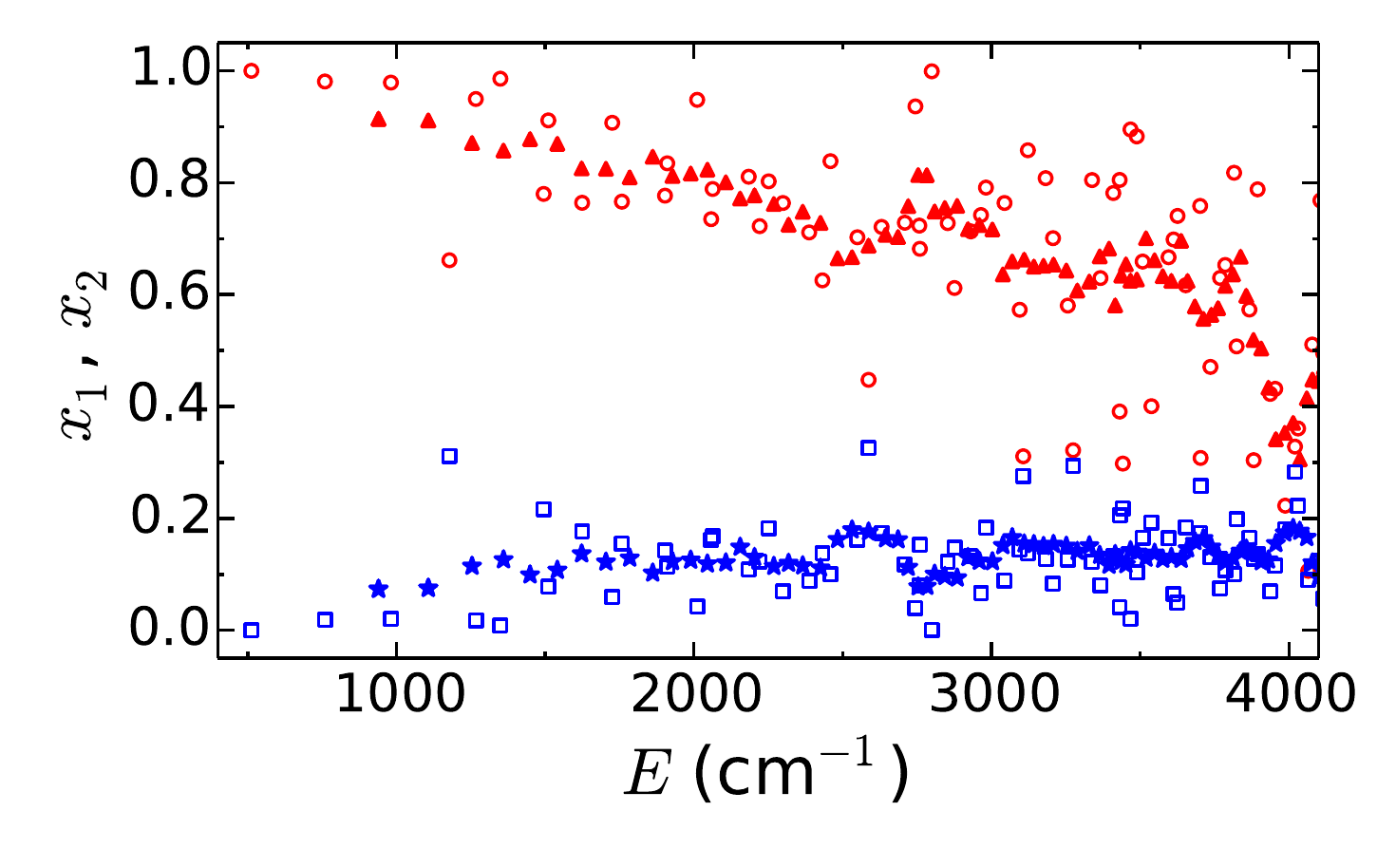}
\caption{Largest localization intensities $x_1$ (red empty circles) 
and $x_2$ (bottom blue empty squares) for the eigenfunctions of 
the LiNC/LiCN system in our basis set of localized functions.
The average, computed as a mobile mean of step~5, is plotted superimposed
with red full triangles and blue full stars, respectively.}
\label{fig.8}
\end{figure}

In Fig.~\ref{fig.8} we present, with empty red circles and 
empty blue squares respectively,
the two largest localization intensities~$x_1$ and~$x_2$ of the 
LiNC/LiCN eigenfunctions computed with our semiclassical basis set, 
as defined in Eq.~\eqref{eq.xn}.
As can be seen, the fluctuation of both quantities is relatively large.
Accordingly, in order to better identify their behaviours, 
we have also plotted superimposed in the figure the corresponding average 
values (full red triangles and full blue stars, respectively), computed 
as a mobile mean of step~5. 
For the low--lying eigenfunctions, the intensity~$x_1$ has a value close to~1,
thus indicating that those eigenfunctions are strongly concentrated over
one single PO. 
Notice that the eigenfunctions that are highly localized over 
unstable POs, i.e.~scar basis functions,
correspond to ``scars'' of the system~\cite{Heller84}.
When this happens,~$x_2$ is smaller than its mean value,
as~$\Sigma_{j=1}^{N_b} x_j = 1$.
The average value of~$x_1$, computed again as a mobile mean,
decreases more or less monotonically with the energy.
Meanwhile, the average value of~$x_2$ increases up to~$\sim 1400$~cm$^{-1}$,
and then remains more or less constant and equal to~$\bar{x}_2 \approx 0.1$.
Let us remark, nevertheless, that~$x_2$ is by definition always smaller than~$x_1$, 
and then it must also decrease for larger values of the energy, 
although this is not noticeable in Fig.~\ref{fig.8}~\cite{Vergini04}.
%

\subsection{Participation ratios and \emph{local} representation
   of the eigenfunctions}
  \label{subsec.PR}
In order to have a more quantitative analysis of the quality of our basis set,
we have also considered participation ratios, $R_N$, 
of the LiNC/LiCN eigenfunctions,~$\vert N \rangle$, 
defined as [cf.~Eq.~\eqref{eq.N}]
\begin{equation} \label{eq.pr}
    R_N=\frac{\sum_{j=1}^{N_b} C_{Nj}^2}{\sum_{j=1}^{N_b} C_{Nj}^4}.
\end{equation}
When examining this magnitude, one has to take into account that the
participation ratios defined in this way are bounded by two limiting cases. 
On the one hand, the optimal basis set is always formed by the eigenfunctions system. 
In this case, all coefficients $C_{Nj}$ appearing in Eq.~\eqref{eq.pr} 
except one would vanish and, consequently,~$R_\text{min}=1$.
On the other hand, the most ill--suited basis set would be one
where all the the coefficients $C_{Nj}$ were equal; 
in this case $R_\text{max}=N_b$.

%
\begin{figure}
\includegraphics[angle=0, width=\columnwidth]{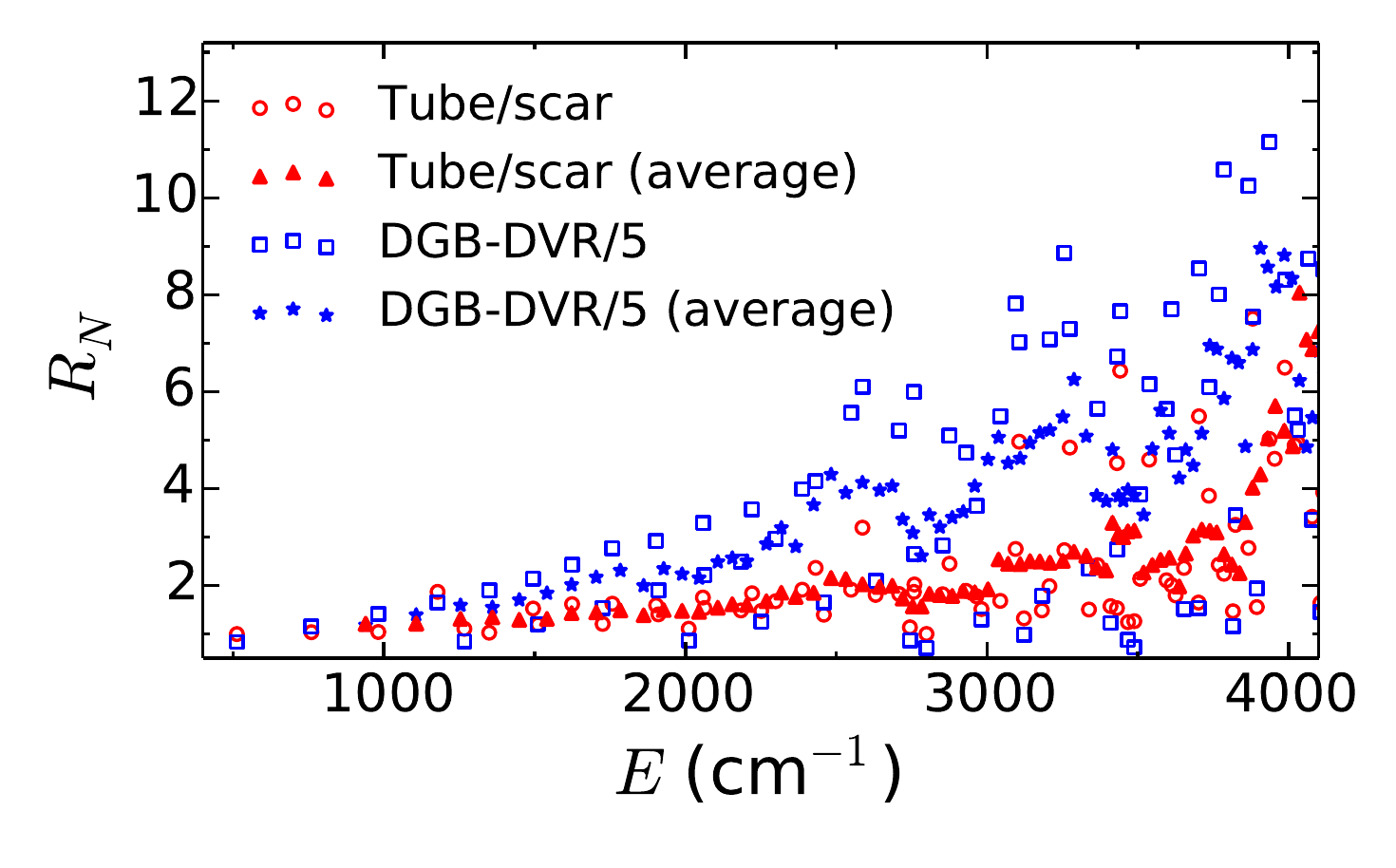}
\caption{
Participation ratios for the LiNC/LiCN eigenfunctions obtained with
our basis set of localized function as a function of the energy 
(red empty circles) and with a DGB--DVR basis set 
as defined in Ref.~\onlinecite{Bacic86} (blue empty squares).
In both cases, the average values, computed as a mobile mean of step 5,
has been also been plotted superimposed (red filled triangles and
blue filled stars, respectively). 
The DGB--DVR results have been divided by~5 in order
to be defined in the same range as our semiclassical results.
}
\label{fig.9}
\end{figure}
We present in Fig.~\ref{fig.9} the participation ratios, $R_N$, 
for the LiNC/LiCN eigenfunctions computed with the basis set 
constructed with our procedure (red empty circles).
As can be seen, most of the low--lying states have a value 
of the participation ratio close to (the optimal) one. 
This results is a consequence of the strong localization of these 
eigenfunctions along the POs considered for the basis construction. 
Thus, the overlap between our semiclassical basis elements and 
the eigenfunctions of the system becomes very large.
As energy increases, more basis elements are necessary for the
computation of the system eigenfunctions, and then the participation 
ratios increase accordingly. 
Although $R_N$ is seen to present large fluctuations with energy, 
specially for~$E \gtrsim 3000$~cm$^{-1}$, the results in Fig.~\ref{fig.9} 
show that its average value increases quite smoothly.
The dramatic increment of the participation ratio
for~$E \gtrsim 3800$~cm$^{-1}$ demonstrates the necessity 
of more basis elements, i.e.~that more POs are required.
For a better observation of the tendency of the participation ratios, 
we have also plotted superimposed in the figure with red triangles 
their average values, computed again as a mobile mean of step~5.

Let us remark that the participation ratios in our localized basis 
are much smaller than those obtained using other standard methods, like, 
for example, the results shown in Fig.~\ref{fig.9} in empty blue squares,
as well as than their average values presented in blue filled stars,
which corresponds to the computation of the LiNC/LiCN eigenstates using
a DGB--DVR basis set.
Notice that these DGB--DVR results have been divided by~5
in order to be defined in the same range as the results rendered 
by the computations of our localized basis set.

Let us finally conclude this section by presenting two examples of 
the structure of the eigenfunctions obtained with our basis set.
For this purpose we have selected the eigenstates~$\vert 53 \rangle$
and $\vert 65 \rangle$ highlighted in red in Fig.~\ref{fig.8}.
%
\begin{figure}
\includegraphics[angle=0, width=\columnwidth]{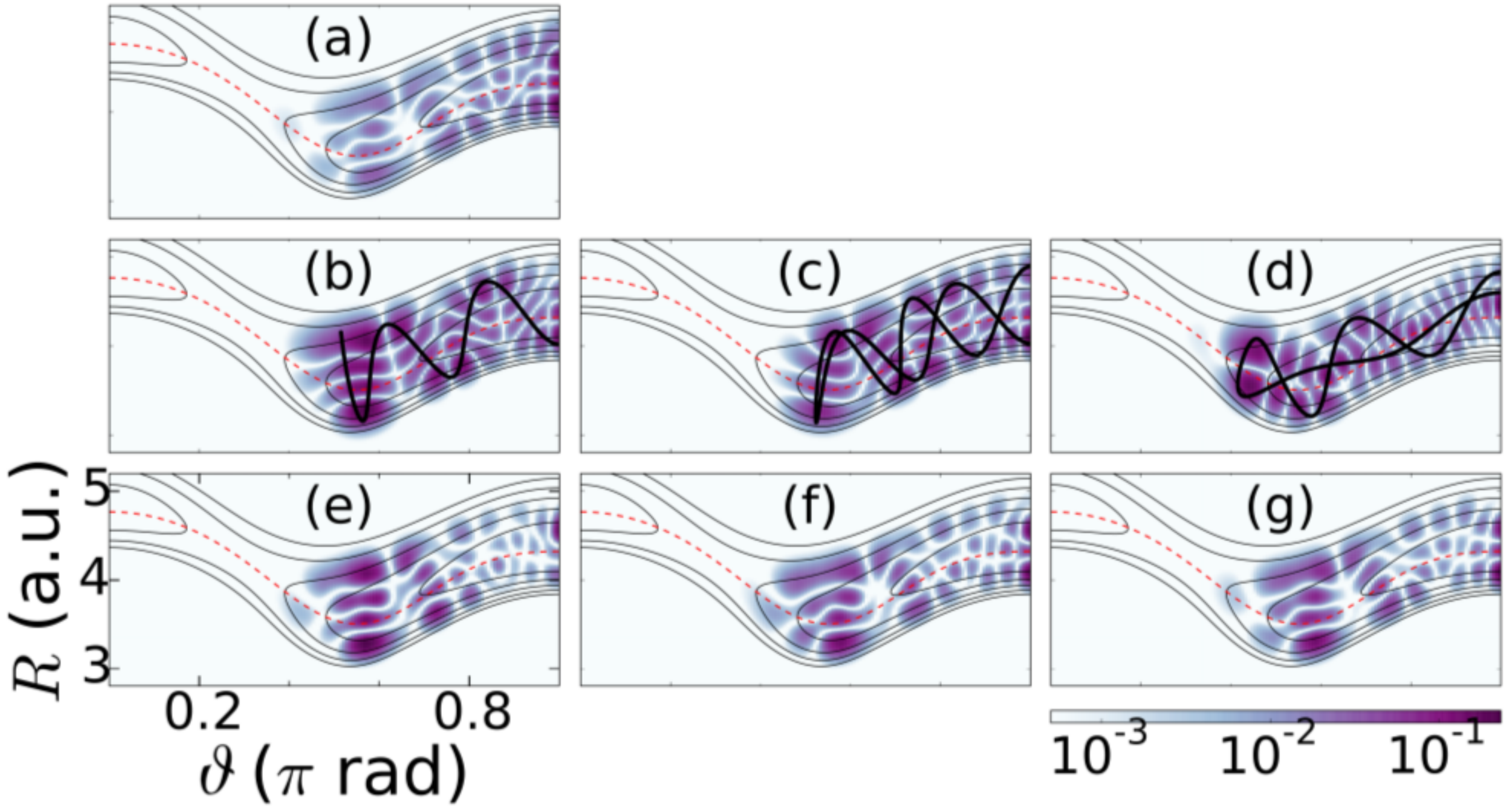}
\caption{Reconstruction of eigenfunction~$\vert 53 \rangle$
of the LiNC/LiCN system [shown in panel (a)].
The local representation is performed using the basis functions
$\vert\textnormal{6B}^{u}_{\pi-0}, 21\rangle$ (b),
$\vert\textnormal{7AB}^{u}_{\pi-0}, 39\rangle$ (c), and
$\vert\textnormal{8AB}^{u}_{\pi-0}, 37\rangle$ (d).
Using the wave function (b) one reconstructs the $65.9\%$
of the exact eigenfunction (e);
combining (b) and (c), one reconstructs $82.5\%$ of it (f), 
and using (b), (c) and (d) $88.5\%$ (g).}
\label{fig.10}
\end{figure}
We present in Fig.~\ref{fig.10} the results corresponding to the first case.
Eigenfunction~$\vert 53 \rangle$ has a participation ratio~$R_{53}=2.16$,
which implies that it can be essentially reconstructed by using only 
2 or at most 3 basis elements. 
Moreover, it has a very irregular nodal pattern, 
something characteristic of classically chaotic systems, 
as shown to be the case here in the Fig.~\ref{fig.10} (a).
The most important contribution to this eigenfunction
is given by the basis scar function $\vert$6B$^{u}_{\pi-0}$, 21$\rangle$,
which is shown in panel~(b) of Fig.~\ref{fig.10}.
Just by using this single basis function $65.9\%$ of the 
(exact) eigenfunction~$\vert 53 \rangle$ can be reconstructed,
as shown in panel~(e).
When the scar function $\vert$7AB$^{u}_{\pi-0}$, 39$\rangle$
[see panel~(c)] is added as a second element to the basis, 
$82.5\%$ of the eigenfunction is reconstructed, see panel~(f). 
Finally, augmenting the basis set with the scar function 
$\vert$8AB$^{u}_{\pi-0}$, 37$\rangle$ [see panel~(d)] as the third element, 
$88.5\%$ of the exact eigenfunction is recovered. 
We believe that this result, namely that by using only the~3 localized functions 
depicted in panels~(b)--(d) one can obtain the state shown in panel~(g), 
which cannot be ascribed to any of the POs shown in Fig.~\ref{fig.4},
it is quite impressive, this giving a clear idea of the quality and 
performance of our basis set construction method.

One last point is worth emphasizing in this discussion on the 
reconstruction of the eigenfunction~$\vert 53 \rangle$ of LiNC/LiCN.
The quantized energies corresponding to the basis elements which we have
considered, i.e.~those shown in panels (b)--(d), lie quite close to the 
eigenenergy $E_{53}=3507.24$~cm$^{-1}$, as can also be seen in Fig.~\ref{fig.6}.
Obviously, when increasing the number of basis elements
this eigenfunction is more accuratelly reconstrated. 
For example, by including~6 basis elements,~$95.1 \%$ of the exact 
eigenfunction is obtained; using~13, 99.1$\%$; with~28, 99.9$\%$, and 
using~39 basis elements an impressive accuracy of~$99.99\%$ of the exact eigenfunction, 
with an error of~0.66~cm$^{-1}$ in the corresponding eigenenergy is obtained.
In general, the accuracy of~$99.9 \%$ in the reconstruction of the 
eigenfunctions is achieved by combination of less than~5 basis elements 
in the case of most of the low--lying states and around~25 of the~90 total 
basis elements for the most excited ones.
Recall that the localized states selected in the reconstruction
of these eigenfunctions are those with the BS quantized energies
that lie closer to the considered eigenenergy.
Let us recall here that we consider \emph{exact} the results obtained 
with the~345 elements basis set needed in the Ba{\u c}i\'{c} and Light 
calculation~\cite{Bacic86}.

\begin{figure}
\includegraphics[width=\columnwidth]{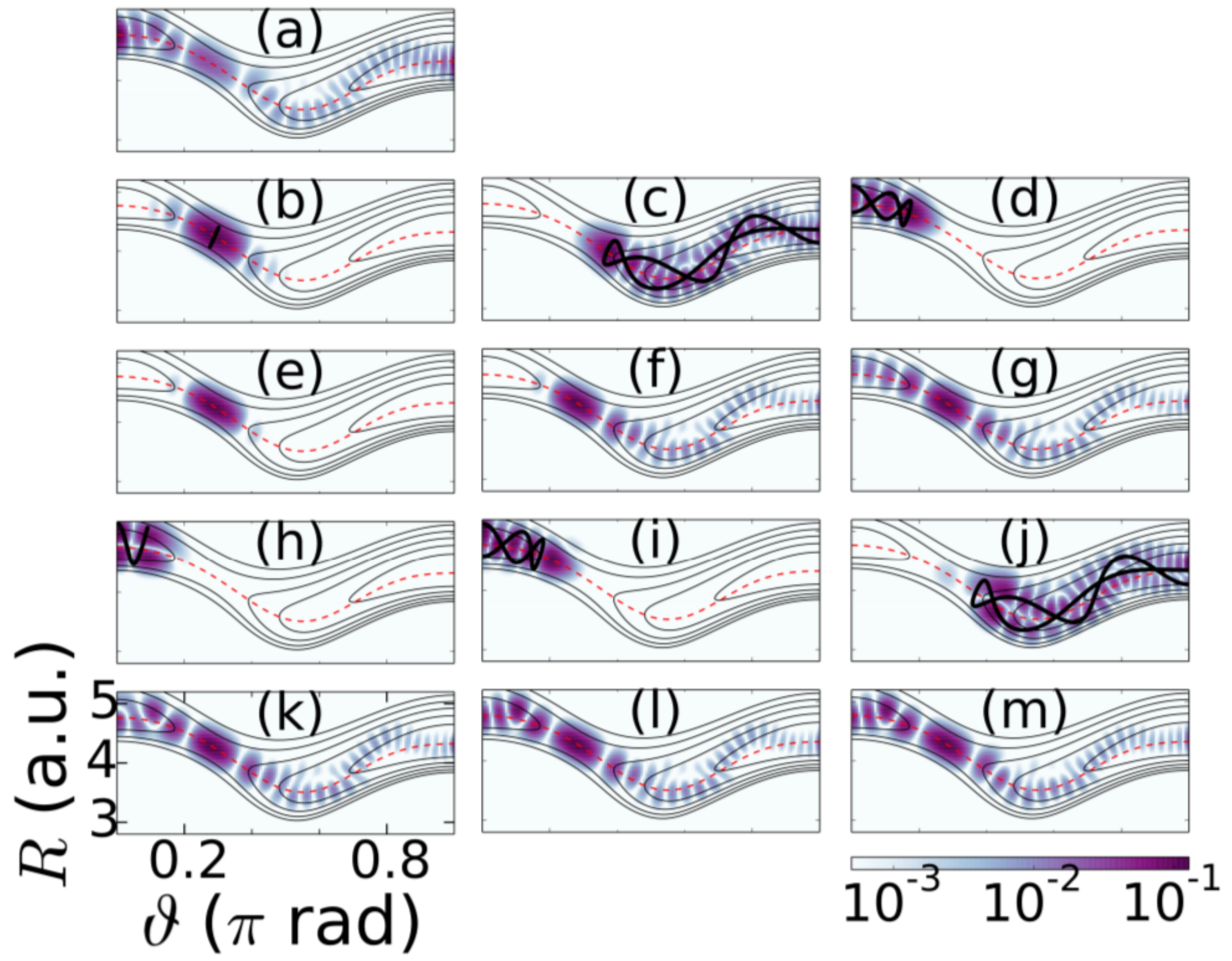}
\caption{Reconstruction of eigenfunction~$\vert 65 \rangle$
of the LiNC/LiCN system [shown in panel (a)].
The local representation is performed using
$\vert\textnormal{TS}^{u}, 0\rangle$ (b),
$\vert\textnormal{2AB}^{u}_{\pi-3}, 38\rangle$ (c), and
$\vert\textnormal{2AB}^{s}_{0-0}, 6\rangle$ (d),
$\vert\textnormal{1A}^{u}_{0-0}, 3\rangle$ (h),
$\vert\textnormal{2AB}^{s}_{0-0}, 8\rangle$ (i),
$\vert\textnormal{2AB}^{u}_{\pi-0}, 39\rangle$ (j).
The reconstruction process is shown in the 
remaining panels, where 
$50.7\%$ (e),
$70.6\%$ (f),
$78.9\%$ (g),
$82.2\%$ (k),
$84.7\%$ (l)
$87.0\%$ (m) of the exact result is obtained.}
\label{fig.11}
\end{figure}

Let us remark that in Fig.~\ref{fig.10}, the squared wave functions shown in 
panels (a)--(d) have been normalized such that their maximum value 
equals one.
Contrary, the partially reconstructed eigenfunctions represented 
in panels (e)--(g) have been normalized in such a way that the maximum 
value of the computed squared eigenfunction using the whole basis set of 
our localized wave functions equals one. As a consequence, the maximum 
value of the partially reconstructed eigenfunctions shown in panels (e)--(g) 
is always smaller than~1. We have decided to present the results with these 
two different normalization criteria because then it is in general easier to 
visualize the contribution of each basis element to the eigenfunction 
reconstruction (cf.~Fig.~\ref{fig.11}). Finally, notice that the scar function 
presented in panel (b) equals the reconstructed eigenfunction shown 
below in panel (e), being the only difference between them the normalization used.
%
%
In Fig.~\ref{fig.11} we show the results of a similar analysis
performed for the structure of eigenfunction~$\vert 65 \rangle$,
which is the first isomerizing state of the system,
i.e.~that having a significant proportion of the quantum density 
simultaneously localized in both isomer wells.
This eigenfunction is shown in panel (a).
It has a participation ratio equal to~$R_{65}=$4.60, and the
corresponding computed eigenenergy of~$3826.84$~cm$^{-1}$ is~3.08~cm$^{-1}$ 
smaller than in the DGB--DVR calculation taken as reference.
Again, the eigenfunction is mostly reconstructed using 
a very small number of basis elements. 
Indeed, by just considering the scar 
function~$\vert\textnormal{TS}^{u}, 0\rangle$ (b), 
$50.7\%$ of the reference eigenfunction is recovered
[see result in~panel~(e)].
Considering the scar functions in panels~(b) and~(c), 
one reconstructs~$70.6\%$ of the reference eigenfunction,
as seen in~panel~(f). 
Finally, combining all basis functions shown in panels~(b)--(d) 
and~(h)--(j) one gets the wave function shown in the panel (m), 
which is very similar to the exact eigenfunction of panel (a), 
despite de fact that it has been calculated using functions 
that are localized over nonisomerizing POs. 
Actually, the overlap between the exact eigenfunction and 
the approximate one computed using this, six elements, basis set 
equals an excellent~$87.0\%$.
By using~12 basis elements, 95.3\% of the exact eigenfunction
is reconstructed, and by combination of~38 basis elements, 99.0\%.
Recall that the localized (tube and scar) wave functions and 
the partially reconstructed eigenfunctions shown in Fig.~\ref{fig.11} 
have been normalized using different criteria 
(see discussion on Fig.~\ref{fig.10} above).
For further information on the structure of all the~66 accurately
computed eigenfunctions obtained with our localized basis set, 
see the Supplemental Material. 

\subsection{Errors in the eigenenergies and the eigenfunctions}
  \label{subsec.errors}

\begin{figure}
\includegraphics[width=\columnwidth]{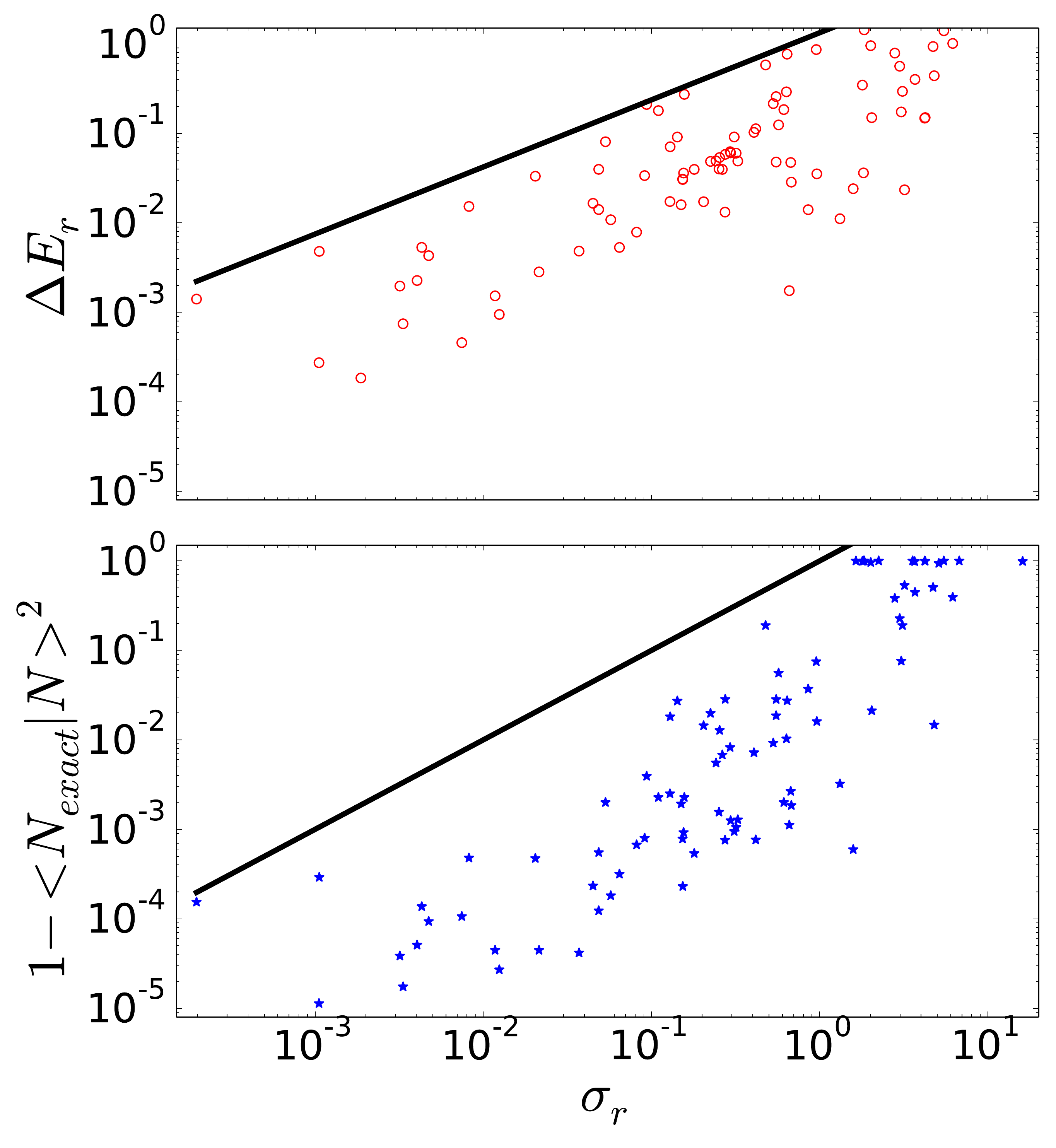}
\caption{Error in the eigenenergies (top red cicles) and 
eigenfunctions (bottom blue asteriscs) of the eigenstates 
using our localized basis set, estimated as described in 
Sec.~\ref{subsec.errors}, as a function of the relative 
dispersion~\eqref{eq.sigmar}. 
The solid lines indicate the upper error bounds given 
by Eq.~\eqref{eq.errors}.}
\label{fig.12}
\end{figure}

Fig.~\ref{fig.12} shows the error in the eigenenergies measured
in mean level spacing units, $\Delta E_r=\vert E-E'\vert \; \rho$,
 (top red circles), and in the corresponding
eigenfunctions, $1-\langle N' \vert N \rangle^2$,  (bottom blue asterisks),
respectively, computed using our localized basis set as a 
function of the relative dispersion,~$\sigma_r$, given by Eq.~\eqref{eq.sigmar}.
As can be inferred from the figure and \emph{a priori} expected,
both errors increase with the relative dispersion.

The black lines in Fig.~\ref{fig.12} correspond to the 
upper bound for the errors in the energies and eigenfunctions of 
our vibrational states calculation given by
\begin{equation}
   \Delta E_r \le \frac{4}{3}\sigma_r^{3/4}, \qquad 
    1-\langle N' \vert N \rangle^2 \le \sigma_r,
 \label{eq.errors}
\end{equation}
which indicates that the error in the eigenenergies 
scales as~$\sigma_r^{3/4}$  with the relative dispersion, 
while that in the eigenfunctions does it linearly.
Let us remark that the Eqns.~\eqref{eq.errors} have been obtained 
\emph{heuristically}, so one could equally well define other (in general more
complicated) functions to estimate the upper bounds.
However, we have decided to use these expressions as they
extremelly simple, and similar to those previously used by some of us
in the study of other classically chaotic systems~\cite{Vergini08, Revuelta13}.

Let us finally remark the usefullness of Eqns.~\eqref{eq.errors} as one can
use them to know \emph{a priori} the errors expected in the calculation 
of highly excitated states~\cite{Vergini08, Revuelta13}
by simply measuring the relative dispersion,
which is a very easy to calculate parameter.

%
\section{Conclusions and outlook}\label{sec.concl}
Summarizing, we have presented a method to efficiently compute 
the vibrational eigenstates of floppy molecular systems, 
in which the classical phase space contains regios of 
regular and irregular motion at the same energy.
The method uses the so--called tube and scar wave functions, 
respectively localized over stable and unstable POs,
which then semiclassically account of these underlying classical 
structures of the system, this including short pieces of the 
invariant manifolds originated in the fixed point in the latter case.
This method was originally introduced in the Ref.~\onlinecite{Revuelta13},
where it was applied to a highly chaotic system consisting of a
homogeneous quartic coupled potential function.
In this paper, we have extended that work by applying it to the study 
of a floppy molecule described by a realistic potential, 
namely the LiNC/LiCN isomerizing system.
Using a basis set formed by~90 localized elements, we have accurately
computed the~66 low--lying eigenenergies and the corresponding 
eigenfunctions of the system.
More importantly, we have demonstrated that each eigenfunction
is essentially reconstructed by a small number of basis elements, 
usually less than~5 in the energy range considered. 
Likewise, in order to demonstrate the efficiency of the method, 
a detailed discussion on the results has been performed,
including an analysis of the structure of the eigenfunctions in terms
of our efficient basis set, localization intensities, participation ratios,
and also the errors of our computations, 
taking as reference the corresponding values as rendered by the 
DGB--DVR method of Ba{\u c}i\'{c} and Light~\cite{Bacic86}.

Finally, let us remark that the extension of our approach to the full
three--degrees--of--freedom calculations of LiNC/LiCN is straightforward, 
since it simply consists of making a direct product basis of the current
functions and functions describing the third degree of freedom, $r$.
However, the results reported by some of us in Ref.~\citenum{GM14}
indicate that significant changes in the conclusions of the 
present work should not be expected.

\section{Acknowledgements}
We acknowledge financial support of the Spanish Ministry of Economy and 
Competitiveness (MINECO)
under Contracts No. MTM2012-39101 and MTM2015-63914-P
and by ICMAT Severo Ochoa under Contract SEV-2015-0554.
We also thank Prof. \`A.~Jorba for having provided us the Shampine--Gordon
routines used in our classical calculations.

\bibliographystyle{apsrev}
\bibliography{base_scar_licn}

\begin{thebibliography}{77}
\expandafter\ifx\csname natexlab\endcsname\relax\def\natexlab#1{#1}\fi
\expandafter\ifx\csname bibnamefont\endcsname\relax
  \def\bibnamefont#1{#1}\fi
\expandafter\ifx\csname bibfnamefont\endcsname\relax
  \def\bibfnamefont#1{#1}\fi
\expandafter\ifx\csname citenamefont\endcsname\relax
  \def\citenamefont#1{#1}\fi
\expandafter\ifx\csname url\endcsname\relax
  \def\url#1{\texttt{#1}}\fi
\expandafter\ifx\csname urlprefix\endcsname\relax\def\urlprefix{URL }\fi
\providecommand{\bibinfo}[2]{#2}
\providecommand{\eprint}[2][]{\url{#2}}

\bibitem[{\citenamefont{Berezin and Shubin}(1991)}]{Berezin91}
\bibinfo{author}{\bibfnamefont{F.~A.} \bibnamefont{Berezin}} \bibnamefont{and}
  \bibinfo{author}{\bibfnamefont{M.}~\bibnamefont{Shubin}},
  \emph{\bibinfo{title}{The Schr\"odinger Equation}}, Astrophysics and Space
  Science Library (\bibinfo{publisher}{Springer Netherlands},
  \bibinfo{year}{1991}), ISBN \bibinfo{isbn}{079231218X, 9780792312185}.

\bibitem[{\citenamefont{Marchildon}(2002)}]{Marchildon02}
\bibinfo{author}{\bibfnamefont{L.}~\bibnamefont{Marchildon}},
  \emph{\bibinfo{title}{Quantum Mechanics: From Basic Principles to Numerical
  Methods and Applications}}, Advanced Texts in Physics
  (\bibinfo{publisher}{Springer-Verlag Berlin Heidelberg},
  \bibinfo{year}{2002}), \bibinfo{edition}{1st} ed., ISBN
  \bibinfo{isbn}{978-3-642-07767-8, 978-3-662-04750-7}.

\bibitem[{\citenamefont{Corey et~al.}(1993)\citenamefont{Corey, Tromp, and
  Lemoine}}]{Corey93}
\bibinfo{author}{\bibfnamefont{G.~C.} \bibnamefont{Corey}},
  \bibinfo{author}{\bibfnamefont{J.~W.} \bibnamefont{Tromp}}, \bibnamefont{and}
  \bibinfo{author}{\bibfnamefont{D.}~\bibnamefont{Lemoine}},
  \emph{\bibinfo{title}{Numerical Grid Methods and Their Application to
  Schr?¶dinger's Equation}}, NATO ASI Series 412
  (\bibinfo{publisher}{Springer Netherlands}, \bibinfo{year}{1993}),
  \bibinfo{edition}{1st} ed., ISBN
  \bibinfo{isbn}{978-90-481-4308-5,978-94-015-8240-7}.

\bibitem[{\citenamefont{Brack and Bhaduri}(1997)}]{Brack97}
\bibinfo{author}{\bibfnamefont{M.}~\bibnamefont{Brack}} \bibnamefont{and}
  \bibinfo{author}{\bibfnamefont{R.~K.} \bibnamefont{Bhaduri}},
  \emph{\bibinfo{title}{Semiclassical Physics}}, The Advanced Book Program
  (\bibinfo{publisher}{Addison--Wesley Publishing Company, Inc.},
  \bibinfo{address}{Reading, Massachussetts}, \bibinfo{year}{1997}), ISBN
  \bibinfo{isbn}{0-201-48351-3}.

\bibitem[{\citenamefont{Gutzwiller}(1990)}]{Gutzwiller90}
\bibinfo{author}{\bibfnamefont{M.~C.} \bibnamefont{Gutzwiller}},
  \emph{\bibinfo{title}{Chaos in Classical and Quantum Mechanics}},
  Interdisciplinary Applied Mathematics (\bibinfo{publisher}{Springer
  Science+Business Media New York}, \bibinfo{address}{New York},
  \bibinfo{year}{1990}), ISBN \bibinfo{isbn}{978-1-4612-6970-0}.

\bibitem[{\citenamefont{Heller}(1984)}]{Heller84}
\bibinfo{author}{\bibfnamefont{E.~J.} \bibnamefont{Heller}},
  \bibinfo{journal}{Phys. Rev. Lett.} \textbf{\bibinfo{volume}{53}},
  \bibinfo{pages}{1515} (\bibinfo{year}{1984}),
  \urlprefix\url{http://link.aps.org/doi/10.1103/PhysRevLett.53.1515}.

\bibitem[{\citenamefont{Shnirelman}(1974)}]{Shnirelman74}
\bibinfo{author}{\bibfnamefont{A.~I.} \bibnamefont{Shnirelman}},
  \bibinfo{journal}{Usp. Mat. Nauk.} \textbf{\bibinfo{volume}{29}},
  \bibinfo{pages}{181} (\bibinfo{year}{1974}),
  \urlprefix\url{http://www.mathnet.ru/php/archive.phtml?wshow=paper&jrnid=rm&paperid=4463&option_lang=eng}.

\bibitem[{\citenamefont{Kwon et~al.}(2011)\citenamefont{Kwon, Lee, and
  An}}]{Kwon11}
\bibinfo{editor}{\bibfnamefont{O.}~\bibnamefont{Kwon}},
  \bibinfo{editor}{\bibfnamefont{B.}~\bibnamefont{Lee}}, \bibnamefont{and}
  \bibinfo{editor}{\bibfnamefont{K.}~\bibnamefont{An}}, eds.,
  \emph{\bibinfo{title}{Trends in nano- and micro-cavities}}
  (\bibinfo{publisher}{Bentham Science Publishers}, \bibinfo{year}{2011}), ISBN
  \bibinfo{isbn}{9781608052363,1608052362}.

\bibitem[{\citenamefont{Ba{\u c}i\'{c} and Light}(1986)}]{Bacic86}
\bibinfo{author}{\bibfnamefont{Z.}~\bibnamefont{Ba{\u c}i\'{c}}}
  \bibnamefont{and} \bibinfo{author}{\bibfnamefont{J.~C.} \bibnamefont{Light}},
  \bibinfo{journal}{J. Chem. Phys.} \textbf{\bibinfo{volume}{85}},
  \bibinfo{pages}{4594} (\bibinfo{year}{1986}),
  \urlprefix\url{http://scitation.aip.org/content/aip/journal/jcp/85/8/10.1063/1.451824}.

\bibitem[{\citenamefont{Tennyson et~al.}(1986)\citenamefont{Tennyson, Brocks,
  and Farantos}}]{Tennyson86}
\bibinfo{author}{\bibfnamefont{J.}~\bibnamefont{Tennyson}},
  \bibinfo{author}{\bibfnamefont{G.}~\bibnamefont{Brocks}}, \bibnamefont{and}
  \bibinfo{author}{\bibfnamefont{S.~C.} \bibnamefont{Farantos}},
  \bibinfo{journal}{Chemical Physics} \textbf{\bibinfo{volume}{104}},
  \bibinfo{pages}{399 } (\bibinfo{year}{1986}), ISSN \bibinfo{issn}{0301-0104},
  \urlprefix\url{http://www.sciencedirect.com/science/article/pii/0301010486850285}.

\bibitem[{\citenamefont{Farantos and Tennyson}(1987)}]{Farantos87}
\bibinfo{author}{\bibfnamefont{S.~C.} \bibnamefont{Farantos}} \bibnamefont{and}
  \bibinfo{author}{\bibfnamefont{J.}~\bibnamefont{Tennyson}},
  \emph{\bibinfo{title}{Stochasticity and Intramolecular Redistribution of
  Energy}} (\bibinfo{publisher}{Springer Netherlands},
  \bibinfo{address}{Dordrecht}, \bibinfo{year}{1987}), chap.
  \bibinfo{chapter}{Chaos in Molecular Systems?}, pp. \bibinfo{pages}{15--30},
  ISBN \bibinfo{isbn}{978-94-009-3837-3},
  \urlprefix\url{http://dx.doi.org/10.1007/978-94-009-3837-3_2}.

\bibitem[{\citenamefont{Benito et~al.}(1989)\citenamefont{Benito, Borondo, Kim,
  Sumpter, and Ezra}}]{Ezra89}
\bibinfo{author}{\bibfnamefont{R.}~\bibnamefont{Benito}},
  \bibinfo{author}{\bibfnamefont{F.}~\bibnamefont{Borondo}},
  \bibinfo{author}{\bibfnamefont{J.-H.} \bibnamefont{Kim}},
  \bibinfo{author}{\bibfnamefont{B.}~\bibnamefont{Sumpter}}, \bibnamefont{and}
  \bibinfo{author}{\bibfnamefont{G.}~\bibnamefont{Ezra}},
  \bibinfo{journal}{Chem. Phys. Lett.} \textbf{\bibinfo{volume}{161}},
  \bibinfo{pages}{60 } (\bibinfo{year}{1989}), ISSN \bibinfo{issn}{0009-2614},
  \urlprefix\url{http://www.sciencedirect.com/science/article/pii/S0009261489870320}.

\bibitem[{\citenamefont{Henderson and Tennyson}(1990)}]{Henderson90}
\bibinfo{author}{\bibfnamefont{J.~R.} \bibnamefont{Henderson}}
  \bibnamefont{and} \bibinfo{author}{\bibfnamefont{J.}~\bibnamefont{Tennyson}},
  \bibinfo{journal}{Molecular Physics} \textbf{\bibinfo{volume}{69}},
  \bibinfo{pages}{639} (\bibinfo{year}{1990}),
  \eprint{http://dx.doi.org/10.1080/00268979000100471},
  \urlprefix\url{http://dx.doi.org/10.1080/00268979000100471}.

\bibitem[{\citenamefont{Arranz et~al.}(1997)\citenamefont{Arranz, Borondo, and
  Benito}}]{Arranz97}
\bibinfo{author}{\bibfnamefont{F.~J.} \bibnamefont{Arranz}},
  \bibinfo{author}{\bibfnamefont{F.}~\bibnamefont{Borondo}}, \bibnamefont{and}
  \bibinfo{author}{\bibfnamefont{R.~M.} \bibnamefont{Benito}},
  \bibinfo{journal}{The Journal of Chemical Physics}
  \textbf{\bibinfo{volume}{107}}, \bibinfo{pages}{2395} (\bibinfo{year}{1997}),
  \urlprefix\url{http://scitation.aip.org/content/aip/journal/jcp/107/7/10.1063/1.474582}.

\bibitem[{\citenamefont{Arranz et~al.}(1998)\citenamefont{Arranz, Borondo, and
  Benito}}]{Arranz98}
\bibinfo{author}{\bibfnamefont{F.~J.} \bibnamefont{Arranz}},
  \bibinfo{author}{\bibfnamefont{F.}~\bibnamefont{Borondo}}, \bibnamefont{and}
  \bibinfo{author}{\bibfnamefont{R.~M.} \bibnamefont{Benito}},
  \bibinfo{journal}{Phys. Rev. Lett.} \textbf{\bibinfo{volume}{80}},
  \bibinfo{pages}{944} (\bibinfo{year}{1998}),
  \urlprefix\url{http://link.aps.org/doi/10.1103/PhysRevLett.80.944}.

\bibitem[{\citenamefont{Arranz et~al.}(2010{\natexlab{a}})\citenamefont{Arranz,
  Seidel, Giralda, Benito, and Borondo}}]{Arranz10}
\bibinfo{author}{\bibfnamefont{F.~J.} \bibnamefont{Arranz}},
  \bibinfo{author}{\bibfnamefont{L.}~\bibnamefont{Seidel}},
  \bibinfo{author}{\bibfnamefont{C.~G.} \bibnamefont{Giralda}},
  \bibinfo{author}{\bibfnamefont{R.~M.} \bibnamefont{Benito}},
  \bibnamefont{and} \bibinfo{author}{\bibfnamefont{F.}~\bibnamefont{Borondo}},
  \bibinfo{journal}{Phys. Rev. E} \textbf{\bibinfo{volume}{82}},
  \bibinfo{pages}{026201} (\bibinfo{year}{2010}{\natexlab{a}}),
  \urlprefix\url{http://link.aps.org/doi/10.1103/PhysRevE.82.026201}.

\bibitem[{\citenamefont{Borondo and Benito}(2006)}]{Borondo06}
\bibinfo{author}{\bibfnamefont{F.}~\bibnamefont{Borondo}} \bibnamefont{and}
  \bibinfo{author}{\bibfnamefont{R.~M.} \bibnamefont{Benito}}, in
  \emph{\bibinfo{booktitle}{Nonlinear Dynamics and Fundamental Interactions}},
  edited by \bibinfo{editor}{\bibfnamefont{F.}~\bibnamefont{Khanna}}
  \bibnamefont{and}
  \bibinfo{editor}{\bibfnamefont{D.}~\bibnamefont{Matrasulov}}
  (\bibinfo{publisher}{Springer}, \bibinfo{address}{Dordrecht},
  \bibinfo{year}{2006}), vol. \bibinfo{volume}{213} of
  \emph{\bibinfo{series}{NATO Science Series}}, pp. \bibinfo{pages}{115--128}.

\bibitem[{\citenamefont{Ben{\'i}tez et~al.}(2013)\citenamefont{Ben{\'i}tez,
  Losada, Benito, and Borondo}}]{Benitez13}
\bibinfo{author}{\bibfnamefont{P.}~\bibnamefont{Ben{\'i}tez}},
  \bibinfo{author}{\bibfnamefont{J.~C.} \bibnamefont{Losada}},
  \bibinfo{author}{\bibfnamefont{R.~M.} \bibnamefont{Benito}},
  \bibnamefont{and} \bibinfo{author}{\bibfnamefont{F.}~\bibnamefont{Borondo}},
  \emph{\bibinfo{title}{Progress and Challenges in Dynamical Systems:
  Proceedings of the International Conference Dynamical Systems: 100 Years
  after Poincar{\'e}, September 2012, Gij{\'o}n, Spain}}
  (\bibinfo{publisher}{Springer Berlin Heidelberg}, \bibinfo{address}{Berlin,
  Heidelberg}, \bibinfo{year}{2013}), chap. \bibinfo{chapter}{Analysis of the
  Full Vibrational Dynamics of the LiNC/LiCN Molecular System}, pp.
  \bibinfo{pages}{77--88}, ISBN \bibinfo{isbn}{978-3-642-38830-9},
  \urlprefix\url{http://dx.doi.org/10.1007/978-3-642-38830-9_6}.

\bibitem[{\citenamefont{P\'arraga et~al.}(2013)\citenamefont{P\'arraga, Arranz,
  Benito, and Borondo}}]{Parraga13}
\bibinfo{author}{\bibfnamefont{H.}~\bibnamefont{P\'arraga}},
  \bibinfo{author}{\bibfnamefont{F.~J.} \bibnamefont{Arranz}},
  \bibinfo{author}{\bibfnamefont{R.~M.} \bibnamefont{Benito}},
  \bibnamefont{and} \bibinfo{author}{\bibfnamefont{F.}~\bibnamefont{Borondo}},
  \bibinfo{journal}{The Journal of Chemical Physics}
  \textbf{\bibinfo{volume}{139}}, \bibinfo{eid}{194304} (\bibinfo{year}{2013}),
  \urlprefix\url{http://scitation.aip.org/content/aip/journal/jcp/139/19/10.1063/1.4830102}.

\bibitem[{\citenamefont{Revuelta et~al.}(2016)\citenamefont{Revuelta, Benito,
  Borondo, Vergini, and \textit{{Scar Functions, barriers for chemical
  reactivity, and vibrational basis sets}}
  \textnormal{(accepted)}}}]{Revuelta15}
\bibinfo{author}{\bibfnamefont{F.}~\bibnamefont{Revuelta}},
  \bibinfo{author}{\bibfnamefont{R.~M.} \bibnamefont{Benito}},
  \bibinfo{author}{\bibfnamefont{F.}~\bibnamefont{Borondo}},
  \bibinfo{author}{\bibfnamefont{E.}~\bibnamefont{Vergini}}, \bibnamefont{and}
  \bibinfo{author}{\bibnamefont{\textit{{Scar Functions, barriers for chemical
  reactivity, and vibrational basis sets}} \textnormal{(accepted)}}},
  \bibinfo{journal}{J. Phys. Chem. A}  (\bibinfo{year}{2016}).

\bibitem[{\citenamefont{Henderson et~al.}(1992)\citenamefont{Henderson, Lam,
  and Tennyson}}]{Henderson92}
\bibinfo{author}{\bibfnamefont{J.~R.} \bibnamefont{Henderson}},
  \bibinfo{author}{\bibfnamefont{H.~A.} \bibnamefont{Lam}}, \bibnamefont{and}
  \bibinfo{author}{\bibfnamefont{J.}~\bibnamefont{Tennyson}},
  \bibinfo{journal}{J. Chem. Soc.{,} Faraday Trans.}
  \textbf{\bibinfo{volume}{88}}, \bibinfo{pages}{3287} (\bibinfo{year}{1992}),
  \urlprefix\url{http://dx.doi.org/10.1039/FT9928803287}.

\bibitem[{\citenamefont{Casati and Chirikov}(1995)}]{Casati95}
\bibinfo{author}{\bibfnamefont{G.}~\bibnamefont{Casati}} \bibnamefont{and}
  \bibinfo{author}{\bibfnamefont{B.}~\bibnamefont{Chirikov}},
  \emph{\bibinfo{title}{Quantum Chaos: Between Order and Disorder}}
  (\bibinfo{publisher}{Cambridge University Press}, \bibinfo{year}{1995}), ISBN
  \bibinfo{isbn}{052143291X, 9780521031660, 0521031664, 9780521432917,
  9780511599989}.

\bibitem[{\citenamefont{Nockel and Stone}(1997)}]{Nockel97}
\bibinfo{author}{\bibfnamefont{J.~U.} \bibnamefont{Nockel}} \bibnamefont{and}
  \bibinfo{author}{\bibfnamefont{A.~D.} \bibnamefont{Stone}},
  \bibinfo{journal}{Nature} \textbf{\bibinfo{volume}{385}}, \bibinfo{pages}{45}
  (\bibinfo{year}{1997}),
  \urlprefix\url{http://www.nature.com/nature/journal/v385/n6611/abs/385045a0.html}.

\bibitem[{\citenamefont{St?¶ckmann}(2006)}]{Stockmann06}
\bibinfo{author}{\bibfnamefont{H.~J.} \bibnamefont{St?¶ckmann}},
  \emph{\bibinfo{title}{{Quantum chaos: an introduction; 1st rev. version}}}
  (\bibinfo{publisher}{Cambridge Univ.}, \bibinfo{address}{Cambridge},
  \bibinfo{year}{2006}), \urlprefix\url{https://cds.cern.ch/record/941387}.

\bibitem[{\citenamefont{Michel et~al.}(2007)\citenamefont{Michel, Doya,
  Legrand, and Mortessagne}}]{Michel07}
\bibinfo{author}{\bibfnamefont{C.}~\bibnamefont{Michel}},
  \bibinfo{author}{\bibfnamefont{V.}~\bibnamefont{Doya}},
  \bibinfo{author}{\bibfnamefont{O.}~\bibnamefont{Legrand}}, \bibnamefont{and}
  \bibinfo{author}{\bibfnamefont{F.}~\bibnamefont{Mortessagne}},
  \bibinfo{journal}{Phys. Rev. Lett.} \textbf{\bibinfo{volume}{99}},
  \bibinfo{pages}{224101} (\bibinfo{year}{2007}),
  \urlprefix\url{http://link.aps.org/doi/10.1103/PhysRevLett.99.224101}.

\bibitem[{\citenamefont{Lee et~al.}(2002)\citenamefont{Lee, Lee, Chang, Moon,
  Kim, and An}}]{Lee02}
\bibinfo{author}{\bibfnamefont{S.-B.} \bibnamefont{Lee}},
  \bibinfo{author}{\bibfnamefont{J.-H.} \bibnamefont{Lee}},
  \bibinfo{author}{\bibfnamefont{J.-S.} \bibnamefont{Chang}},
  \bibinfo{author}{\bibfnamefont{H.-J.} \bibnamefont{Moon}},
  \bibinfo{author}{\bibfnamefont{S.~W.} \bibnamefont{Kim}}, \bibnamefont{and}
  \bibinfo{author}{\bibfnamefont{K.}~\bibnamefont{An}}, \bibinfo{journal}{Phys.
  Rev. Lett.} \textbf{\bibinfo{volume}{88}}, \bibinfo{pages}{033903}
  (\bibinfo{year}{2002}),
  \urlprefix\url{http://link.aps.org/doi/10.1103/PhysRevLett.88.033903}.

\bibitem[{\citenamefont{Kwak et~al.}(2015)\citenamefont{Kwak, Shin, Moon, Lee,
  Yang, and An}}]{Kwak15}
\bibinfo{author}{\bibfnamefont{H.}~\bibnamefont{Kwak}},
  \bibinfo{author}{\bibfnamefont{Y.}~\bibnamefont{Shin}},
  \bibinfo{author}{\bibfnamefont{S.}~\bibnamefont{Moon}},
  \bibinfo{author}{\bibfnamefont{S.-B.} \bibnamefont{Lee}},
  \bibinfo{author}{\bibfnamefont{J.}~\bibnamefont{Yang}}, \bibnamefont{and}
  \bibinfo{author}{\bibfnamefont{K.}~\bibnamefont{An}}, \bibinfo{journal}{Sci.
  Rep.} \textbf{\bibinfo{volume}{5}}, \bibinfo{pages}{9010}
  (\bibinfo{year}{2015}),
  \urlprefix\url{http://www.nature.com/articles/srep09010}.

\bibitem[{\citenamefont{Wilkinson et~al.}(1996)\citenamefont{Wilkinson,
  Fromhold, Eaves, Sheard, Miura, and Takamasu}}]{Wilkinson96}
\bibinfo{author}{\bibfnamefont{P.~B.} \bibnamefont{Wilkinson}},
  \bibinfo{author}{\bibfnamefont{T.~M.} \bibnamefont{Fromhold}},
  \bibinfo{author}{\bibfnamefont{L.}~\bibnamefont{Eaves}},
  \bibinfo{author}{\bibfnamefont{F.~W.} \bibnamefont{Sheard}},
  \bibinfo{author}{\bibfnamefont{N.}~\bibnamefont{Miura}}, \bibnamefont{and}
  \bibinfo{author}{\bibfnamefont{T.}~\bibnamefont{Takamasu}},
  \bibinfo{journal}{Nature} \textbf{\bibinfo{volume}{380}},
  \bibinfo{pages}{608} (\bibinfo{year}{1996}), ISSN \bibinfo{issn}{0028-0836},
  \urlprefix\url{http://www.nature.com/nature/journal/v380/n6575/abs/380608a0.html}.

\bibitem[{\citenamefont{Huang et~al.}(2009)\citenamefont{Huang, Lai, Ferry,
  Goodnick, and Akis}}]{Huang09}
\bibinfo{author}{\bibfnamefont{L.}~\bibnamefont{Huang}},
  \bibinfo{author}{\bibfnamefont{Y.-C.} \bibnamefont{Lai}},
  \bibinfo{author}{\bibfnamefont{D.~K.} \bibnamefont{Ferry}},
  \bibinfo{author}{\bibfnamefont{S.~M.} \bibnamefont{Goodnick}},
  \bibnamefont{and} \bibinfo{author}{\bibfnamefont{R.}~\bibnamefont{Akis}},
  \bibinfo{journal}{Phys. Rev. Lett.} \textbf{\bibinfo{volume}{103}},
  \bibinfo{pages}{054101} (\bibinfo{year}{2009}),
  \urlprefix\url{http://link.aps.org/doi/10.1103/PhysRevLett.103.054101}.

\bibitem[{\citenamefont{Xu et~al.}(2013)\citenamefont{Xu, Huang, Lai, and
  Grebogi}}]{Xu13}
\bibinfo{author}{\bibfnamefont{H.}~\bibnamefont{Xu}},
  \bibinfo{author}{\bibfnamefont{L.}~\bibnamefont{Huang}},
  \bibinfo{author}{\bibfnamefont{Y.-C.} \bibnamefont{Lai}}, \bibnamefont{and}
  \bibinfo{author}{\bibfnamefont{C.}~\bibnamefont{Grebogi}},
  \bibinfo{journal}{Phys. Rev. Lett.} \textbf{\bibinfo{volume}{110}},
  \bibinfo{pages}{064102} (\bibinfo{year}{2013}),
  \urlprefix\url{http://link.aps.org/doi/10.1103/PhysRevLett.110.064102}.

\bibitem[{\citenamefont{Larson et~al.}(2013)\citenamefont{Larson, Anderson, and
  Altland}}]{lar13}
\bibinfo{author}{\bibfnamefont{J.}~\bibnamefont{Larson}},
  \bibinfo{author}{\bibfnamefont{B.~M.} \bibnamefont{Anderson}},
  \bibnamefont{and} \bibinfo{author}{\bibfnamefont{A.}~\bibnamefont{Altland}},
  \bibinfo{journal}{Phys. Rev. A} \textbf{\bibinfo{volume}{87}},
  \bibinfo{pages}{013624} (\bibinfo{year}{2013}),
  \urlprefix\url{http://link.aps.org/doi/10.1103/PhysRevA.87.013624}.

\bibitem[{\citenamefont{Bogomolny}(1988)}]{Bogomolny88}
\bibinfo{author}{\bibfnamefont{E.~B.} \bibnamefont{Bogomolny}},
  \bibinfo{journal}{Phys. D} \textbf{\bibinfo{volume}{31}},
  \bibinfo{pages}{169} (\bibinfo{year}{1988}), ISSN \bibinfo{issn}{0167-2789},
  \urlprefix\url{http://www.sciencedirect.com/science/article/pii/0167278988900759}.

\bibitem[{\citenamefont{de~Polavieja et~al.}(1994)\citenamefont{de~Polavieja,
  Borondo, and Benito}}]{Polavieja94}
\bibinfo{author}{\bibfnamefont{G.~G.} \bibnamefont{de~Polavieja}},
  \bibinfo{author}{\bibfnamefont{F.}~\bibnamefont{Borondo}}, \bibnamefont{and}
  \bibinfo{author}{\bibfnamefont{R.~M.} \bibnamefont{Benito}},
  \bibinfo{journal}{Phys. Rev. Lett.} \textbf{\bibinfo{volume}{73}},
  \bibinfo{pages}{1613} (\bibinfo{year}{1994}),
  \urlprefix\url{http://link.aps.org/doi/10.1103/PhysRevLett.73.1613}.

\bibitem[{\citenamefont{Berry}(1989)}]{Berry89}
\bibinfo{author}{\bibfnamefont{M.~V.} \bibnamefont{Berry}},
  \bibinfo{journal}{Proc. R. Soc. Lon. A} \textbf{\bibinfo{volume}{243}},
  \bibinfo{pages}{219} (\bibinfo{year}{1989}), ISSN \bibinfo{issn}{1471-2946},
  \urlprefix\url{http://rspa.royalsocietypublishing.org/content/423/1864/219.article-info}.

\bibitem[{\citenamefont{Keating and Prado}(2001)}]{Prado01}
\bibinfo{author}{\bibfnamefont{J.~P.} \bibnamefont{Keating}} \bibnamefont{and}
  \bibinfo{author}{\bibfnamefont{S.~D.} \bibnamefont{Prado}},
  \bibinfo{journal}{Proc. R. Soc. Lon. A} \textbf{\bibinfo{volume}{457}},
  \bibinfo{pages}{1855} (\bibinfo{year}{2001}), ISSN \bibinfo{issn}{1471-2946},
  \urlprefix\url{http://rspa.royalsocietypublishing.org/content/457/2012/1855}.

\bibitem[{\citenamefont{Tomsovic and Heller}(1993)}]{Tomsovic93}
\bibinfo{author}{\bibfnamefont{S.}~\bibnamefont{Tomsovic}} \bibnamefont{and}
  \bibinfo{author}{\bibfnamefont{E.~J.} \bibnamefont{Heller}},
  \bibinfo{journal}{Phys. Rev. Lett.} \textbf{\bibinfo{volume}{70}},
  \bibinfo{pages}{1405} (\bibinfo{year}{1993}),
  \urlprefix\url{http://link.aps.org/doi/10.1103/PhysRevLett.70.1405}.

\bibitem[{\citenamefont{Tomsovic and Lefebvre}(1997)}]{Tomsovic97}
\bibinfo{author}{\bibfnamefont{S.}~\bibnamefont{Tomsovic}} \bibnamefont{and}
  \bibinfo{author}{\bibfnamefont{J.~H.} \bibnamefont{Lefebvre}},
  \bibinfo{journal}{Phys. Rev. Lett.} \textbf{\bibinfo{volume}{79}},
  \bibinfo{pages}{3629} (\bibinfo{year}{1997}),
  \urlprefix\url{http://link.aps.org/doi/10.1103/PhysRevLett.79.3629}.

\bibitem[{\citenamefont{Wisniacki et~al.}(2001)\citenamefont{Wisniacki,
  Borondo, Vergini, and Benito}}]{Wisniacki01}
\bibinfo{author}{\bibfnamefont{D.~A.} \bibnamefont{Wisniacki}},
  \bibinfo{author}{\bibfnamefont{F.}~\bibnamefont{Borondo}},
  \bibinfo{author}{\bibfnamefont{E.}~\bibnamefont{Vergini}}, \bibnamefont{and}
  \bibinfo{author}{\bibfnamefont{R.~M.} \bibnamefont{Benito}},
  \bibinfo{journal}{Phys. Rev. E} \textbf{\bibinfo{volume}{63}},
  \bibinfo{pages}{066220} (\bibinfo{year}{2001}),
  \urlprefix\url{http://link.aps.org/doi/10.1103/PhysRevE.63.066220}.

\bibitem[{\citenamefont{Wisniacki et~al.}(2004)\citenamefont{Wisniacki,
  Vergini, Benito, and Borondo}}]{Wisniacki04}
\bibinfo{author}{\bibfnamefont{D.~A.} \bibnamefont{Wisniacki}},
  \bibinfo{author}{\bibfnamefont{E.}~\bibnamefont{Vergini}},
  \bibinfo{author}{\bibfnamefont{R.~M.} \bibnamefont{Benito}},
  \bibnamefont{and} \bibinfo{author}{\bibfnamefont{F.}~\bibnamefont{Borondo}},
  \bibinfo{journal}{Phys. Rev. E} \textbf{\bibinfo{volume}{70}},
  \bibinfo{pages}{035202} (\bibinfo{year}{2004}),
  \urlprefix\url{http://link.aps.org/doi/10.1103/PhysRevE.70.035202}.

\bibitem[{\citenamefont{Wisniacki et~al.}(2005)\citenamefont{Wisniacki,
  Vergini, Benito, and Borondo}}]{Wisniacki05}
\bibinfo{author}{\bibfnamefont{D.~A.} \bibnamefont{Wisniacki}},
  \bibinfo{author}{\bibfnamefont{E.}~\bibnamefont{Vergini}},
  \bibinfo{author}{\bibfnamefont{R.~M.} \bibnamefont{Benito}},
  \bibnamefont{and} \bibinfo{author}{\bibfnamefont{F.}~\bibnamefont{Borondo}},
  \bibinfo{journal}{Phys. Rev. Lett.} \textbf{\bibinfo{volume}{94}},
  \bibinfo{pages}{054101} (\bibinfo{year}{2005}),
  \urlprefix\url{http://link.aps.org/doi/10.1103/PhysRevLett.94.054101}.

\bibitem[{\citenamefont{Wisniacki et~al.}(2006)\citenamefont{Wisniacki,
  Vergini, Benito, and Borondo}}]{Wisniacki06}
\bibinfo{author}{\bibfnamefont{D.~A.} \bibnamefont{Wisniacki}},
  \bibinfo{author}{\bibfnamefont{E.}~\bibnamefont{Vergini}},
  \bibinfo{author}{\bibfnamefont{R.~M.} \bibnamefont{Benito}},
  \bibnamefont{and} \bibinfo{author}{\bibfnamefont{F.}~\bibnamefont{Borondo}},
  \bibinfo{journal}{Phys. Rev. Lett.} \textbf{\bibinfo{volume}{97}},
  \bibinfo{pages}{094101} (\bibinfo{year}{2006}),
  \urlprefix\url{http://link.aps.org/doi/10.1103/PhysRevLett.97.094101}.

\bibitem[{\citenamefont{Wisniacki and Carlo}(2008)}]{Wisniacki08}
\bibinfo{author}{\bibfnamefont{D.}~\bibnamefont{Wisniacki}} \bibnamefont{and}
  \bibinfo{author}{\bibfnamefont{G.~G.} \bibnamefont{Carlo}},
  \bibinfo{journal}{Phys. Rev. E} \textbf{\bibinfo{volume}{77}},
  \bibinfo{pages}{045201} (\bibinfo{year}{2008}),
  \urlprefix\url{http://link.aps.org/doi/10.1103/PhysRevE.77.045201}.

\bibitem[{\citenamefont{Novaes et~al.}(2009)\citenamefont{Novaes, Pedrosa,
  Wisniacki, Carlo, and Keating}}]{Novaes09}
\bibinfo{author}{\bibfnamefont{M.}~\bibnamefont{Novaes}},
  \bibinfo{author}{\bibfnamefont{J.~M.} \bibnamefont{Pedrosa}},
  \bibinfo{author}{\bibfnamefont{D.}~\bibnamefont{Wisniacki}},
  \bibinfo{author}{\bibfnamefont{G.~G.} \bibnamefont{Carlo}}, \bibnamefont{and}
  \bibinfo{author}{\bibfnamefont{J.~P.} \bibnamefont{Keating}},
  \bibinfo{journal}{Phys. Rev. E} \textbf{\bibinfo{volume}{80}},
  \bibinfo{pages}{035202} (\bibinfo{year}{2009}),
  \urlprefix\url{http://link.aps.org/doi/10.1103/PhysRevE.80.035202}.

\bibitem[{\citenamefont{Vergini}(2000)}]{Vergini00a}
\bibinfo{author}{\bibfnamefont{E.~G.} \bibnamefont{Vergini}},
  \bibinfo{journal}{J. Phys. A} \textbf{\bibinfo{volume}{33}},
  \bibinfo{pages}{4709} (\bibinfo{year}{2000}),
  \urlprefix\url{http://stacks.iop.org/0305-4470/33/i=25/a=311}.

\bibitem[{\citenamefont{Vergini and Carlo}(2000)}]{Vergini00b}
\bibinfo{author}{\bibfnamefont{E.~G.} \bibnamefont{Vergini}} \bibnamefont{and}
  \bibinfo{author}{\bibfnamefont{G.~G.} \bibnamefont{Carlo}},
  \bibinfo{journal}{J. Phys. A} \textbf{\bibinfo{volume}{33}},
  \bibinfo{pages}{4717} (\bibinfo{year}{2000}),
  \urlprefix\url{http://stacks.iop.org/0305-4470/33/i=25/a=312}.

\bibitem[{\citenamefont{Vergini and Carlo}(2001)}]{Vergini01}
\bibinfo{author}{\bibfnamefont{E.~G.} \bibnamefont{Vergini}} \bibnamefont{and}
  \bibinfo{author}{\bibfnamefont{G.~G.} \bibnamefont{Carlo}},
  \bibinfo{journal}{J. Phys. A} \textbf{\bibinfo{volume}{34}},
  \bibinfo{pages}{4525} (\bibinfo{year}{2001}),
  \urlprefix\url{http://stacks.iop.org/0305-4470/34/i=21/a=308}.

\bibitem[{\citenamefont{Sibert~III et~al.}(2008)\citenamefont{Sibert~III,
  Vergini, Benito, and Borondo}}]{Sibert08}
\bibinfo{author}{\bibfnamefont{E.~L.} \bibnamefont{Sibert~III}},
  \bibinfo{author}{\bibfnamefont{E.}~\bibnamefont{Vergini}},
  \bibinfo{author}{\bibfnamefont{R.~M.} \bibnamefont{Benito}},
  \bibnamefont{and} \bibinfo{author}{\bibfnamefont{F.}~\bibnamefont{Borondo}},
  \bibinfo{journal}{New J. Phys.} \textbf{\bibinfo{volume}{10}},
  \bibinfo{pages}{053016} (\bibinfo{year}{2008}),
  \urlprefix\url{http://stacks.iop.org/1367-2630/10/i=5/a=053016}.

\bibitem[{\citenamefont{Revuelta et~al.}(2012)\citenamefont{Revuelta, Vergini,
  Benito, and Borondo}}]{Revuelta12}
\bibinfo{author}{\bibfnamefont{F.}~\bibnamefont{Revuelta}},
  \bibinfo{author}{\bibfnamefont{E.~G.} \bibnamefont{Vergini}},
  \bibinfo{author}{\bibfnamefont{R.~M.} \bibnamefont{Benito}},
  \bibnamefont{and} \bibinfo{author}{\bibfnamefont{F.}~\bibnamefont{Borondo}},
  \bibinfo{journal}{Phys. Rev. E} \textbf{\bibinfo{volume}{85}},
  \bibinfo{pages}{026214} (\bibinfo{year}{2012}),
  \urlprefix\url{http://link.aps.org/doi/10.1103/PhysRevE.85.026214}.

\bibitem[{\citenamefont{Vagov et~al.}(2009)\citenamefont{Vagov, Schomerus, and
  Zalipaev}}]{Vagov09}
\bibinfo{author}{\bibfnamefont{A.}~\bibnamefont{Vagov}},
  \bibinfo{author}{\bibfnamefont{H.}~\bibnamefont{Schomerus}},
  \bibnamefont{and} \bibinfo{author}{\bibfnamefont{V.~V.}
  \bibnamefont{Zalipaev}}, \bibinfo{journal}{Phys. Rev. E}
  \textbf{\bibinfo{volume}{80}}, \bibinfo{pages}{056202}
  (\bibinfo{year}{2009}),
  \urlprefix\url{http://link.aps.org/doi/10.1103/PhysRevE.80.056202}.

\bibitem[{\citenamefont{Revuelta et~al.}(2013)\citenamefont{Revuelta, Benito,
  Borondo, and Vergini}}]{Revuelta13}
\bibinfo{author}{\bibfnamefont{F.}~\bibnamefont{Revuelta}},
  \bibinfo{author}{\bibfnamefont{R.~M.} \bibnamefont{Benito}},
  \bibinfo{author}{\bibfnamefont{F.}~\bibnamefont{Borondo}}, \bibnamefont{and}
  \bibinfo{author}{\bibfnamefont{E.}~\bibnamefont{Vergini}},
  \bibinfo{journal}{Phys. Rev. E} \textbf{\bibinfo{volume}{87}},
  \bibinfo{pages}{042921} (\bibinfo{year}{2013}),
  \urlprefix\url{http://link.aps.org/doi/10.1103/PhysRevE.87.042921}.

\bibitem[{\citenamefont{Garc\'{\i}a-M\"uller
  et~al.}(2008)\citenamefont{Garc\'{\i}a-M\"uller, Borondo, Hernandez, and
  Benito}}]{GM08}
\bibinfo{author}{\bibfnamefont{P.~L.} \bibnamefont{Garc\'{\i}a-M\"uller}},
  \bibinfo{author}{\bibfnamefont{F.}~\bibnamefont{Borondo}},
  \bibinfo{author}{\bibfnamefont{R.}~\bibnamefont{Hernandez}},
  \bibnamefont{and} \bibinfo{author}{\bibfnamefont{R.~M.}
  \bibnamefont{Benito}}, \bibinfo{journal}{Phys. Rev. Lett.}
  \textbf{\bibinfo{volume}{101}}, \bibinfo{pages}{178302}
  (\bibinfo{year}{2008}),
  \urlprefix\url{http://link.aps.org/doi/10.1103/PhysRevLett.101.178302}.

\bibitem[{\citenamefont{Murgida et~al.}(2010)\citenamefont{Murgida, Wisniacki,
  Tamborenea, and Borondo}}]{Murgida10}
\bibinfo{author}{\bibfnamefont{G.~E.} \bibnamefont{Murgida}},
  \bibinfo{author}{\bibfnamefont{D.~A.} \bibnamefont{Wisniacki}},
  \bibinfo{author}{\bibfnamefont{P.~I.} \bibnamefont{Tamborenea}},
  \bibnamefont{and} \bibinfo{author}{\bibfnamefont{F.}~\bibnamefont{Borondo}},
  \bibinfo{journal}{Chem. Phys. Lett.} \textbf{\bibinfo{volume}{496}},
  \bibinfo{pages}{356 } (\bibinfo{year}{2010}), ISSN \bibinfo{issn}{0009-2614},
  \urlprefix\url{http://www.sciencedirect.com/science/article/pii/S0009261410009899}.

\bibitem[{\citenamefont{Garcia-Muller et~al.}(2014)\citenamefont{Garcia-Muller,
  Hernandez, Benito, and Borondo}}]{GM14}
\bibinfo{author}{\bibfnamefont{P.~L.} \bibnamefont{Garcia-Muller}},
  \bibinfo{author}{\bibfnamefont{R.}~\bibnamefont{Hernandez}},
  \bibinfo{author}{\bibfnamefont{R.~M.} \bibnamefont{Benito}},
  \bibnamefont{and} \bibinfo{author}{\bibfnamefont{F.}~\bibnamefont{Borondo}},
  \bibinfo{journal}{J. Chem. Phys.} \textbf{\bibinfo{volume}{141}},
  \bibinfo{pages}{074312} (\bibinfo{year}{2014}),
  \urlprefix\url{http://scitation.aip.org/content/aip/journal/jcp/141/7/10.1063/1.4892921}.

\bibitem[{\citenamefont{Murgida et~al.}(2015)\citenamefont{Murgida, Arranz, and
  Borondo}}]{Murgida15}
\bibinfo{author}{\bibfnamefont{G.~E.} \bibnamefont{Murgida}},
  \bibinfo{author}{\bibfnamefont{F.~J.} \bibnamefont{Arranz}},
  \bibnamefont{and} \bibinfo{author}{\bibfnamefont{F.}~\bibnamefont{Borondo}},
  \bibinfo{journal}{J. Chem. Phys.} \textbf{\bibinfo{volume}{143}},
  \bibinfo{pages}{214305} (\bibinfo{year}{2015}),
  \urlprefix\url{http://scitation.aip.org/content/aip/journal/jcp/143/21/10.1063/1.4936424}.

\bibitem[{\citenamefont{Essers et~al.}(1982)\citenamefont{Essers, Tennyson, and
  Wormer}}]{Essers82}
\bibinfo{author}{\bibfnamefont{R.}~\bibnamefont{Essers}},
  \bibinfo{author}{\bibfnamefont{J.}~\bibnamefont{Tennyson}}, \bibnamefont{and}
  \bibinfo{author}{\bibfnamefont{P.~E.~S.} \bibnamefont{Wormer}},
  \bibinfo{journal}{Chem. Phys. Lett.} \textbf{\bibinfo{volume}{89}},
  \bibinfo{pages}{223 } (\bibinfo{year}{1982}), ISSN \bibinfo{issn}{0009-2614},
  \urlprefix\url{http://www.sciencedirect.com/science/article/pii/0009261482800468}.

\bibitem[{\citenamefont{Lichtenberg and Lieberman}(2010)}]{LL10}
\bibinfo{author}{\bibfnamefont{A.~J.} \bibnamefont{Lichtenberg}}
  \bibnamefont{and} \bibinfo{author}{\bibfnamefont{M.~A.}
  \bibnamefont{Lieberman}}, \emph{\bibinfo{title}{Regular and chaotic
  dynamics}}, Applied Mathematical Sciences (\bibinfo{publisher}{Springer},
  \bibinfo{address}{New York, Berlin, Heidelberg}, \bibinfo{year}{2010}), ISBN
  \bibinfo{isbn}{3-540-97745-7},
  \urlprefix\url{http://opac.inria.fr/record=b1120113}.

\bibitem[{\citenamefont{Shampine and Gordon}(1975)}]{Shampine75}
\bibinfo{author}{\bibfnamefont{L.~F.} \bibnamefont{Shampine}} \bibnamefont{and}
  \bibinfo{author}{\bibfnamefont{M.~K.} \bibnamefont{Gordon}},
  \emph{\bibinfo{title}{Computer solution of ordinary differential equations :
  the initial value problem}} (\bibinfo{publisher}{W. H. Freeman},
  \bibinfo{address}{W. H. Freeman / San Francisco : USA},
  \bibinfo{year}{1975}), ISBN \bibinfo{isbn}{07-167-0461-7}.

\bibitem[{\citenamefont{Arnolʹd et~al.}(1989)\citenamefont{Arnolʹd, Vogtmann,
  and Weinstein}}]{Arnold78}
\bibinfo{author}{\bibfnamefont{V.~I.} \bibnamefont{Arnolʹd}},
  \bibinfo{author}{\bibfnamefont{K.}~\bibnamefont{Vogtmann}}, \bibnamefont{and}
  \bibinfo{author}{\bibfnamefont{A.}~\bibnamefont{Weinstein}},
  \emph{\bibinfo{title}{Mathematical methods of classical mechanics}}, Graduate
  texts in mathematics (\bibinfo{publisher}{Springer-Verlag},
  \bibinfo{address}{New York}, \bibinfo{year}{1989}), ISBN
  \bibinfo{isbn}{978-0-387-96890-2}.

\bibitem[{\citenamefont{Borondo et~al.}(1995)\citenamefont{Borondo, Zembekov,
  and Benito}}]{Zembekov95}
\bibinfo{author}{\bibfnamefont{F.}~\bibnamefont{Borondo}},
  \bibinfo{author}{\bibfnamefont{A.~A.} \bibnamefont{Zembekov}},
  \bibnamefont{and} \bibinfo{author}{\bibfnamefont{R.~M.}
  \bibnamefont{Benito}}, \bibinfo{journal}{Chem. Phys. Lett.}
  \textbf{\bibinfo{volume}{246}}, \bibinfo{pages}{421 } (\bibinfo{year}{1995}),
  ISSN \bibinfo{issn}{0009-2614},
  \urlprefix\url{http://www.sciencedirect.com/science/article/pii/000926149501147X}.

\bibitem[{\citenamefont{Borondo et~al.}(1996)\citenamefont{Borondo, Zembekov,
  and Benito}}]{Zembekov96}
\bibinfo{author}{\bibfnamefont{F.}~\bibnamefont{Borondo}},
  \bibinfo{author}{\bibfnamefont{A.~A.} \bibnamefont{Zembekov}},
  \bibnamefont{and} \bibinfo{author}{\bibfnamefont{R.~M.}
  \bibnamefont{Benito}}, \bibinfo{journal}{J. Chem. Phys.}
  \textbf{\bibinfo{volume}{105}}, \bibinfo{pages}{5068} (\bibinfo{year}{1996}),
  \urlprefix\url{http://scitation.aip.org/content/aip/journal/jcp/105/12/10.1063/1.472351}.

\bibitem[{\citenamefont{Zembekov et~al.}(1997)\citenamefont{Zembekov, Borondo,
  and Benito}}]{Zembekov97}
\bibinfo{author}{\bibfnamefont{A.~A.} \bibnamefont{Zembekov}},
  \bibinfo{author}{\bibfnamefont{F.}~\bibnamefont{Borondo}}, \bibnamefont{and}
  \bibinfo{author}{\bibfnamefont{R.~M.} \bibnamefont{Benito}},
  \bibinfo{journal}{J. Chem. Phys.} \textbf{\bibinfo{volume}{107}},
  \bibinfo{pages}{7934} (\bibinfo{year}{1997}),
  \urlprefix\url{http://scitation.aip.org/content/aip/journal/jcp/107/19/10.1063/1.475147}.

\bibitem[{\citenamefont{Heller}(1976)}]{Heller76}
\bibinfo{author}{\bibfnamefont{E.~J.} \bibnamefont{Heller}},
  \bibinfo{journal}{J. Chem. Phys.} \textbf{\bibinfo{volume}{65}},
  \bibinfo{pages}{4979} (\bibinfo{year}{1976}),
  \urlprefix\url{http://scitation.aip.org/content/aip/journal/jcp/65/11/10.1063/1.432974}.

\bibitem[{\citenamefont{Littlejohn}(1986)}]{Littlejohn86}
\bibinfo{author}{\bibfnamefont{R.~G.} \bibnamefont{Littlejohn}},
  \bibinfo{journal}{Phys. Rep.} \textbf{\bibinfo{volume}{138}},
  \bibinfo{pages}{193 } (\bibinfo{year}{1986}), ISSN \bibinfo{issn}{0370-1573},
  \urlprefix\url{http://www.sciencedirect.com/science/article/pii/0370157386901031}.

\bibitem[{\citenamefont{Eckhardt and Wintgen}(1991)}]{Eckhardt91}
\bibinfo{author}{\bibfnamefont{B.}~\bibnamefont{Eckhardt}} \bibnamefont{and}
  \bibinfo{author}{\bibfnamefont{D.}~\bibnamefont{Wintgen}},
  \bibinfo{journal}{J. Phys. A} \textbf{\bibinfo{volume}{24}},
  \bibinfo{pages}{4335} (\bibinfo{year}{1991}),
  \urlprefix\url{http://stacks.iop.org/0305-4470/24/i=18/a=020}.

\bibitem[{\citenamefont{Maslov and Fedoriuk}(1991)}]{Maslov91}
\bibinfo{author}{\bibfnamefont{V.~P.} \bibnamefont{Maslov}} \bibnamefont{and}
  \bibinfo{author}{\bibfnamefont{M.~V.} \bibnamefont{Fedoriuk}},
  \emph{\bibinfo{title}{Semi--Classical Approximation in Quantum Mechanics.}},
  Mathematical Physics and Applied Mathematics (\bibinfo{publisher}{D. Reidel
  Publishing Company}, \bibinfo{address}{Dordrecht: Holland / Boston: USA},
  \bibinfo{year}{1991}), ISBN \bibinfo{isbn}{90-277-1219-0}.

\bibitem[{\citenamefont{Creagh et~al.}(1990)\citenamefont{Creagh, Robbins, and
  Littlejohn}}]{Creagh90}
\bibinfo{author}{\bibfnamefont{S.~C.} \bibnamefont{Creagh}},
  \bibinfo{author}{\bibfnamefont{J.~M.} \bibnamefont{Robbins}},
  \bibnamefont{and} \bibinfo{author}{\bibfnamefont{R.~G.}
  \bibnamefont{Littlejohn}}, \bibinfo{journal}{Phys. Rev. A}
  \textbf{\bibinfo{volume}{42}}, \bibinfo{pages}{1907} (\bibinfo{year}{1990}),
  \urlprefix\url{http://link.aps.org/doi/10.1103/PhysRevA.42.1907}.

\bibitem[{\citenamefont{Robbins}(1991)}]{Robbins91}
\bibinfo{author}{\bibfnamefont{J.~M.} \bibnamefont{Robbins}},
  \bibinfo{journal}{Nonlinearity} \textbf{\bibinfo{volume}{4}},
  \bibinfo{pages}{343} (\bibinfo{year}{1991}),
  \urlprefix\url{http://stacks.iop.org/0951-7715/4/i=2/a=007}.

\bibitem[{\citenamefont{Vergini et~al.}(2008)\citenamefont{Vergini, Schneider,
  and Rivas}}]{Vergini08}
\bibinfo{author}{\bibfnamefont{E.~G.} \bibnamefont{Vergini}},
  \bibinfo{author}{\bibfnamefont{D.}~\bibnamefont{Schneider}},
  \bibnamefont{and} \bibinfo{author}{\bibfnamefont{A.~M.~F.}
  \bibnamefont{Rivas}}, \bibinfo{journal}{J. Phys. A}
  \textbf{\bibinfo{volume}{41}}, \bibinfo{pages}{405102}
  (\bibinfo{year}{2008}),
  \urlprefix\url{http://stacks.iop.org/1751-8121/41/i=40/a=405102}.

\bibitem[{\citenamefont{Sparks and Johnson}(2006)}]{Sparks06}
\bibinfo{author}{\bibfnamefont{D.~K.} \bibnamefont{Sparks}} \bibnamefont{and}
  \bibinfo{author}{\bibfnamefont{B.~R.} \bibnamefont{Johnson}},
  \bibinfo{journal}{J. Chem. Phys.} \textbf{\bibinfo{volume}{125}},
  \bibinfo{pages}{114104} (\bibinfo{year}{2006}),
  \urlprefix\url{http://scitation.aip.org/content/aip/journal/jcp/125/11/10.1063/1.2338318}.

\bibitem[{\citenamefont{Lang}(2002)}]{Lang02}
\bibinfo{author}{\bibfnamefont{S.}~\bibnamefont{Lang}},
  \emph{\bibinfo{title}{Algebra}} (\bibinfo{publisher}{Springer},
  \bibinfo{year}{2002}), ISBN \bibinfo{isbn}{978-1-4613-0041-0}.

\bibitem[{\citenamefont{Press and Metcalf}(1996)}]{NR96}
\bibinfo{author}{\bibfnamefont{W.~H.} \bibnamefont{Press}} \bibnamefont{and}
  \bibinfo{author}{\bibfnamefont{M.}~\bibnamefont{Metcalf}},
  \emph{\bibinfo{title}{Numerical recipes in Fortran 90 : the art of parallel
  scientific computing : volume 2 of Fortran numerical recipes}}, Fortran
  numerical recipes (\bibinfo{publisher}{Cambridge university press},
  \bibinfo{address}{Cambridge}, \bibinfo{year}{1996}), ISBN
  \bibinfo{isbn}{0-521-57440-4}, \bibinfo{note}{edition originale : 1986},
  \urlprefix\url{http://opac.inria.fr/record=b1120138}.

\bibitem[{\citenamefont{Arranz et~al.}(2010{\natexlab{b}})\citenamefont{Arranz,
  Safi, Benito, and Borondo}}]{Arranz10b}
\bibinfo{author}{\bibfnamefont{F.~J.} \bibnamefont{Arranz}},
  \bibinfo{author}{\bibfnamefont{Z.~S.} \bibnamefont{Safi}},
  \bibinfo{author}{\bibfnamefont{R.~M.} \bibnamefont{Benito}},
  \bibnamefont{and} \bibinfo{author}{\bibfnamefont{F.}~\bibnamefont{Borondo}},
  \bibinfo{journal}{The European Physical Journal D}
  \textbf{\bibinfo{volume}{60}}, \bibinfo{pages}{279}
  (\bibinfo{year}{2010}{\natexlab{b}}),
  \urlprefix\url{http://link.springer.com/article/10.1140/epjd/e2010-00228-y}.

\bibitem[{\citenamefont{Ba{\u c}i\'{c}}(1991)}]{Bacic91}
\bibinfo{author}{\bibfnamefont{Z.}~\bibnamefont{Ba{\u c}i\'{c}}},
  \bibinfo{journal}{J. Chem. Phys.} \textbf{\bibinfo{volume}{95}},
  \bibinfo{pages}{3456} (\bibinfo{year}{1991}),
  \urlprefix\url{http://scitation.aip.org/content/aip/journal/jcp/95/5/10.1063/1.461798}.

\bibitem[{\citenamefont{Ba{\u c}i\'{c} et~al.}(1988)\citenamefont{Ba{\u
  c}i\'{c}, Watt, and Light}}]{Bacic88}
\bibinfo{author}{\bibfnamefont{Z.}~\bibnamefont{Ba{\u c}i\'{c}}},
  \bibinfo{author}{\bibfnamefont{D.}~\bibnamefont{Watt}}, \bibnamefont{and}
  \bibinfo{author}{\bibfnamefont{J.~C.} \bibnamefont{Light}},
  \bibinfo{journal}{The Journal of Chemical Physics}
  \textbf{\bibinfo{volume}{89}}, \bibinfo{pages}{947} (\bibinfo{year}{1988}),
  \urlprefix\url{http://scitation.aip.org/content/aip/journal/jcp/89/2/10.1063/1.455163}.

\bibitem[{\citenamefont{Bramley et~al.}(1994)\citenamefont{Bramley, Tromp,
  Carrington, and Corey}}]{Bramley94}
\bibinfo{author}{\bibfnamefont{M.~J.} \bibnamefont{Bramley}},
  \bibinfo{author}{\bibfnamefont{J.~W.} \bibnamefont{Tromp}},
  \bibinfo{author}{\bibfnamefont{T.}~\bibnamefont{Carrington}},
  \bibnamefont{and} \bibinfo{author}{\bibfnamefont{G.~C.} \bibnamefont{Corey}},
  \bibinfo{journal}{J. Chem. Phys.} \textbf{\bibinfo{volume}{100}},
  \bibinfo{pages}{6175} (\bibinfo{year}{1994}),
  \urlprefix\url{http://scitation.aip.org/content/aip/journal/jcp/100/9/10.1063/1.467273}.

\bibitem[{\citenamefont{Ma et~al.}(1999)\citenamefont{Ma, Chen, and
  Guo}}]{Ma99}
\bibinfo{author}{\bibfnamefont{G.}~\bibnamefont{Ma}},
  \bibinfo{author}{\bibfnamefont{R.}~\bibnamefont{Chen}}, \bibnamefont{and}
  \bibinfo{author}{\bibfnamefont{H.}~\bibnamefont{Guo}}, \bibinfo{journal}{J.
  Chem. Phys.} \textbf{\bibinfo{volume}{110}}, \bibinfo{pages}{8408}
  (\bibinfo{year}{1999}),
  \urlprefix\url{http://scitation.aip.org/content/aip/journal/jcp/110/17/10.1063/1.478749}.

\bibitem[{\citenamefont{Vergini}(2004)}]{Vergini04}
\bibinfo{author}{\bibfnamefont{E.~G.} \bibnamefont{Vergini}},
  \bibinfo{journal}{Journal of Physics A: Mathematical and General}
  \textbf{\bibinfo{volume}{37}}, \bibinfo{pages}{6507} (\bibinfo{year}{2004}),
  \urlprefix\url{http://stacks.iop.org/0305-4470/37/i=25/a=006}.

\end{thebibliography}

\newpage
\section{Supplemental Material}

In this Supplemental Material, we report full details 
of the characteristics of the LiNC/LiCN eigenstates, 
specially on the structure of the corresponding eigenfunctions, 
obtained with our semiclassical basis set of functions highly
localized on periodic orbits (POs). 

As previously discussed,
we first construct a set of 508 wave functions localized 
over~30 POs of the LiNC/LiCN isomerizing system.
From this whole set, our Gram--Schmidt Selective
Method (GSSM) selects the~90 most suited functions,
being~7 of them tube functions and the remaining~83 scar
functions, localized over~21 POs, and we then diagonalize 
the corresponding Hamiltonian matrix. 
The reason for having more scar than tube functions 
is the smaller dispersion of the formers, 
fact that is taken into account in the GSSM actual application.
Using this much reduced basis set, we are able to
reproduce the~66 low--lying eigenenegies and corresponding 
eigenfunctions of the system with great accuracy.

\begin{figure*}
    \centering
    \includegraphics[width=1.6\columnwidth]{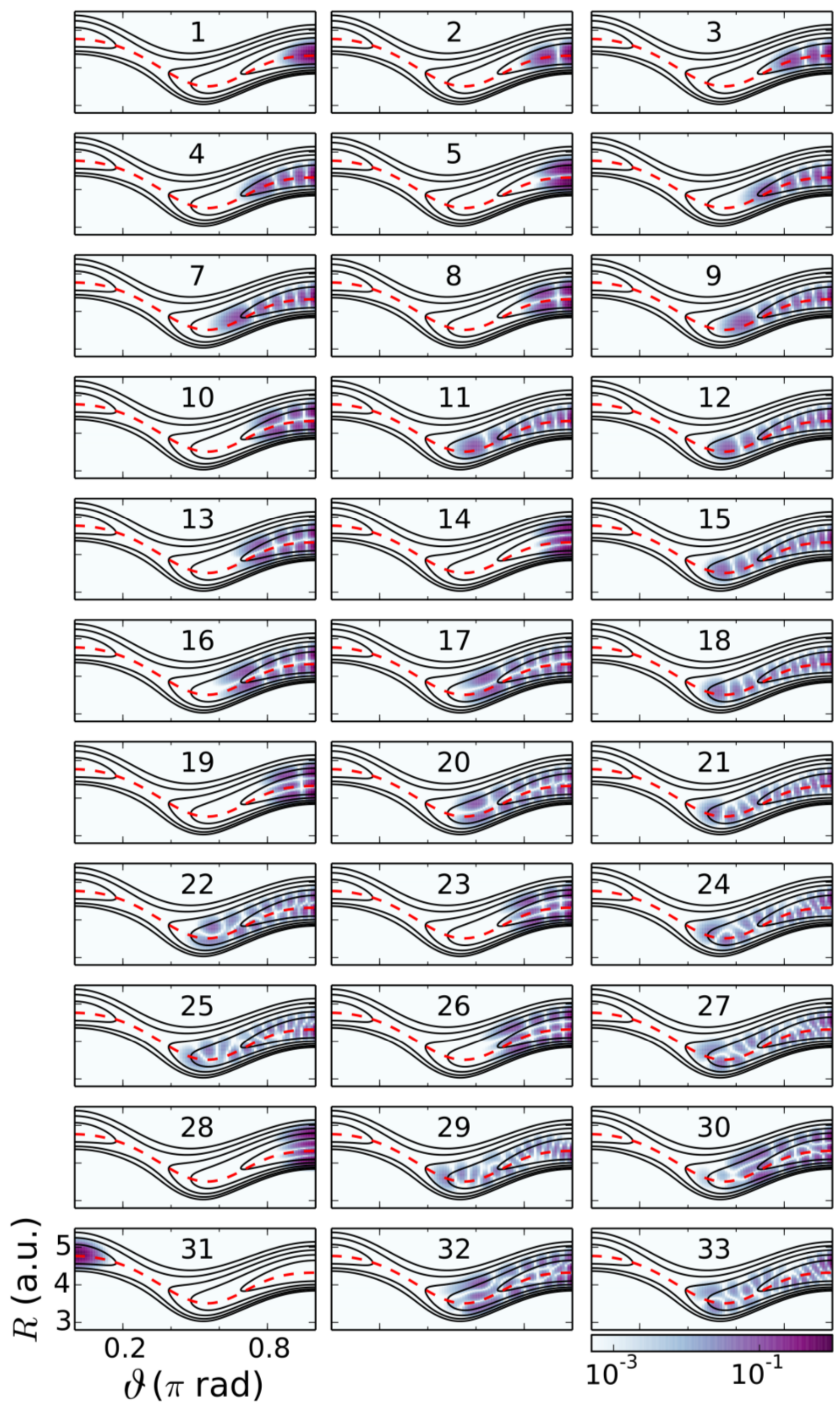}
    \caption{Eigenfunctions~1 to~33 of the LiNC/LiCN molecular isomerizing system.}
    \label{fig.SM1}
 \end{figure*}

\begin{figure*}
    \centering
    \includegraphics[width=1.6 \columnwidth]{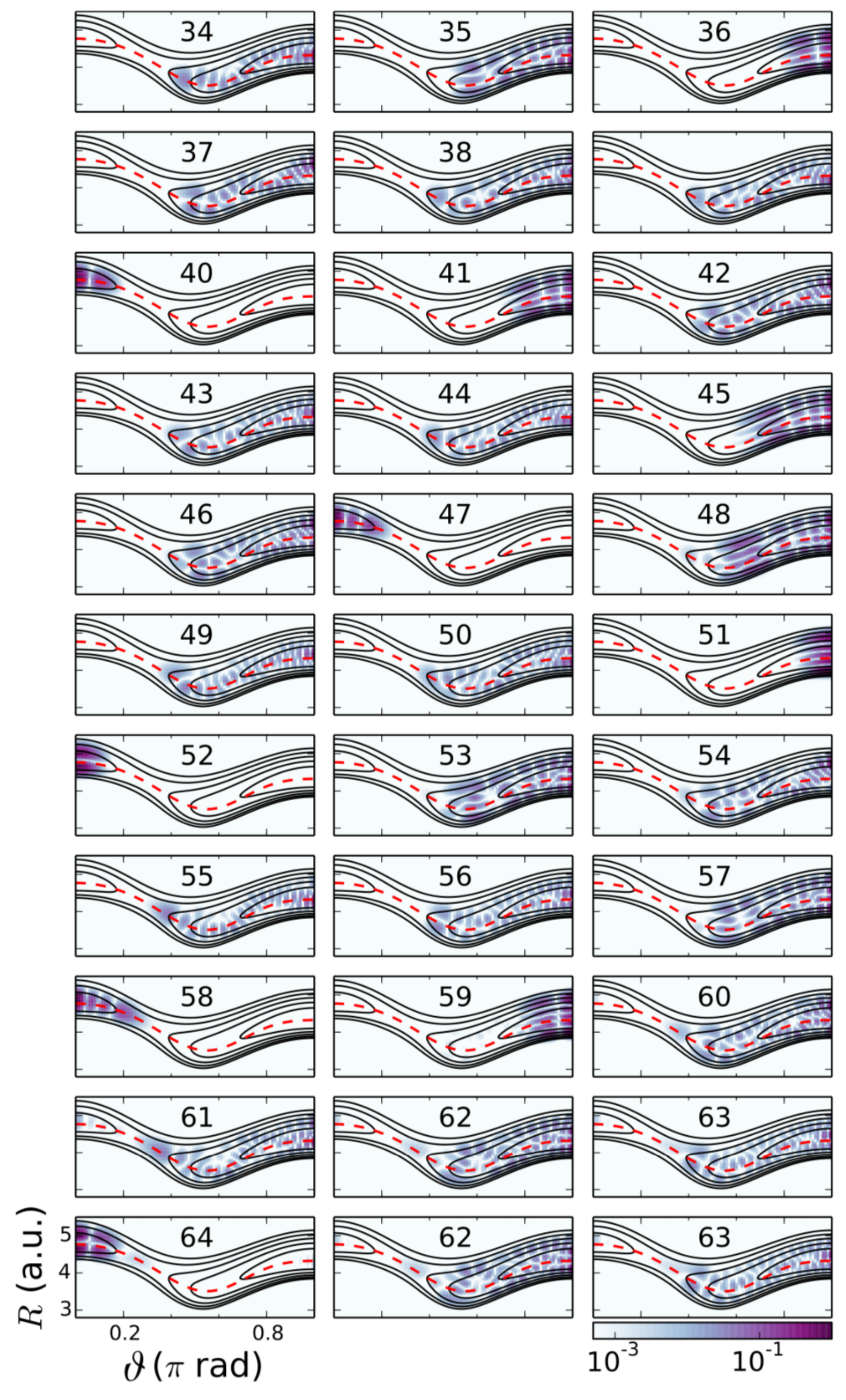}
    \caption{Eigenfunctions~34 to~66 of the LiNC/LiCN molecular isomerizing system.}
    \label{fig.SM2}
\end{figure*}

The structure of the computed eigenfunctions can be 
seen in Figs.~\ref{fig.SM1} and~\ref{fig.SM2}, and the
corresponding details are given in Table~\ref{Tab.I}.
This Table consists of seven columns.
The first two give~$N$ and~$E$, the number and the energy 
of the eigenfunctions obtained using our method.
To check the accuracy of our computation, we present in
the next two columns the reference results~$N'$ and~$E'$,  
obtained using the discrete variable representation (DVR) 
in the~$\vartheta$ coordinate and distributed Gaussian basis 
(DGB) in the radial coordinate $R$ method of
Ba{\u c}i\'{c} and Light~\cite{Bacic86}.
The fifth column shows the overlap (given as a percentage),
$\Sigma' = 100 \langle N \vert N' \rangle^2$,
between states~$\vert N \rangle$ and~$\vert N' \rangle$.
In the sixth column we show the participation ratio,~$R_N$,
of~$\vert N \rangle$ in our semiclassical basis set,
given by the Eq.~(20).
The last column contains the structure of the eigenfunction
in our localized basis set folowing the notation reported in 
the Sec.~II.C. 
Here, we do not only give the PO 
(see Figs.~3 and~4) and the 
quantum number,~$n_i$, fulfilling the Bohr--Sommerfeld 
quantization condition~(4). 
but we also include the percentage of the exact eigenfunction
that is reconstructed using the basis elements
$\vert \textnormal{PO}_1, n_1 \rangle, \ldots,
\vert \textnormal{PO}_i, n_i \rangle$.
In these data we have included all localized states needed 
to reproduce not less than $85 \%$ of the exact eigenfunction.
Notice the small number of localized states necessary 
in all cases for this purpose.
Let us remark that in our actual calculations the 
basis sets consisted of more elements, this
rendering eigenfunctions with the overlaps
also reported as~$\Sigma'$ in the Table.
Recall that the basis elements localized over
stable PO are the tube functions
given by Eq.~(2). 
while for unstable ones the basis elements equal 
the scar function defined in Eq.~(7) of the same reference.

The eigenenergies reported in Table~\ref{Tab.I} are in very 
good agreement with the reference (taken as exact) ones, 
being the errors always smaller than~0.3 times the mean 
level spacing.
Also, the eigenfunctions themselves are
well converged, as it can be seen from the 
overlap with the exact eigenfunction
given by the parameter~$\Sigma'$.
For example, eigenfunction~$\vert 26 \rangle$ has
the smallest value of this overlap, but this
still leads to a quite large value of $\Sigma_{26}' = 92.5\%$.
Also, notice that~89.5\% of this eigenfunction
is reconstructed by combination of only two scar functions, 
$\vert$3A$_{\pi-0}^u$,11$\rangle$ accounting for 72.1 \% 
of the exact eigenfunction, and~$\vert$1AB$_{\pi-1}^u$,24$\rangle$,
which brings an additional 17.4 \% of the eigenfunction~$\vert 26 \rangle$.
Notice that the number of basis elements required in the 
reconstruction of each single eigenfunction increases with energy.
As can be inferred from Table~\ref{Tab.I},
most of the~10 low--lying eigenfunctions are essentially equal to 
one single scar functions. 
For example, the three low--lying eigenfunctions have an overlap of more than
97.9 \% with~$\vert$1A$_{\pi-0}^u$,0$\rangle$,~$\vert$1A$_{\pi-0}^u$,1$\rangle$,
and~$\vert$3A$_{\pi-0}^u$,2$\rangle$, respectively, and, 
as a consequence, they have a participation ratio very close to~1.
Most of the eigenfunctions~$\vert 11 \rangle$ to $\vert 19 \rangle$
have a participation ratio closer to~2, i.e.~only two basis elements
are required for their reconstruction.
For higher energies, an increasing number of basis elements is required 
for the calculation of each eigenfunction. 
Still, the number of basis elements necessary for the computation of 
each eigenfunction remains small in comparison
to other standard methods, such as the DVR~\cite{Bacic86}, 
as discussed above. 

A minor drawback of our method is the following.
The eigenfunctions associated with levels $N=29$ and~30,
which have very close energies 
$E_{29}=2752.93$~cm$^{-1}$ and~$E_{30}=2757.27$~cm$^{-1}$, 
are obtained in the wrong order,
i.e.~our eigenfunction~$\vert 30 \rangle$ has an overlap
of 94.6~\% with~$\vert 29' \rangle$, while
eigenfunction~$\vert 29 \rangle$ has an overlap
of 98.4~\% with~$\vert 30' \rangle$. 
Notice that these overlaps are still very remarkable, while the error in
the energies is smaller than~5~cm$^{-1}$.
%
\begin{table*}
\caption{Structure of the eigenfunctions~$\vert 1 \rangle$ 
to~$\vert 66 \rangle$ of the isomerizing molecular system {\reacLiCN} 
shown in Figs.~\ref{fig.SM1} and~\ref{fig.SM2} obtained with the basis of (tube and scar) 
localized wave functions,  $\vert \OP,n \rangle$.
$N$ is the eigenfunction number, 
$E$ its energy, 
$\Sigma '=100 \langle N' \vert N \rangle^2$ is the overlap between the 
computed eigenfunction, $\vert N \rangle$, and the exact 
one,~$\vert N' \rangle$, whose energy equals $E'$.
$R_N$ is the participation ratio,
%
PO is the stable ($s$) or unstable ($u$) periodic orbit along which the tube or scar function
is respectively constructed with $n$ excitations, 
%
and $\Sigma_{i}$ is the percentage of the (exact) eigenfunction that 
is reconstructed by combination of the localized functions 
$\vert \OP_1,n_1 \rangle$, 
$\vert \OP_2,n_2 \rangle$, 
$\vert \OP_3,n_3 \rangle$, 
$\ldots$, 
$\vert \OP_i,n_i \rangle$. 
}
\label{Tab.I}
\setlength{\tabcolsep}{2.pt}
\begin{center}
\begin{tabular}{|c|c|c|c|c|c|rrrrr|}
\hline
$N$  &  $E$ & $N'$ & $E'$  & $\Sigma '$ & $R_N$ & 
PO{\footnotesize$_1$}, $n${\footnotesize$_1$}, $\Sigma${\footnotesize$_{1}$} &  
PO{\footnotesize$_2$}, $n${\footnotesize$_2$}, $\Sigma${\footnotesize$_{2}$} &  
PO{\footnotesize$_3$}, $n${\footnotesize$_3$}, $\Sigma${\footnotesize$_{3}$} &
PO{\footnotesize$_4$}, $n${\footnotesize$_4$}, $\Sigma${\footnotesize$_{4}$} & 
PO{\footnotesize$_5$}, $n${\footnotesize$_5$}, $\Sigma${\footnotesize$_{5}$}\\ \hline
1   & 512.357 & 1 & 512.436 & 100 & 1.00 & 1A\tLiNCBu, 0, 100      &                                      && &                                    \\ \hline
2   & 759.448 & 2 & 759.669 &100 & 1.04 & 1A\tLiNCBu, 1, 98.1     &                                      & &&                                    \\ \hline
3   & 981.466 & 3 & 981.477 &100 & 1.04 & 3A\tLiNCBu, 2, 97.9     &                                       &&&                                    \\ \hline
4   & 1177.52 & 4 & 1178.09 &100 & 1.87 & 1A\tLiNCBu, 3, 66.2     & S$_\pi$, 1, 97.3              & &&                                    \\ \hline
5   & 1266.81 & 5 & 1266.84 &100 & 1.11 & S$_\pi$, 1, 95.0                 &                                  &     &&                                    \\ \hline
6   & 1349.10 & 6 & 1349.28 &100 & 1.03 & 5A\tLiNCBu, 4, 98.6     &                                       &&&                                     \\ \hline
7   & 1494.67 & 7 & 1494.74 &100 & 1.53 & 5A\tLiNCBu, 5, 78.0     & 5A\tLiNCBu, 4, 99.6   &&  & \\ \hline
8   & 1510.38 & 8 & 1510.52 & 100  & 1.20 & 1A\tLiNCBu, 4, 91.1     &                                       &&&                                      \\ \hline
9   & 1624.09 & 9 & 1624.15 & 100 & 1.62 & 6A\tLiNCBu, 6, 76.4     & 5A\tLiNCBu, 5, 94.0    &&    &                                 \\ \hline
10 & 1718.86 & 10 & 1725.29 & 99.9 & 1.21 & 3A\tLiNCBu, 6, 90.7     &                                       &&&                                      \\ \hline
11 & 1757.71 & 11 & 1757.73 & 100 & 1.63 & 7AB\tLiNCBu, 14, 76.6   & 6A\tLiNCBu, 6, 92.1   & &&                                      \\ \hline
12 & 1902.70 &12 & 1902.71 & 100 & 1.60 & 7AB\tLiNCBu, 16, 77.7   & 7AB\tLiNCBu, 14, 91.9    &&&                                      \\ \hline
13 & 1909.18 & 13 & 1910.16 & 99.9 & 1.41 & 5A\tLiNCBu, 8, 83.5     & 3A\tLiNCBu, 6, 94.8    && &                                     \\ \hline
14 & 2009.86 &14 & 2010.80 & 100 & 1.11 & 3A\tLiNCBu, 6, 94.8       &                                       &&         &                              \\ \hline
15 & 2057.86 & 15 & 2057.87 & 100 & 1.75 & 7AB\tLiNCBu, 18, 73.5    & 7AB\tLiNCBu, 16, 89.6    &&&                                        \\ \hline
16 & 2062.50 &16 & 2062.81 & 100 & 1.54 & 5A\tLiNCBu, 9, 78.9     & 5A\tLiNCBu, 8, 95.7    &     &    &                                \\ \hline
17 & 2182.83 &17 & 2183.91 & 99.9 & 1.49 & 1BA\tLiNCBIu, 21, 81.1 & 5A\tLiNCBu, 9, 91.9    && & \\ \hline
18 & 2220.77 &18 & 2220.81 & 100 & 1.84 & 0\tLiNCBu, 10, 72.2      & 7AB\tLiNCBu, 18, 84.6 &                                         7AB\tLiNCBu, 16, 90.6 &&\\ \hline
19 & 2246.31 &19 & 2251.11 & 99.9 & 1.47 & 1A\tLiNCBu, 7, 80.2     & 1A\tLiNCBu, 6, 98.5                 &&                    &                      \\ \hline
20 & 2298.58 &20 & 2299.02 & 100 & 1.68 & 6A\tLiNCBu, 11, 76.4   & 1BA\tLiNCBIu, 21, 82.9  & 0\tLiNCBu, 11, 89.8  &  &\\ \hline
21 & 2387.19 &21 & 2387.26 & 100 & 1.92 & 0\tLiNCBu, 11, 71.2      & 0\tLiNCBu 10, 79.9 & 6A\tLiNCBu 11, 86.8   & &\\ \hline
22 & 2431.21 & 22 & 2431.66 & 98.6 & 2.37 & 6B\tLiNCBu, 12, 62.5  & 0\tLiNCBu, 11, 76.3      &  6A\tLiNCBu, 11, 86.0&&    \\ \hline
23 & 2439.15 & 23 & 2458.82 &97.3 & 1.39 & 1AB\tLiNCBIu, 24, 83.8 & 1A\tLiNCBu, 7, 93.8     &                                   &&      \\ \hline
24 & 2549.10 &24 & 2549.24 & 100 & 1.91 & 0\tLiNCBu, 12, 70.2      &  6B\tLiNCBu, 12, 86.4   & && \\ \hline
25 & 2583.73 & 25  & 2586.83 & 94.4 & 3.20 & 9AB\tLiNCBu, 26, 44.8  & 0\tLiNCBu, 12, 77.3 &                                        3A\tLiNCBu,11,82.5 & 6A\tLiNCBu, 1, 85.6  & \\ \hline
26 & 2609.17 & 26 & 2630.57 & 92.5 & 1.80 & 3A\tLiNCBu, 11, 72.1    & 1AB\tLiNCBIu, 24, 89.5    &                                  &&        \\ \hline
27 & 2708.51 & 27 & 2708.70 & 99.9 & 1.83 & 0\tLiNCBu, 13, 72.8      & 9AB\tLiNCBu, 26, 84.6    & 0\tLiNCBu, 12, 87.6   &&\\ \hline
28 & 2742.85 & 28 & 2744.26 &  97.2 & 1.14 & 1A\tLiNCs, 9, 93.6     &                                        &                                    &&      \\ \hline
29 & 2752.93 & 30 & 2759.22 & 94.6 & 2.02 & 1AB\tLiNCBIu, 30, 68.2 & 3A\tLiNCBu, 11, 83.4  & 1AB\tLiNCBIu, 24, 88.7  && \\ \hline
30 & 2757.27 & 29 & 2757.39 & 98.4 & 1.87 & 1A\tLiNCBIIu, 13, 72.3  & 
1AB\tLiNCBIu, 30, 77.1 & 1A\tLiNCBIIu,14,80.3 & 9AB\tLiNCBu, 26, 82.7 & 0\tLiNCBIu, 12, 90.7 \\ \hline
31 & 2799.11 & 31 & 2799.23 & 100 & 1.00 & 1A\tLiCNBu, 0, 100       &                                         & &&                                          \\ \hline
32 & 2846.73 & 32 & 2852.86 & 98.1 & 1.82 & 1BA\tLiNCBu, 32, 72.8 & 0\tLiNCBu, 14, 85.0     &   && \\ \hline
33 & 2874.29 & 33 & 2875.56& 98.7 & 2.45 & 0\tLiNCBu, 14, 61.2      & 1BA\tLiNCBIu, 32, 75.9 & 0\tLiNCBu, 13, 85.8  && \\ \hline
\end{tabular}
\end{center}
\end{table*}

%
\begin{table*}
\setlength{\tabcolsep}{2.pt}
\begin{center}
\begin{tabular}{|c|c|c|c|c|c|rrrrr|}
\hline 
\multirow{2}{*}{$N$}  &  \multirow{2}{*}{$E$} & \multirow{2}{*}{$N'$} & \multirow{2}{*}{$E'$} & \multirow{2}{*}{$\Sigma '$} & \multirow{2}{*}{$R_N$} &
PO{\footnotesize$_1$}, $n${\footnotesize$_1$}, $\Sigma${\footnotesize$_{1}$}     &  
PO{\footnotesize$_2$}, $n${\footnotesize$_2$}, $\Sigma${\footnotesize$_{2}$}     &  
PO{\footnotesize$_3$}, $n${\footnotesize$_3$}, $\Sigma${\footnotesize$_{3}$} &
PO{\footnotesize$_4$}, $n${\footnotesize$_4$}, $\Sigma${\footnotesize$_{4}$}         & 
PO{\footnotesize$_5$}, $n${\footnotesize$_5$}, $\Sigma${\footnotesize$_{5}$}\\
  &   &  & &  & &
PO{\footnotesize$_6$}, $n${\footnotesize$_6$}, $\Sigma${\footnotesize$_{6}$}   &
PO{\footnotesize$_7$}, $n${\footnotesize$_7$}, $\Sigma${\footnotesize$_{7}$}    & 
PO{\footnotesize$_8$}, $n${\footnotesize$_8$}, $\Sigma${\footnotesize$_{8}$}    & 
PO{\footnotesize$_9$}, $n${\footnotesize$_9$}, $\Sigma${\footnotesize$_{9}$}  & \\  \hline
34 & 2930.23 & 34 & 2930.61&99.8 & 1.89 & 1A\tLiNCBIIu, 14, 71.3  & 1A\tLiNCBIIu, 13, 84.6 & 8AB\tLiNCBu, 31, 87.1 & &  \\ \hline
35 & 2959.71 & 35 & 2964.73& 99.1 & 1.78 & 1BA\tLiNCBIu, 34, 74.2 & 1BA\tLiNCBIu, 32, 80.2      & 1AB\tLiNCBIu, 30, 86.9 & & \\ \hline
36 & 2981.02 & 36 & 2981.42&99.7 & 1.52 & 1A\tLiCNBu, 10, 79.1      & 1A\tLiNCs, 9, 97.4                 &                                & &          \\ \hline
37 & 3042.84 & 37 & 3043.76& 99.8 & 1.68 & 8AB\tLiNCBu, 30, 76.4   & 0\tLiNCBu, 14, 85.3   & & &  \\ \hline
38 & 3092.41 & 38 & 3094.75 & 99.3 & 2.76 & 8AB\tLiNCBu, 31, 57.3   & 6B\tLiNCBu, 17, 71.7 & 1BA\tLiNCBIu, 34, 82.0 & 8AB\tLiNCBu, 30, 85.4   &   \\ \hline
39 & 3105.52 & 39 & 3106.65 & 99.4 & 4.97 & 6B\tLiNCBu, 17, 31.1    & 8AB\tLiNCBu, 31, 58.7   & 8AB\tLiNCBu, 30, 69.6 & 1A\tLiNCBIIu, 14, 80.4     & 8AB\tLiNCBIIu, 32, 85.1  \\ \hline
40 & 3121.57 & 40 & 3122.27& 100 & 1.32 & 1A\tLiCNBu, 1, 85.8        &                                         && &                                         \\ \hline
41 & 3181.59 & 41 & 3181.91& 100 & 1.49 & 3A\tLiNCBu, 14, 80.8      & 1A\tLiNCBu, 10, 93.5    &   & &                                      \\ \hline
42 & 3205.71 & 42 & 3206.81 & 99.9 & 1.99 & 8AB\tLiNCBu, 32, 70.1        & 8AB\tLiNCBu, 30, 78.4    & 1B\tLiNCBIIu,16,83.1 & 8AB\tLiNCBu, 34, 86.1 & \\ \hline
43 & 3255.06 & 43 & 3256.41 & 99.9 & 2.73 & 8AB\tLiNCBu, 33, 58.0   & 1B\tLiNCBIIu, 16, 70.7  & 8AB\tLiNCBu, 31, 79.5 & 6B\tLiNCBu, 17, 84.1      & 8AB\tLiNCBu, 30, 88.8  \\ \hline
\multirow{2}{*}{44} & \multirow{2}{*}{3273.41} & \multirow{2}{*}{44} & \multirow{2}{*}{3274.29} & \multirow{2}{*}{99.9} & \multirow{2}{*}{4.85} & 1B\tLiNCBIIu, 16, 32.2   & 8AB\tLiNCBu, 33, 61.6 & 2AB\tLiNCBIIIu, 34, 70.7 & 8AB\tLiNCBu, 34, 75.6    & 8AB\tLiNCBu, 31, 79.7        \\
&&&&&&   1A\tLiNCBIIu, 14, 83.0 & 1BA\tLiNCBIu, 41, 85.7    &   &&       \\ \hline
45 & 3331.87 & 45 & 3337.87& 99.8 & 1.50 & 1AB\tLiNCBIu, 39, 80.5      & 3A\tLiNCBu, 14, 92.7    &&&                                          \\ \hline
46 & 3364.22 & 46 & 3366.22 & 99.9 & 2.42 & 8AB\tLiNCBu, 34, 63.0      & 1B\tLiNCBIIu, 17, 71.0  & 8AB\tLiNCBu, 32, 78.4 & 8AB\tLiNCBu, 33, 83.1        & 6B\tLiNCBu, 17, 86.3     \\ \hline
47 & 3409.89 & 47 & 3409.37& 99.9 & 1.58 & 2AB\tLiCNBs, 4, 78.1      & 1A\tLiCNBu, 1, 91.4      &                                          &&\\ \hline
48 & 3429.17 & 48 & 3430.71 & 98.2 & 1.53 & 1BA\tLiNCBIu, 41, 80.5      & 1B\tLiNCBIIu, 17, 84.6    & 6A\tLiNCBu, 14, 87.7    && \\ \hline
49 & 3429.81 & 49 & 3430.86 & 98.0 & 4.53 & 2AB\tLiNCBIIIu, 34, 39.1   & 8AB\tLiNCBu, 34, 50.5    & 1B\tLiNCBIIu,19,71.0 & 8AB\tLiNCBu, 33, 77.4   & 9AB\tLiNCBu, 41, 85.8       \\ \hline
\multirow{2}{*}{50} & \multirow{2}{*}{3440.10} &\multirow{2}{*}{50} & \multirow{2}{*}{3441.45} & \multirow{2}{*}{99.2} & \multirow{2}{*}{6.43} & 1B\tLiNCBIIu, 17, 29.8    & 2AB\tLiNCBIIIu, 34, 51.6  & 8AB\tLiNCBu, 34, 59.2  & 1B\tLiNCBIIu, 16, 64.1    & 8AB\tLiNCBu, 33, 70.2  \\
     &               & && &         & 1BA\tLiNCBIu, 41, 73.6 & 9AB\tLiNCBu, 41, 78.9    & 2AB\tLiNCBIIIu, 36, 81.9  &                                       8AB\tLiNCBIIIu,37,85.1 &\\ \hline
51 & 3466.42 & 51 & 3467.20 & 99.9 & 1.25 & 1A\tLiNCBu, 12, 89.5          &                                  &                      &&                  \\ \hline
52 & 3485.89 & 52 & 3488.32 & 99.9 & 1.27 & 1A\tLiCNBu, 2, 88.3          &                                  &                        &&           \\ \hline
53 & 3507.24 & 53 & 3507.90 & 99.9 & 2.14 & 6B\tLiNCBu, 21, 65.9      & 7AB\tLiNCBu, 39, 82.5    & 8AB\tLiNCBu, 37, 88.5  &&   \\ \hline
\multirow{2}{*}{54} & \multirow{2}{*}{3536.21} &\multirow{2}{*}{54} & \multirow{2}{*}{3537.49} & \multirow{2}{*}{99.9} & \multirow{2}{*}{4.60} & 9AB\tLiNCBu, 41, 40.1      & 6B\tLiNCBu, 21, 59.3    & 1BA\tLiNCBIu, 41, 69.3 & 8AB\tLiNCBu, 34, 76.5    & 2AB\tLiNCBIIIu, 36, 81.4   \\
     &               &  &&     &          & 8AB\tLiNCBu, 33, 83.9 & 1B\tLiNCBIIu, 17, 85.7 &&&  \\ \hline
55 & 3594.78 & 55 & 3595.06 & 99.9 & 2.11 & 2AB\tLiNCBIIIu, 36, 66.7   & 2AB\tLiNCBIIIu, 34, 83.1   & 2AB\tLiNCBIIIu, 38, 85.7  && \\ \hline
56 & 3605.75 & 56 & 3611.91 & 99.0 & 2.01 & 8AB\tLiNCBu, 37, 69.9    & 9AB\tLiNCBu, 41, 74.9   & 1B\tLiNCBIIu, 17, 81.3 & 2AB\tLiNCBIIIu, 34, 83.7   & 8AB\tLiNCBIIIu, 34, 87.1                                   \\ \hline
57 & 3624.39 & 57 & 3625.23 & 99.3 & 1.80 & 7AB\tLiNCBu,39,74.1      & 6B\tLiNCBu, 22, 78.8     & 9AB\tLiNCBu, 41, 82.4  & 1BA\tLiNCBIu, 41, 87.4   &                                     \\ \hline
58 & 3662.20 &58 & 3652.95 & 98.5 & 2.34 & 2AB\tLiCNBs, 6, 61.7          & 2AB\tLiCNBs, 8, 80.1      & 2AB\tLiCNBu, 4, 87.6 &&  \\ \hline
59 & 3699.39 &59 & 3701.29 & 99.3 & 1.65 & 1A\tLiNCBu, 13, 75.8      & 1A\tLiNCBu, 12, 93.2          &                                  &&       \\ \hline
\multirow{2}{*}{60} & \multirow{2}{*}{3701.71} &\multirow{2}{*}{60} &\multirow{2}{*}{3702.71}& \multirow{2}{*}{97.2} & \multirow{2}{*}{5.49} & 6B\tLiNCBu, 22, 30.8      & 7AB\tLiNCBu, 39, 56.6         & 1B\tLiNCBIIu, 19, 66.3 & 1A\tLiNCBu, 13, 72.2       & 2AB\tLiNCBIIIu, 38, 77.9  \\
     &               &    &&   &                & 8AB\tLiNCBu, 37, 80.8 & 2AB\tLiNCBIIIu, 36, 83.6  & 6B\tLiNCBu, 23, 86.4       &&                \\ \hline 
\multirow{2}{*}{61} & \multirow{2}{*}{3735.88} &\multirow{2}{*}{61} &\multirow{2}{*}{3735.84}& \multirow{2}{*}{99.9} & \multirow{2}{*}{3.85} & 2AB\tLiNCBIIIu, 38, 47.1   & 2AB\tLiNCBIIIu, 36, 60.1    & 6B\tLiNCBu, 22, 70.4  &2AB\tLiNCBIIIu,40,77.2    & 7AB\tLiNCBIIIu, 39, 82.9   \\
     &&&          &&    & TS$^u$, 0, 86.3      &&&&    \\ \hline 
62 & 3767.55 & 62 & 3768.14 & 99.8 & 2.43 & 1B\tLiNCBIIu, 19, 63.0     & 2AB\tLiNCBIIIu, 39, 70.4 & 7AB\tLiNCBu, 39, 75.2 & 6B\tLiNCBu, 21, 82.1     & 2AB\tLiNCBIIIu, 40, 85.1 \\ \hline 
63 & 3784.73 & 63 & 3785.71 & 99.7 & 2.24 & 2AB\tLiNCBIIIu, 39, 65.3    & 8AB\tLiNCBu, 37, 76.0 & 6B\tLiNCBu, 23, 83.6 &1B\tLiNCBIIu, 19, 85.0     &  6B\tLiNCBu, 22, 86.3  \\ \hline 
64 & 3815.63 &64 & 3815.40 & 99.7 & 1.47 & 1A\tLiCNBu, 3, 81.8        &  1A\tLiNCBu, 2, 91.8  &      &&  \\ \hline 
\multirow{2}{*}{65} & \multirow{2}{*}{3826.84} & \multirow{2}{*}{65} &\multirow{2}{*}{3823.76}&\multirow{2}{*}{97.9} & \multirow{2}{*}{3.23} & TS$^u$, 0, 50.7    & 2AB\tLiNCBIIIu, 38, 70.6        & 2AB\tLiCNBs, 6, 78.9  &1A\tLiCNBu, 3, 82.2   & 2AB\tLiCNBs, 8, 84.7    \\
     &&&         & && 2AB\tLiNCBIIIu, 39, 87.0  &&&& \\ \hline 
\multirow{2}{*}{66} & \multirow{2}{*}{3866.51} &\multirow{2}{*}{66}&\multirow{2}{*}{3866.79}& \multirow{2}{*}{99.3} & \multirow{2}{*}{2.76} & 6B\tLiNCBu, 23, 57.3 &  2AB\tLiNCBIIIu, 39, 73.8      & 1B\tLiNCBIIu, 19, 77.3 & 8AB\tLiCNBs, 37, 79.2    & 6B\tLiCNBs, 24, 81.1    \\
     &&&         &&&   7AB\tLiNCBu, 39, 82.5 & 66B\tLiNCBu, 22, 85.6          & & &              \\ \hline 
\end{tabular}
\end{center}
\end{table*}
%
 
\end{document}